\documentclass[graybox, nosecnum]{svmult}

\usepackage{a4wide}

\usepackage{mathptmx}       
\usepackage{helvet}         
\usepackage{courier}        
\usepackage{type1cm}        
\usepackage{multirow}
\usepackage{tabularx}
\usepackage{array}
\usepackage{siunitx}
\usepackage{makeidx}         
\usepackage{graphicx}        
\usepackage{caption}
\usepackage{subcaption}
\usepackage{multicol}        
\usepackage[bottom]{footmisc}
\usepackage{hyperref}        
\usepackage{soul}            
\usepackage{chemfig}         
\hypersetup{colorlinks=true,urlcolor=blue}
\usepackage[square,numbers]{natbib}
\usepackage{aas_macros}     

\makeindex             

\begin{document}
\title*{Silicon Pore Optics}
\author{
Nicolas M. Barri{\`e}re \thanks{corresponding author},
Marcos Bavdaz,
Maximilien J. Collon,
Ivo Ferreira,
David Girou,
Boris Landgraf, and
Giuseppe Vacanti
}
\authorrunning{Barri{\`e}re, Bavdaz, Collon, Ferreira, Girou, Landgraf, and Vacanti}

\institute{
N.M. Barri{\`e}re, M. Collon, D. Girou, B. Landgraf and G. Vacanti
at cosine, Warmonderweg 14, 2171 AH Sassenheim, The Netherlands,
\email{f.lastname@cosine.nl}, and
M. Bavdaz and I. Ferreira
at European Space Agency, ESTEC, Keplerlaan 1, PO Box 299, 2200 AG Noordwijk, The Netherlands, \email{ivo.ferreira@esa.int}
}
%
%
\maketitle
\abstract{
Silicon Pore Optics (SPO) uses commercially available monocrystalline double-sided super-polished silicon wafers as a basis to produce mirrors that form lightweight high-resolution X-ray optics.
The technology has been invented by cosine Measurement Systems and the European Space Agency (ESA) and developed together with scientific and industrial partners to mass production levels.
It leverages techniques and processes developed over decades by the semiconductor industry to handle, process, and clean silicon wafers and plates. 
SPO is an enabling technology for large space-borne X-ray telescopes such as Athena and ARCUS, operating in the 0.2 to 12~keV band, with angular resolution aiming for 5~arc~seconds. 
SPO has also shown to be a versatile technology that can be further developed for gamma-ray optics, medical applications and for material research.
}

\section{Keywords} 
X-ray optics,
X-ray telescope,
Athena,
Wolter I,
monocrystaline silicon,
silicon wafer,
mass production

\section{Introduction}
\label{sec:intro}
The launch of the high-energy astrophysics observatories XMM-Newton \cite{jansen.2001vn} and Chandra \cite{Weisskopf:2002wb} in 1999 marked the culmination of two major optics development efforts, and made observations of the X-ray sky with unprecedented sensitivity possible.
Significant investments were required, and two different optics technologies had to be developed to make these missions possible.
The optics of XMM-Newton are based on electro-formed nickel, replicated from precision mandrels.
They maximise effective area to make high-resolution spectroscopy of distant sources possible \cite{Gondoin:1994wo}.
The Chandra optics, on the other hand, optimise the angular resolution at the cost of mass and effective area to provide high-resolution imaging.
After more than two decades of active operations, these two missions remain in high demand for their stand-alone capabilities and the synergies with observations at other wavelengths.
Their systems are however ageing, and the need for higher performance continues to increase.

The next generation high-energy astrophysics observatory will have to reach deeper into the Universe, keeping pace with the present and future ground and space based observatories like the James Webb Space Telescope (JWST), the Square Kilometer Array (SKA), and the Extremely Large Telescope (ELT).
More photons will have to be collected from very distant sources, requiring larger effective area, good angular resolution to avoid source confusion, and more sophisticated detector instruments.

In 2014, the European Space Agency (ESA) selected the Advanced Telescope for High-ENergy Astrophysics (Athena) as the second large-class mission designed to address the Cosmic Vision science theme 'The Hot and Energetic Universe' \cite{Nandra:2013ue}.
This X-ray space telescope will rely on a novel type of optics, which was specifically invented and developed to enable large area telescopes with few-arc-second resolution: the Silicon Pore Optics (SPO)  \cite{Bavdaz2005,Beijersbergen:2004vc}.

The challenge of the Athena optics is the need to simultaneously comply with three technical requirements:
\begin{itemize}
    \item Provide a large effective area (1.4~m$^2$ at 1~keV),
    \item Provide a good angular resolution (5~arcsec HEW),
    \item Remain in the mass allocation (around 1000~kg).
\end{itemize}

XMM-Newton and Chandra required different optics technologies, because they had to deliver different combinations of these three parameters.
This is true also for the other X-ray missions flown to date, from Exosat \cite{de-Korte:1981ts}, Einstein \cite{van-Speybroeck:1979wp}, ASCA  \cite{Serlemitsos:1995ti} and Rosat \cite{Truemper:1982uu}, to BeppoSAX \cite{Conti:1994uc}, Hitomi (Astro-H) \cite{Okajima:2016ts}, NuSTAR \cite{Harrison:2013wl}, and eRosita \cite{Merloni:2012ug}.
The optics for each of these missions are an engineering masterpiece, each requiring substantial effort, skill, time, and funding to be developed and flown.

\begin{figure}[tbp]
    \centering
    \includegraphics[width=0.9\textwidth]{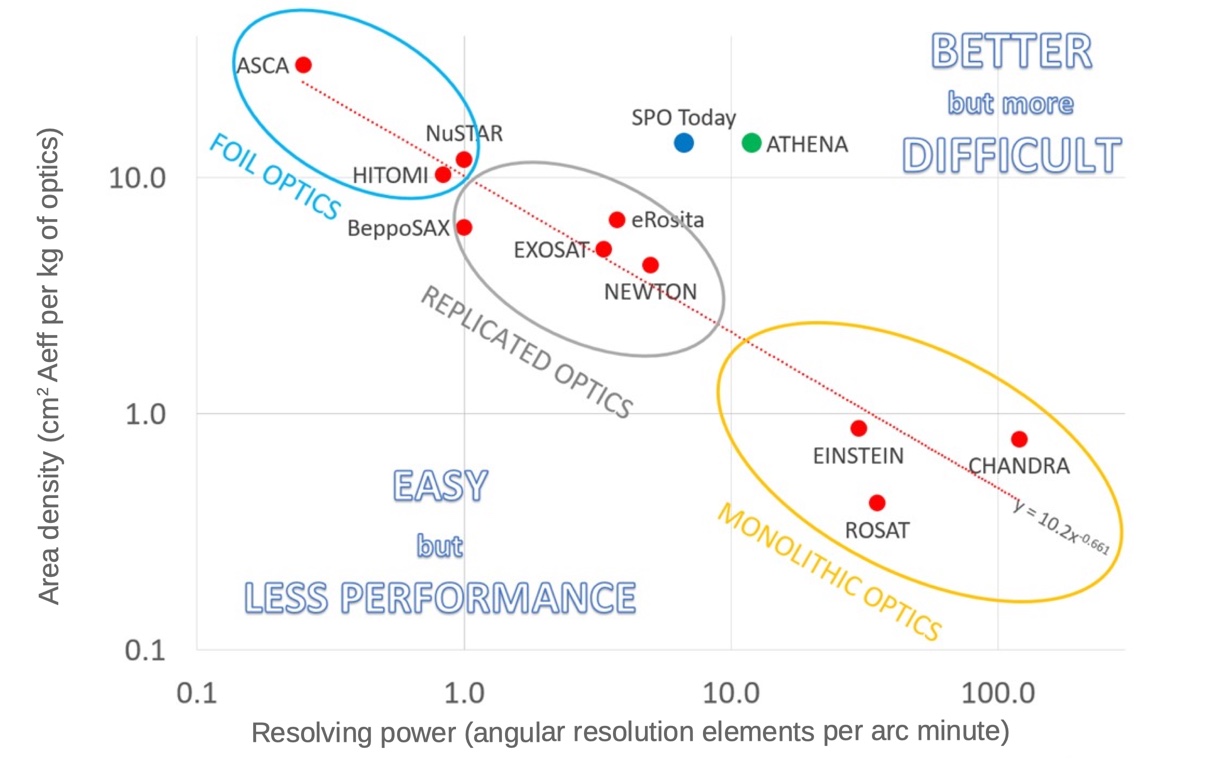}
    \caption 
    { 
    Characteristics of the optics flown on X-ray missions to date (red dots), plotting the area density of the optics (i.e. cm$^2$  of effective area per kg of optics) as function of the resolving power of the optics (i.e. number of angular resolution elements resolved by the optics in one arcminute). The characteristics of the X-ray optics flown to date show a clear correlation, described by a power law. Athena requires optics which cannot be achieved with the X-ray optics technologies flown to date.
    Adapted from \cite{Bavdaz_SPIE_2021}.
    }
    \label{fig:graph_missions} 
\end{figure}

In Figure~\ref{fig:graph_missions} the optics of these missions are compared, considering the aforementioned three parameters.
The area density of the respective optics, expressed as effective area per mass of the optics is plotted against the resolving power of the optics, expressed as angular resolution elements per arcminute \cite{Bavdaz_SPIE_2021}.
A clear correlation is observed despite the fact that the optics technologies are very different, ranging from foil-based optics to replicated optics and monolithic optics.
Each of these systems was developed and built with large effort, and each represents a major achievement, but they all follow a power law correlation.
Even the Chandra optics, which are the result of extraordinary efforts in engineering and funding, manage to move only very slightly to the upper right side of the correlation line.
The same is true for eRosita, which represents the highly optimised third generation of electroformed nickel technology.

The graph illustrates that lightweight optics with large effective area but lower angular resolution, or high resolution but heavy optics with a limited effective area have been built until now.
Current optics-making technologies have so far allowed for the production of systems where only two of the three parameters can be effectively optimized at one time.
The optics for the Athena mission (green dot in Figure~\ref{fig:graph_missions}) however require a technological change to move away from this
empirical power law into the ‘difficult corner’ to the upper right of the graph.
Linear extrapolation of all the X-ray missions flown to date using the three previous conventional technologies suggests that it would be possible to deliver the resolution and effective area, but it would require a mass of about 7000~kg, which is the mass of the complete Athena mission.
Or it could respect the mass allocation of 1000~kg and deliver the effective area, but with a severely degraded angular resolution of about 1.6~arcminutes.

In this chapter we describe the status of SPO, an X-ray optics technology developed to make it possible to build a system that departs from the empirical power law defined by existing technologies, and make missions like Athena possible.
The SPO technology has already demonstrated the compliance with the Athena effective area and mass requirements. 
The angular resolution is approaching the Athena requirement as well, being already one order of magnitude better than what the established technologies could deliver for the same mass and effective area, and it has been continuously improving over the past 10 years (e.g. \cite{Collon:2011tx, Collon:2013tj, Collon:2015wv, Collon:2017vj, Collon:2019ww, Collon:2021tk}).
This technology will undoubtedly also find other applications in demanding future space missions (the ARCUS mission studied by NASA is an example \cite{Randall_SPIE}) and ground-based applications, extending from opportunities in medicine \cite{Girou_2021} to material testing and security applications.

We introduce the SPO technology, its potential, and how it is currently realized in the first part of the chapter. 
Although the development of SPO is driven by Athena, we purposely keep the description of the concept and technology as general as possible. 
In the second part, we discuss more specifically the optics for Athena with its current design and performance estimates.

\section{SPO concept}
\label{sec:concept}

SPO technology uses commercially available monocrystalline double-sided super-polished silicon wafers as a basis to produce X-ray mirrors.
In addition to its direct bonding property \cite{Shimbo1986,Maszara1988}, silicon is rigid, has a relatively low density, and a very good thermal conductivity. Silicon wafers have excellent surface finish, both in terms of figure and surface roughness, making them a superb base material to build X-ray mirrors.

\begin{figure}[tpb]
    \centering
    \includegraphics[width=0.7\textwidth]{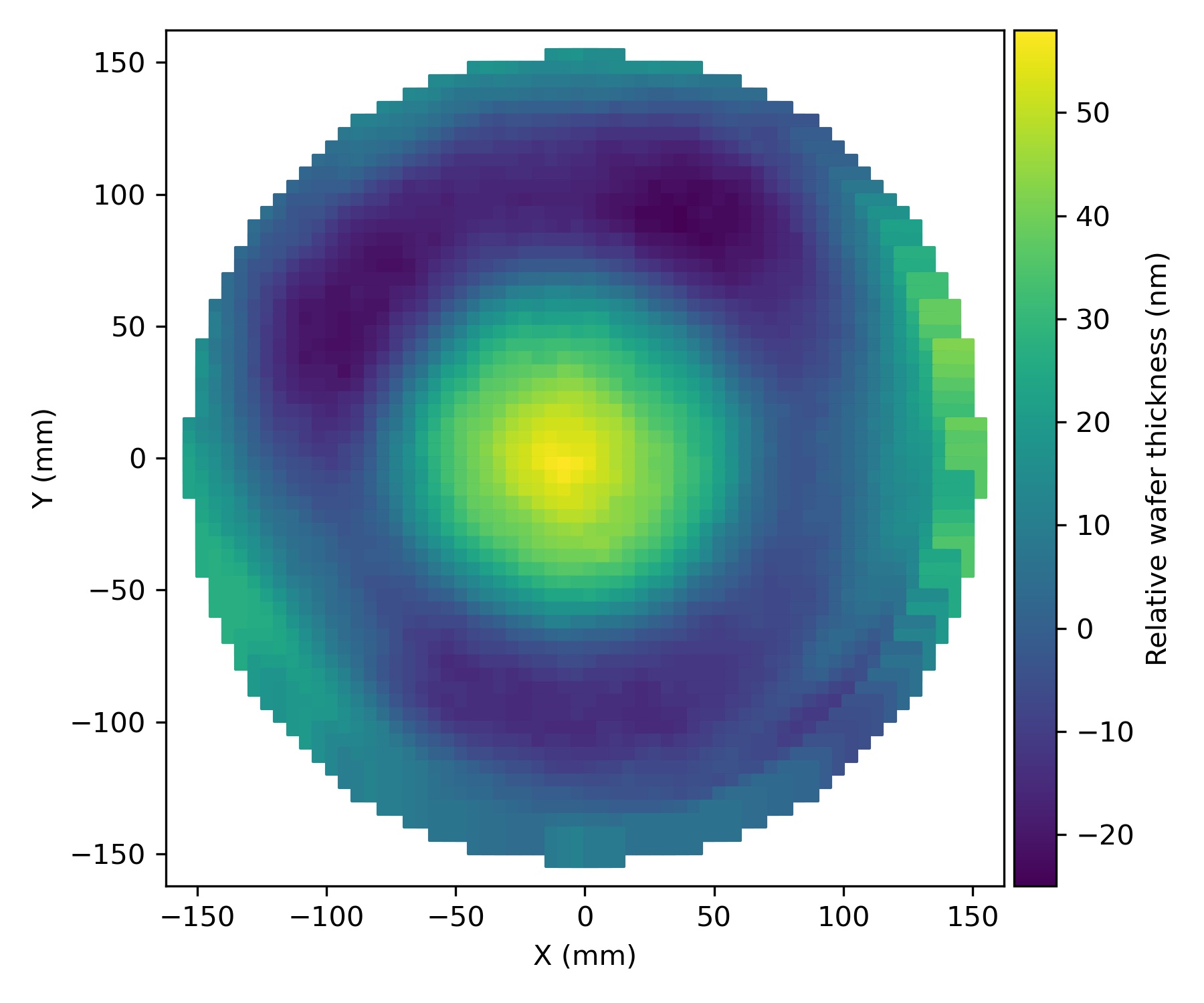}
    \caption{
    Typical relative thickness map of a silicon wafer for SPO plate production.
    The data is plotted relative to the median thickness of 775110~nm of the wafer.
    The thickness is measured by Fizeau interferometry. 
    Credit: cosine.
    }
    \label{fig:wafer-ttv}
\end{figure}

SPO development leverages the massive investments done in the semiconductor and automotive industries, which made available high-performance equipment, tools, materials and processes allowing high-quality mass production. 
The work-horse of the semi-conductor industry are thin 300-mm wafers sawn out of monocrystalline ingots, grown from poly-crystalline silicon using the Czochralski method. 
These wafers are produced with outstanding quality for a modest cost:
prime grade double-sided super-polished monocrystalline 300 mm silicon wafers have specifications of 0.1~nm root mean square (RMS) surface roughness and total thickness variation (TTV) of less than 0.2~$\mu$m \cite{Landgraf:2019uz} (see Figure~\ref{fig:wafer-ttv}).

SPO substrates, also called mirror plates or simply plates, are cut out of wafers and carved out on one side in a regular pattern to leave thin parallel walls called ribs. 
In other words, plates consist of a thin membrane with a smooth reflective surface on one side and ribs along the plate length on the other.
Plates are stacked to form a lightweight and stiff structure with a large number of rectangular pores that allow X-ray photons to be reflected: hence the name silicon pore optics.

\begin{figure}[tpb]
    \centering
    \includegraphics[width=0.8\textwidth]{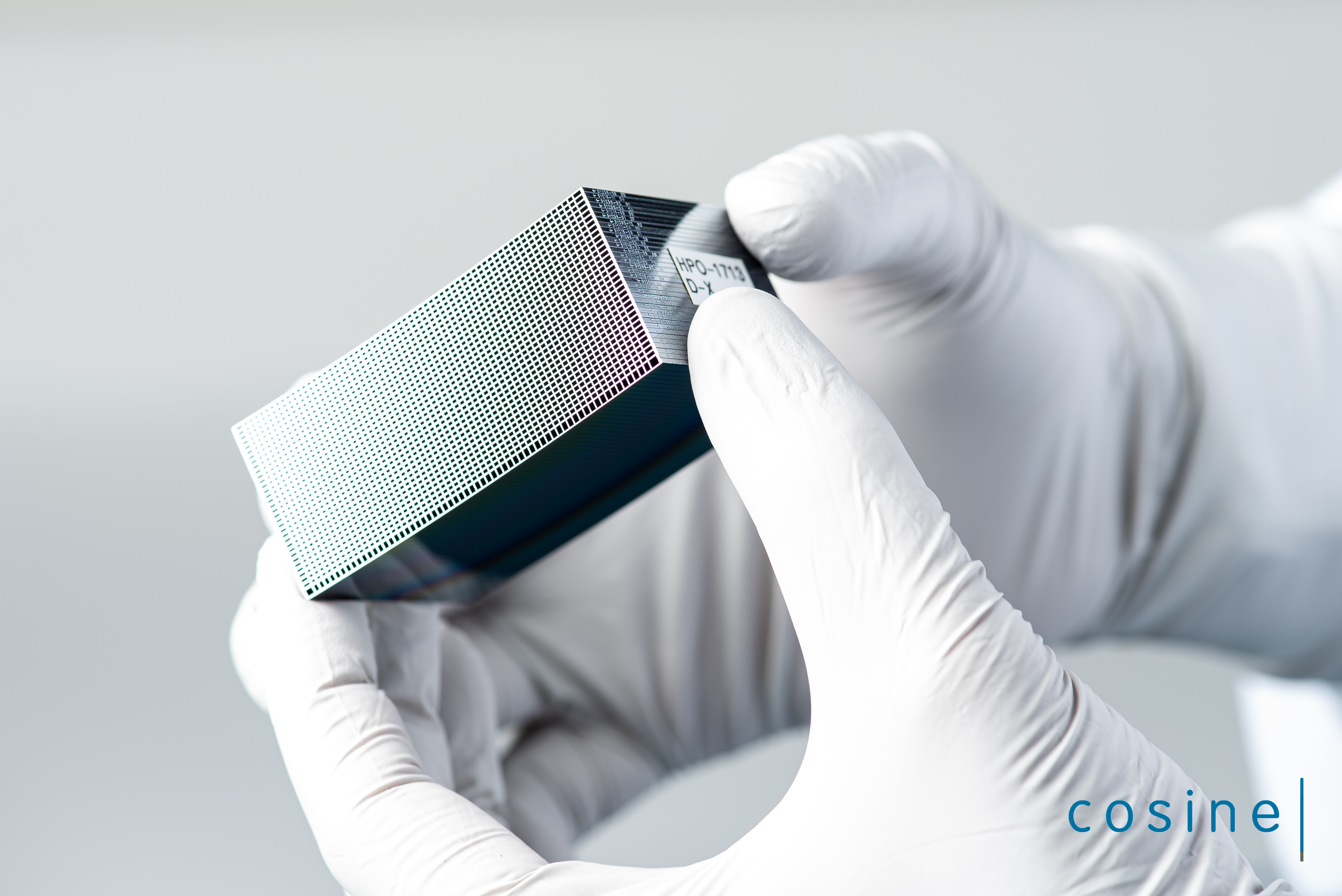}
    \caption{
    A stack of 34 SPO plates, with an outer radius of 737~mm. The plate diemsion is 65.7~mm $\times$ 40.0~mm $\times$ 0.775~mm.
    Credit: cosine.
    }
    \label{fig:34-plate-stack}
\end{figure}

The plates are bonded on top of each other thanks to direct silicon bonding, which requires no adhesives. 
Starting from a mandrel that defines the optical design, plates are stacked and the mandrel figure is preserved owing to the plan-planarity of the mirror plates and the lack of any adhesive between plates.
It results from the process a self-supporting stack of elastically deformed mirror plates that reproduce the shape of the mandrel. 
In such a stack, X-ray photons enter pores at low-grazing angles, are reflected on the reflective side of each plate, and exit the optics at the opposite end.
These stacks consistute lightweight, moderate-to-high resolution, X-ray optics (see Figure~\ref{fig:34-plate-stack}).

Typical X-ray focusing optics use nested shells to fill the available aperture and maximize effective area because low grazing incidence angles are needed for X-ray reflections.
Because the ribs act as spacers between mirror plates, SPO can achieve an extremely high packing density of shells. 
The stiffness obtained by the bonded plates makes it possible to reduce the thickness of the membranes without impacting the figure accuracy, in a lightweight structure leading to large open area ratio.

SPO can be made to comply to many optical geometries, with the Wolter I \cite{Wolter:1952vf} configuration being in general the most attractive for high-energy astrophysics applications.
Two reflections are required to form an imaging system at grazing angles of incidence, and this is achieved by placing two stacks, a primary and a secondary, in series along the path of the incoming photons. Two brackets, to which the stacks are bonded, are used to fix the relative orientation of the primary and the secondary, and they provide a mechanical interface for integration in a larger structure.

\subsection{Potential and limitations of SPO}
\label{sec:design-space}
In this section, we discuss the potential and limitations of the SPO technology in general terms. For Wolter I geometry, three aspects are discussed: the plate minimum and maximum lengths, the radial spacing between mirror plates, and the smallest sagittal radius that can be achieved.

The maximum dimensions of a plate are constrained by the size of the wafer that is used. 
Standard silicon wafers come in diameters ranging from 50 to 300~mm, with thickness increasing from 275~$\mu$m to 775~$\mu$m, respectively.
At present, the semiconductor industry is widely making use of 300~mm diameter wafers.
Larger wafers with 450~mm diameter and 925~$\mu$m thickness have been discussed for many years, but are not yet widespread due to the significant investments needed to transition mass-processing facilities and tools from 300~mm to 450~mm.
Therefore, at present, SPO mirror plates are limited to lengths shorter than 300~mm. 
At the other end of the range, although there is no clear minimum, it would become difficult to handle plates shorter than $\sim\,$10~mm.

The pore height $h$, which is related to the plate length $l$ by equation \ref{eq:length} where $\alpha$ is the incidence angle, can be decreased by using smaller and hence thinner wafers.
\begin{equation}
  \label{eq:length}
 l = \frac{h}{\tan{\alpha}}
\end{equation}
Thinner wafers are produced in smaller series and present higher roughness and larger thickness variations than the standard 300~mm ones, which has until recently made them less attractive to use.
However, some of these defects can be eliminated with the use of
ion beam figuring (IBF), which now offers a way to reduce thickness variations and reduce roughness (see Section~\ref{subsec:plates}). 
Thus, it is possible to think of SPO with mirror plate pitch smaller than 0.775~mm. 
At the other end, it would also be possible to bond wafers together and then process them. 
This opens the possibility to create stacks of mirror plates with double-pitch of 1.55~mm. 
This would be useful in case of large incidence angle, where the plate length would become too short otherwise.

SPO stacks replicate the shape of the mandrel, whether it is flat, cylindrical or conical. 
It can also feature a meridional curvature, parabolic, hyperbolic, circular, or any other shape.
The minimum radius of sagittal curvature is only linked to the material properties, and in turn to the membrane thickness.
Membrane thickness of 100~$\mu$m has been tested successfully, while 110~$\mu$m has been manufactured routinely for Athena.
With these thicknesses the plates can be bent to about 150 mm diameter while remaining comfortably far from critical stress that would lead to breakage.

The width and spacing of the ribs determine the open area ratio of the optics, affecting on-axis effective area and vignetting. 
These parameters also affect the final stiffness of the system, and have impact on the optical performance. 
The higher the density of ribs, the smaller the open area ratio, the stronger the support of the mirror membrane.

SPO can also be used to manufacture medium to large series of other type of optics, where only the top plate is reflective, and the lower plates are structural to maintain the shape. 
This can be used, for instance, for Kirckpatrick-Baez optics  \cite{1948JOSA...38..766K}.
SPO technology has also the potential to create advanced gamma-ray focusing elements via the use of diffraction in the volume of the crystalline plates. 
Self-standing single or double-curvature stacks of plates can be used as Laue lens elements providing improved focusing capability compared to other methods \cite{Girou_2021}.

\section{SPO realization}
\label{sec:realization}

\subsection{Production of SPO mirror plates}
\label{subsec:plates}
\begin{figure}[tbp]
    \begin{subfigure}{0.6\textwidth}
        \centering
        \includegraphics[width=0.95\textwidth]{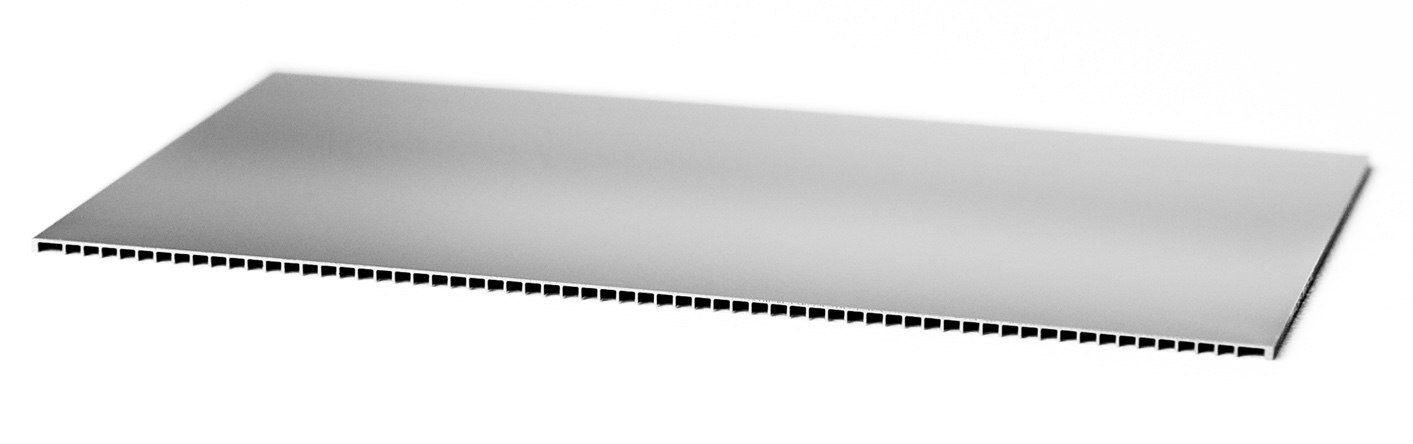}
        \caption{}
    \end{subfigure}%
    \begin{subfigure}{0.4\textwidth}
        \centering
        \includegraphics[width=0.95\textwidth]{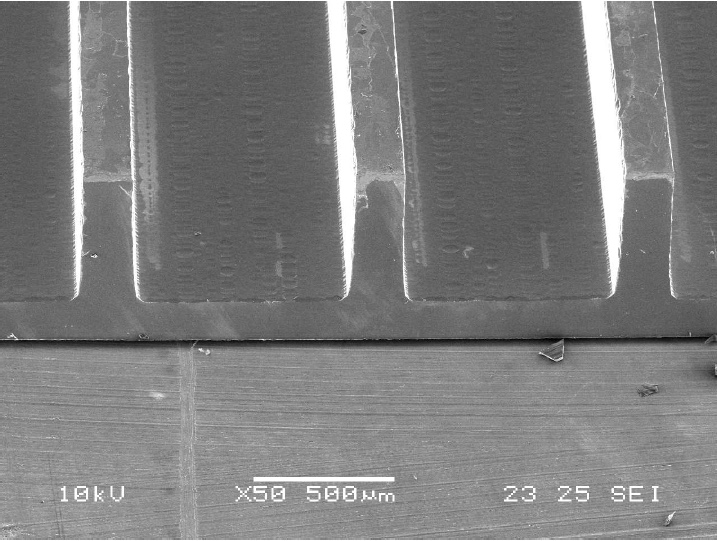}
        \caption{}
    \end{subfigure}%
    \caption{
    (a) Example of a SPO mirror plate with its reflective side on top, and ribbed side facing down.
    (b) Scanning electron microscope image of the ribs of a SPO mirror plate.
    Credit: cosine.
    }
    \label{fig:SPO-plate}
\end{figure}

With the technology contributions from the semiconductor industry, mass production is being set up and standardized processes are being developed in order to manufacture SPO mirror plates in high quality, large quantity, and at relatively low cost \cite{Landgraf:2019uz, Landgraf:2021tf}
The production of SPO mirror plates starts with carving out ribs into 300~mm wafers. 
This process is effectively performed by removing material from the wafer down to the desired membrane thickness.
Subsequently, rectangular plates are diced off the wafer, with the ribs along the direction of travel of the X-rays.
The orientation of plate ribs and edges is aligned with a crystal plane of the silicon to ensure correct bending, as well as chemical processing properties of the material in subsequent production steps.
The top side of the mirror plate is the reflective side (Figure~\ref{fig:SPO-plate}, a)) and the bottom is the ribbed side (Figure~\ref{fig:SPO-plate} b)) with the ribs parallel to each other. 
The plates currently manufactured have rib pitch varying from 1.0~mm to 2.4~mm with rib width of 0.17~mm.

The SPO mirror plates feature a thickness gradient along the rib direction to ensure that all the plates within a stack are confocal (see section \ref{subsec:Optical_design}). 
The plane of the ribs is at an angle with the reflective side, referred to as the wedge angle. The challenge is to implement this angle with a linear profile, and without increasing the roughness of the surfaces.
Two approaches have been developed to realize the wedge angle. 
A thickness gradient is either created by wet-chemical etching, or by IBF.
A wedge accuracy better than 1\% ensures limited contribution to the optical quality budget \cite{Vacanti:2021wh}. A wedge accuracy better than 1\% has been obtained with the wet-chemical processing \cite{Landgraf:2021tf}, while the IBF is still under development.
The advantage of IBF is that the TTV of wafers can be improved and reduced to potentially less than 10~nm in the same production step as the wedge processing.

Cleanliness is a crucial parameter for the production stacks.
Particles on the plate surface can reduce bonding area and distort the optical figure.
For this reason, SPO mirror plate production is performed in high-quality clean room environments of ISO~6 and ISO~5, and with specific equipment reducing also the risk of organic contamination, which can have a negative impact on the bondability of SPO mirror plates.

\subsection{Development of coatings}
\label{subsec:development-of-coatings}

The reflectivity of SPO mirror plates can be enhanced by deposition of thin-film coatings.
The effectiveness of the coating is thereby described by the effective area of the optics.
It depends on the collecting area and the efficiency of the mirrors in reflecting high-energy photons.
The broadest energy window can be achieved with combinations of high-Z (e.g. Au, W, Mo, Pt and Ir) and low-Z materials (e.g. Si, SiC, C, DLC and B4C).
Figure~\ref{fig:A_eff_coatings} in Section~\ref{subsec:MA_effective_area} shows an example of the computed on-axis effective areas for different reflective coatings for the Athena telescope energy range.
The effective area is significantly larger with the addition of high-Z/low-Z bilayer coatings compared to the effective area of un-coated SPO mirror plates.
In addition, multilayer coatings, consisting of superimposed bilayers of high-Z/low-Z material, can be applied to further increase the X-ray response of the optics at higher energies \cite{10.1117/1.JATIS.6.3.034005}.

\begin{figure}[tpb]
    \centering
    \includegraphics[width=0.6\textwidth]{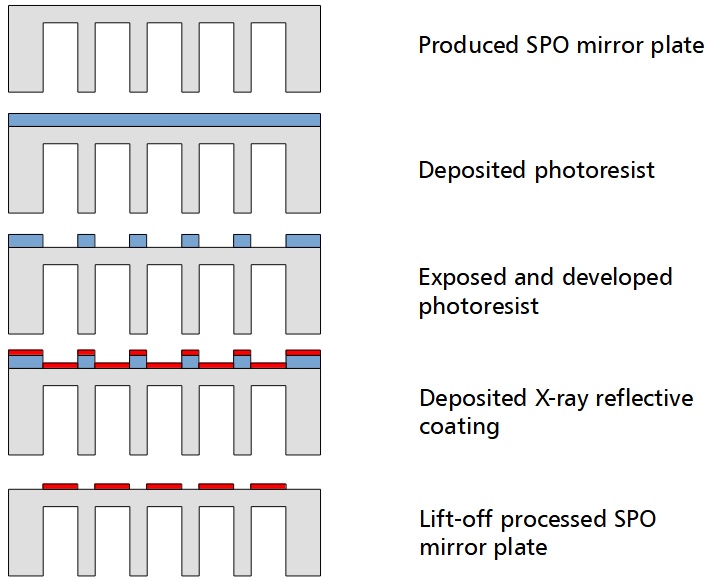}
    \caption{
    Schematic of optical lithography for SPO mirror plate with patterned X-ray reflective coating.
    Credit: cosine.
    }
    \label{fig:litho}
\end{figure}

The SPO mirror plates feature a patterned coating on their reflective side in order to enable direct silicon bonding \cite{Shimbo1986,Maszara1988} (see Section~\ref{sec:stacking}).
The patterning of the coating is performed via optical lithography.
This process, sketched in Figure~\ref{fig:litho}, consists of the following steps. 
Firstly, a photosensitive polymer is deposited on the reflective side of the SPO mirror plate.
In a second step, a patterned photoresist layer is created by exposing the photosensitive polymer to ultraviolet light and developing it by immersing the sample in a chemical solution.
Subsequently, an X-ray reflective coating is deposited using a sputtering deposition technique. 
Finally, chemical lift-off is performed to remove the coated photoresist resulting in a patterned coated SPO mirror plate. 

Direct current magnetron sputter deposition (MSD) has been applied to produce X-ray reflective coatings on mirrors for a number of X-ray telescopes, and as such, is a method with a good track record. 
It is a low-cost and well-controlled process. 
Particularly, iridium films with a thickness of 30~nm have been sputtered onto the Chandra telescope mirrors \cite{Gorenstein2010}, and Pt/C and W/Si bilayer systems have been used for the NuSTAR telescope mirrors successfully  \cite{10.1117/12.894615}.

\begin{figure}[htpb]
    \begin{subfigure}{0.5\textwidth}
        \centering
        \includegraphics[width=0.95\textwidth]{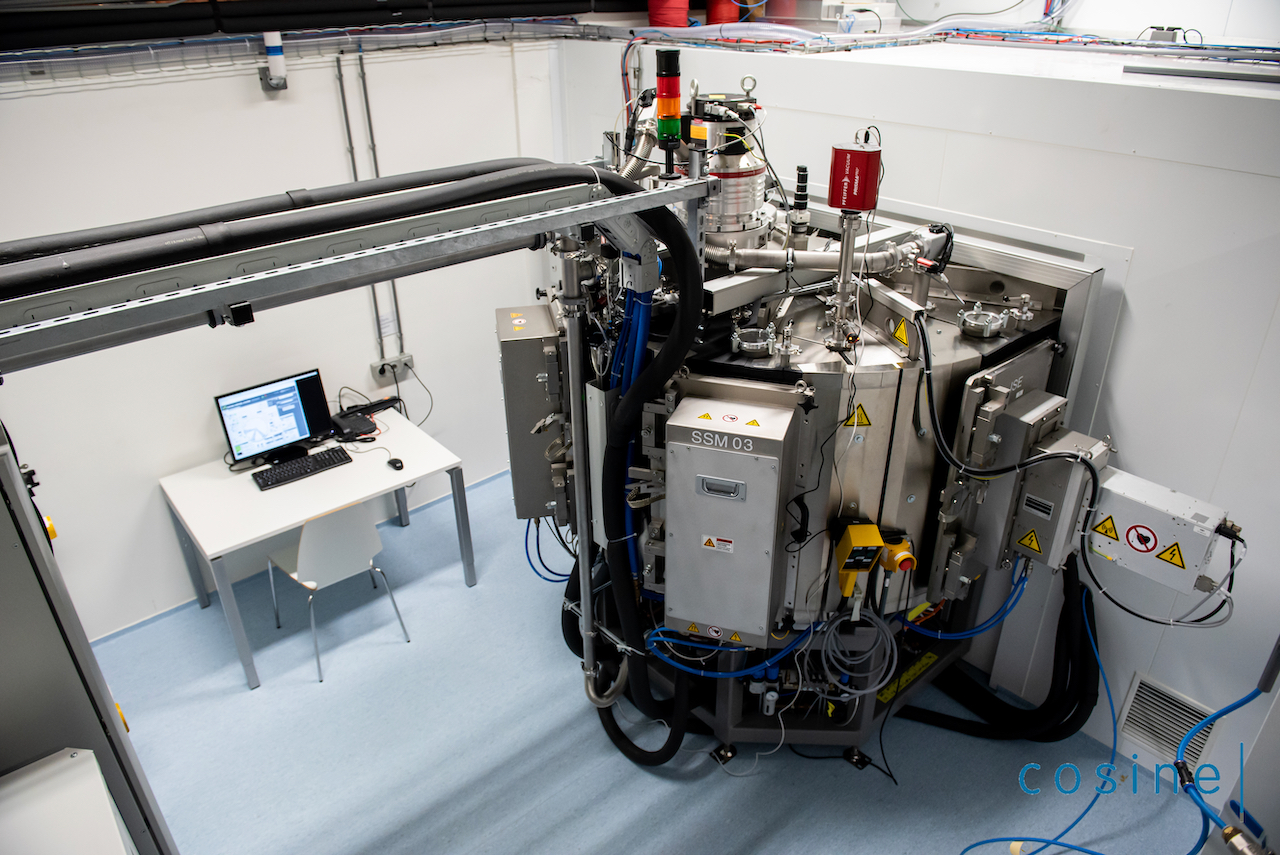}
        \caption{}
    \end{subfigure}%
    \begin{subfigure}{0.5\textwidth}
        \centering
        \includegraphics[width=0.95\textwidth]{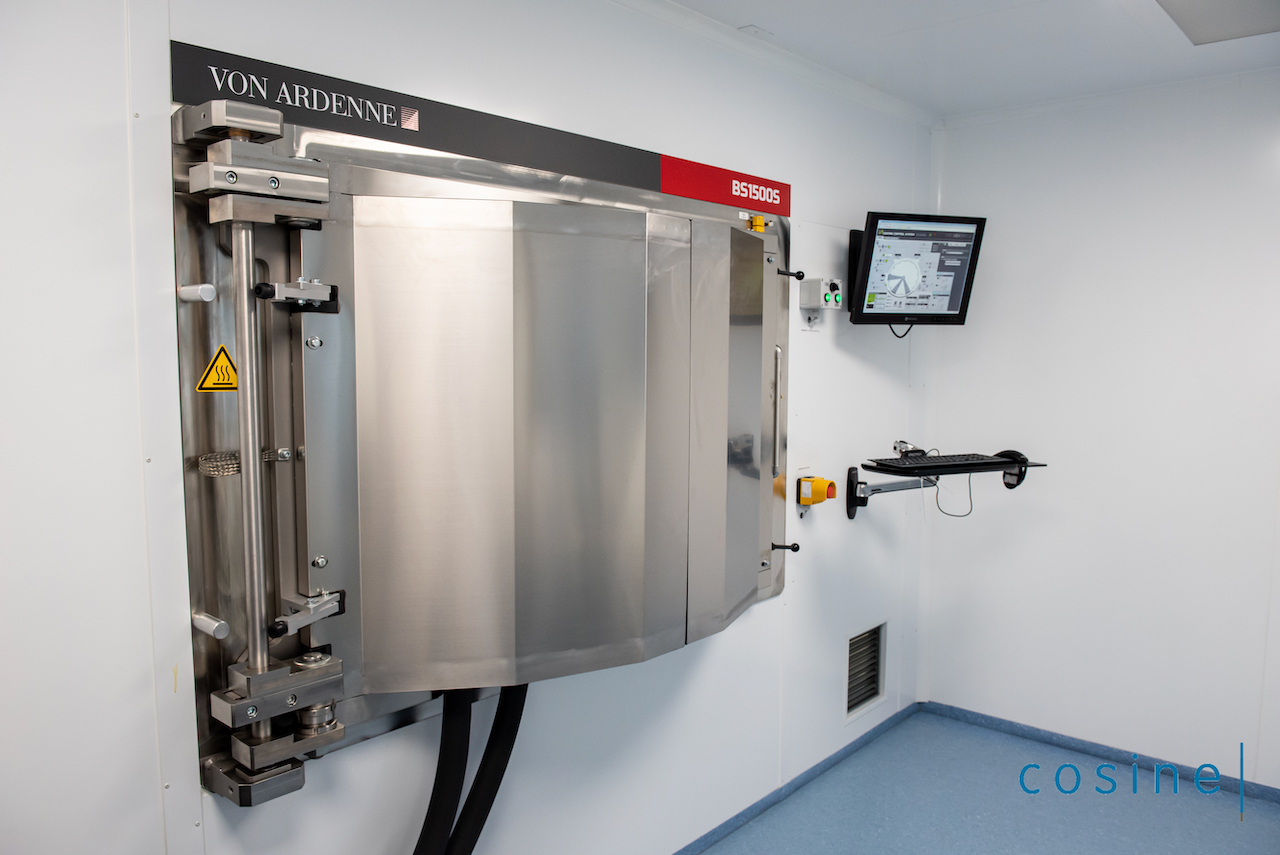}
        \caption{}
    \end{subfigure}%
    \caption{
    Coating machine at cosine:
    (a) back side and
    (b) clean room side of the system. Credit: cosine.
    }
    \label{fig:coating-machine}
\end{figure}

The SPO mirror plate coating process takes place at cosine using an industrial, high-throughput magnetron sputtering system  \cite{10.1117/12.2528351} (Figure \ref{fig:coating-machine}).
It features three magnetrons for coating target materials and a plasma cleaning system in order to clean substrates from organic materials before a coating deposition run \cite{Girou2020}. It is also possible to accommodate other types of deposition methods, such as thermoionic vacuum
arc deposition that may be more suitable to improve the coating quality for particular materials.
With this system, substrates as large as full 300~mm wafers can be coated  with various high-Z materials such as tungsten, molybdenum, platinum and iridium and low-Z materials such as silicon, silicon carbide, carbon, diamond-like carbon, and boron carbide.

\begin{figure}[htpb]
    \begin{subfigure}{0.5\textwidth}
        \centering
        \includegraphics[width=0.95\textwidth]{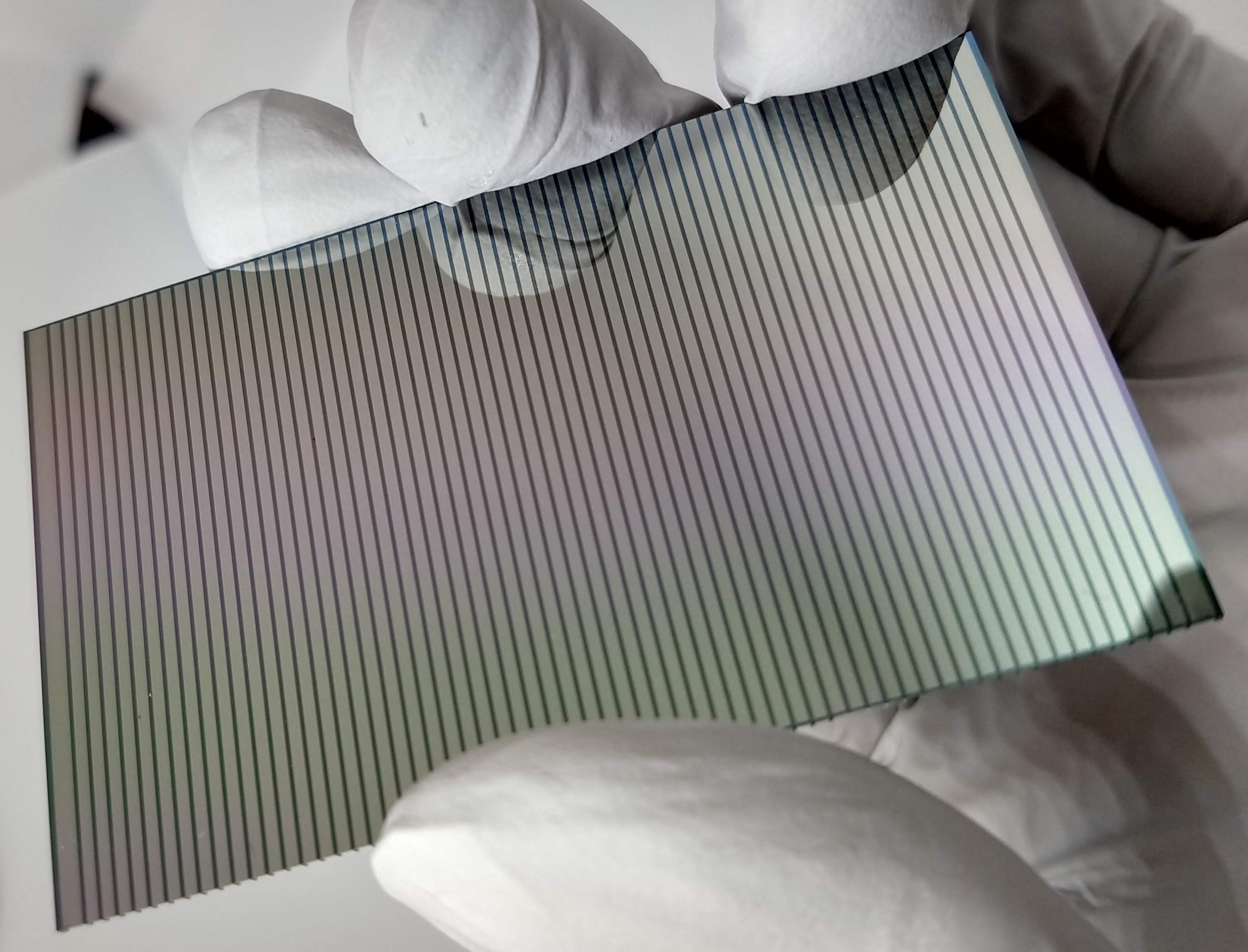}
        \caption{}
    \end{subfigure}%
    \begin{subfigure}{0.5\textwidth}
        \centering
        \includegraphics[width=0.95\textwidth]{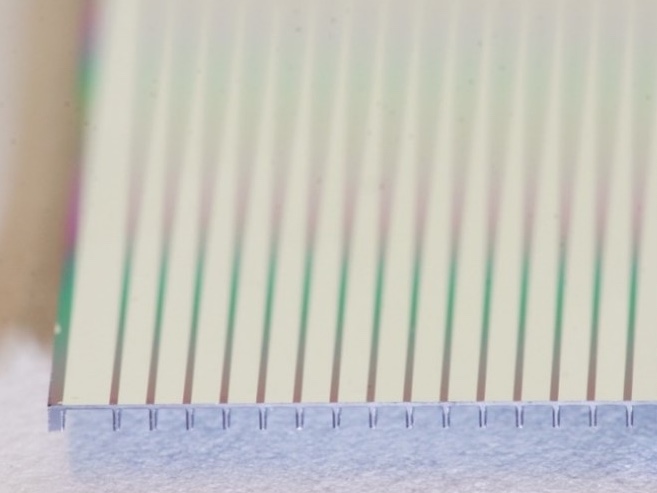}
        \caption{}
    \end{subfigure}%
    \caption{
    (a) SPO mirror plate with patterned coating of 10~nm-thick iridium. (b) close-up view of a SPO plate with patterned iridium coating. The uncoated tracks are clearly visible above the ribs. 
    Credit: cosine.
    }
    \label{fig:patterned}
\end{figure}

The left photograph in Figure~\ref{fig:patterned} shows an example of the reflective side of a produced patterned iridium-coated SPO mirror plate.
In the right photograph, the coating stripes can clearly be distinguished from the silicon bonding areas.

\subsection{Cleaning and activation}
\label{sec:cleaning}
\begin{figure}[tbp]
    \begin{subfigure}{0.5\textwidth}
        \centering
        \includegraphics[width=0.95\linewidth]{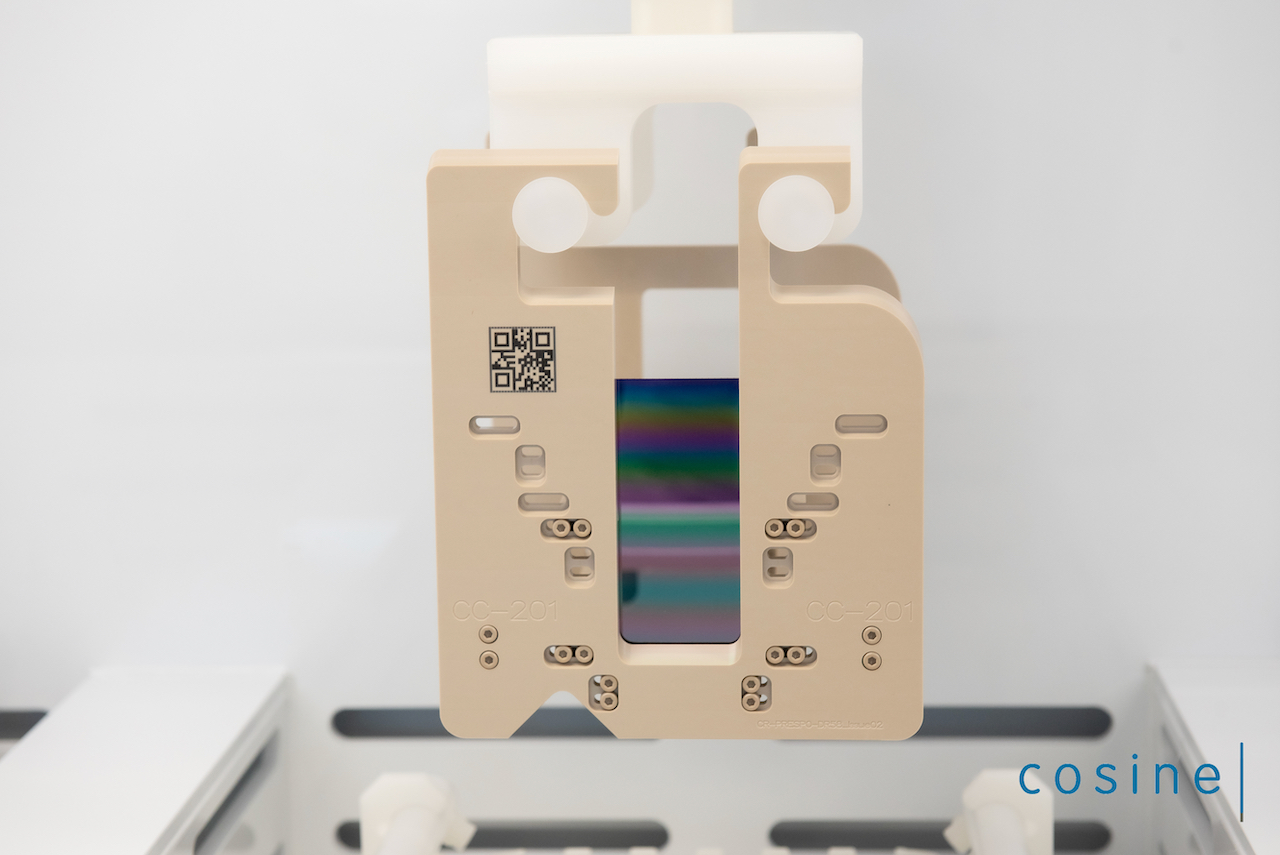}
        \caption{}
    \end{subfigure}%
    \begin{subfigure}{0.5\textwidth}
        \centering
        \includegraphics[width=0.95\linewidth]{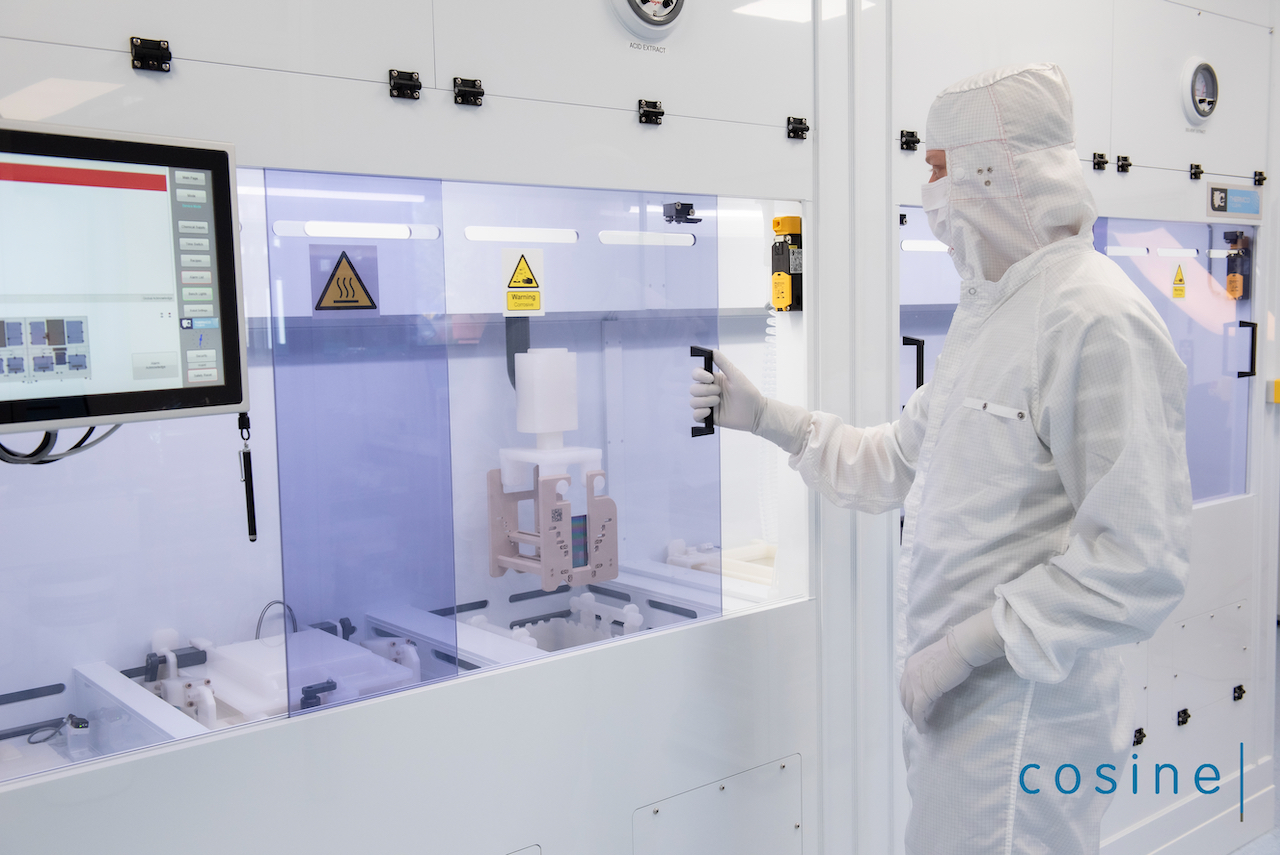}
        \caption{}
    \end{subfigure}%
    \caption{
    Plate cleaning container (a) and fully automated wet bench (b) for chemical cleaning and activation of SPO mirror plates at cosine.
    Credit: cosine.
    }
    \label{fig:wetbench}
\end{figure}

Cleanliness is paramount to the SPO technology. 
The mirror plates are cleaned in industrial wet benches with a SC-1 cleaning \cite{10003367909} and then dried.
This results in hydrophilic (‘activated’) surfaces, free of particulate contamination or drying stains.
The hydrophilic surfaces are key to direct silicon bonding, as explained in the next section. 
The photographs in Figure~\ref{fig:wetbench} show the fully automated cleaning wet bench at cosine, which is used to prepare the SPO mirror plates for stacking.

\subsection{Stacking of mirror plates}
\label{sec:stacking}

SPO stacks hold together and maintain their shape without adhesive between the mirror plates. 
This direct bonding process can take place at room temperature and is widely used to join glass or silicon surfaces to each other \cite{Shimbo1986,Maszara1988}.
The two mating surfaces need to be chemically prepared (i.e. activated) and need to be clean, smooth, and conforming for a direct bond to take place.
After activation, a thin layer of OH- groups is present on both the ribbed and the reflective sides of each mirror, forming silanol (Si-OH) groups \cite{gosele}.
These absorbed water molecules form a bridge (Si-OH+HO-Si) across the two bonding surfaces that is maintained via Van-der-Waals forces for temperatures ranging from 10$^{\circ}$C to 110$^{\circ}$C \cite{gosele}.
Furthermore, post-stacking thermal treatment, such as annealing above 110$^{\circ}$C, enables the silanol bonds to transform into stronger siloxane (Si-O-Si) bonds \cite{gosele}.
During annealing, the interface dries steadily and siloxane bridges form across the bonding surfaces. 
For temperatures below 1000$^{\circ}$C, the bonding energy is limited by the area in contact, which is itself limited by surface roughness \cite{gosele}.
Annealing at temperatures above 1000$^{\circ}$C can increase the contact area through viscous flow of the oxide, which then fills up the micro-gaps at the interface to form additional siloxane bonds \cite{gosele}.
SPO mirror surfaces have a micro-roughness lower than 0.5~nm RMS and the flatness over length-scales of several tens of millimetres is better than 100~nm, which means that annealing stacks at temperatures between 150$^{\circ}$C and 1000$^{\circ}$C is sufficient. 
This is confirmed by environmental testing (see Section~\ref{sec:ruggedization}).
Note that the excess water can also exude over time and siloxane linkage can form after months of storage \cite{gosele}.

The plates are stacked in a concave mandrel, starting with a plate placed with the reflective side in contact with the mandrel and ribs pointing inwards, called the base plate.
The mandrel is polished with the design shape of the stack. 
It can be cylindrical or conical, and can feature additional meridional curvature. 
The next plate is then brought into a similar shape as the base plate by using a ’die’, a tool with the same figure as the mandrel. 
The plate is attached with the mirror surface to the die, resulting in the ribs pointing outwards.
The two plates are brought into close contact such that the two sets of ribs make contact and that direct bonding starts. 
In that moment, the plate copies the shape of the mandrel. This is the first reflective plate of the stack. 
The die then releases the plate and the process can be repeated, with the only difference being that the next plates are bonded with their ribs to the mirror surface of the upper plate in the stack. 
At the end of the process the stack is released from the mandrel and keeps the desired shape through the acting bond forces and the stiffening effect of the ribs.

To achieve good angular resolution, it is important to replicate the initial figure of the mandrels with minimum systematic and stochastic errors. 
Cleanliness of the plates, mandrel, and die is of prime importance to avoid degrading the figure with bumps created by trapped particles plate after plate. 
Systematic slope errors can be minimized with optimized plate design and stacking approach.

Note that when it is curved, the mandrel defines only the outer starting radius of a stack.
With each plate added, the stacks' sagittal radius of curvature becomes smaller by the thickness of the mirror plate.
In the case of double-reflection optical design such as Wolter I, it is important to align the ribs to several microns. 
This minimizes loss of open area as the two stacks are aligned one behind the other.

\subsubsection{Stacking robots}
\label{sec:stacking-robot}

\begin{figure}[tpb]
    \centering
    \includegraphics[width=0.9\textwidth]{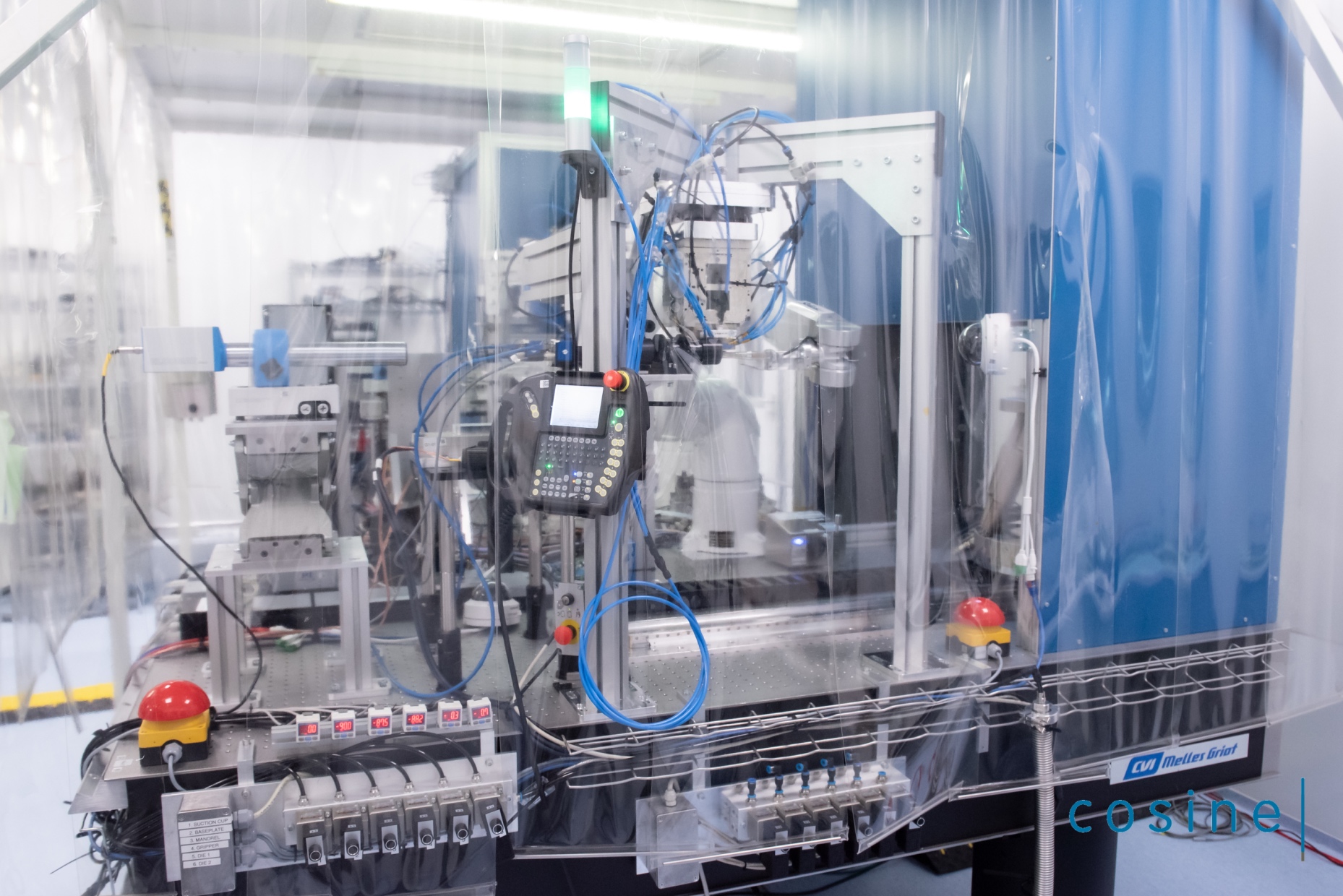}
    \caption{
    Stacking robot inside a cleanroom, as installed at cosine.
    It is equipped with a robotic arm to extract plates for stacking.
    A hexapod allows aligning the mandrel to the die.
    The blue box on the right is an optical metrology system that can measure the figure and cleanliness of each plate after having been stacked.
    Credit: cosine.
    }
    \label{fig:robot}
\end{figure}

Automated stacking robots have been developed to meet the mass production and quality requirements of SPO technology: automation provides cost-efficiency, improves repeatability for large series, and ensures the highest standard of cleanliness. 
The robots are composed of a robotic arm equipped with a plate gripper, a die attached upside down, and a mandrel which rests onto an hexapod (Figure \ref{fig:robot}).
The hexapod is on a linear translation stage that allows moving the mandrel and the stack from the die to an optical metrology system.

The stacking process starts with a batch of several tens of plates loaded in a container that are cleaned and activated (Figure \ref{fig:wetbench}).
The container is loaded into the robot, which is placed inside an ISO-class 5 clean area to keep the plates from accruing particulate contamination.
The robot extracts the first plate from the container, aligns it to the gripper, and places it on the mandrel, as described in Section~\ref{sec:stacking}.
A second plate is extracted, aligned to the gripper and attached to the die with the ribs pointing downwards.
The mandrel and die are brought into close contact until the ribs align and the bond is created. 
Then the die releases the plate.
The first reflective plate is now bonded to the base-plate.
The stack is translated to the in-situ metrology, which measures the figure of the bonded plate, determining not only the shape, but also the cleanliness and therefore the quality of the bond.
The robot then proceeds to extracting the next plate, attaches it to the die and the stacking process repeats, this time with the ribs bonding to the mirror surface of the stack on the mandrel.

The process repeats until the desired stack height is reached. Stacks up to 70 plates have been manufactured, 35 plates being the most common so far.
Once the stack is finished, it is released from the mandrel and a final inspection is performed to measure the outer geometry and to visually check for defects.

Prior to stacking, the mandrel axis is aligned with the die axis.
The plate gripper, which is the reference for the plates alignment, is also aligned with respect to the die and mandrel.
A robot is configured for one specific set of stacks defined by its mandrels and plate types.
The robot needs to be reconfigured to produce stacks of different radii, plate geometry or figure. 

The stacking process takes about 5~minutes per plate and can run completely autonomously. This is the result of more than 15 years of development \cite{Gunther:2006uy}. The degree of automation, the reliability and speed of the stacking robots is expected to further increase as production of the flight models for Athena are being setup.

\subsection{Mirror modules}
\label{sec:mm}
\begin{figure}[tbp]
    \centering
    \includegraphics[width=0.7\textwidth]{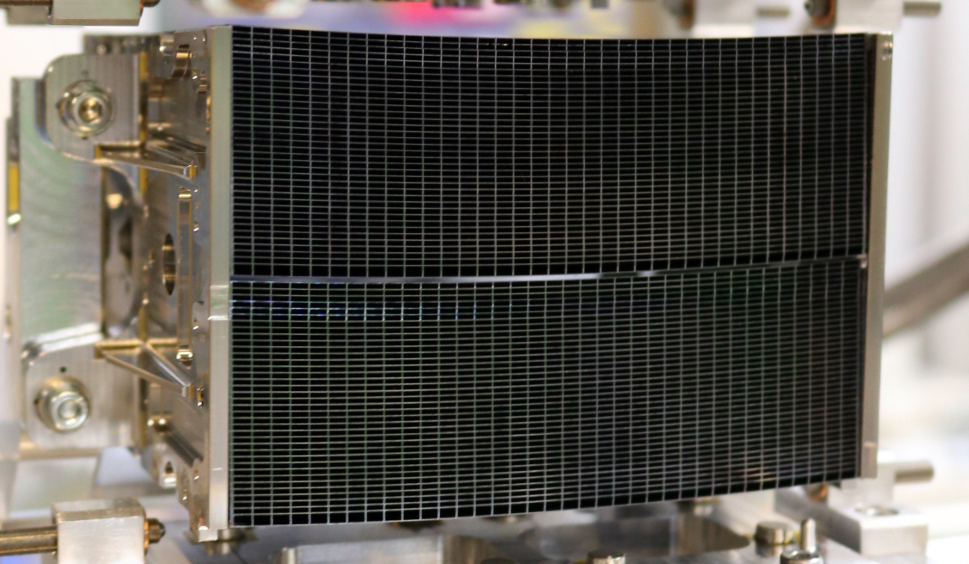}
    \caption{
    Mirror module (MM) for Athena row-08 (see Section \ref{subsec:MA_design}). The exit side of the two secondary stacks are visible, each with 37 mirror plates in addition of their base plate. The ribs are 0.170 mm wide, the pores 2.2 mm wide and the membranes 0.110 thick. Bracket A, with its two ears, is visible in this view. Credit: cosine.
    }
    \label{fig:mm-0046}
\end{figure}

SPO consists of stacks of mirror plates with a shape that is driven by the mandrel. 
In the case of a Wolter I optical design, two different stack shapes (and consequently, two mandrels) are needed: primary and secondary stacks for the two reflections. 
When aligned after each other, these stacks form a so-called X-ray Optic Unit (XOU). 
The fact that the primary and secondary stacks are initially disjoined represents a challenge as their relative alignment can influence greatly the optical quality of the XOU. 
Their optical axis must be co-aligned along with the mirror plates ribs and membranes.
Then this alignment must be frozen. 
This is done by gluing two side plates to the stacks, commonly referred to as brackets, which then form a mirror module (MM). 
A MM is thus a stand alone X-ray focusing optics assembly with mechanical interfaces.
In the current design, Athena MMs are made of two XOUs co-aligned to be confocal, glued in between a pair of brackets (Figure \ref{fig:mm-0046}).

The brackets are made of an austenitic nickel-iron alloy containing 36\% nickel (Invar 36), a material with very low coefficient of thermal expansion of $1.2 \times 10^{-6}$~K$^{-1}$, close to that of silicon $2.6 \times 10^{-6}$~K$^{-1}$.
Their role is to freeze the alignment of the stacks, protect the stacks, provide interfaces between the MM and any support structure or mount, and provide a mechanical reference for the optical axis of the MM. 
The brackets can also provide interfaces to baffles or aperture mask. Low-outgassing and low-shrinkage epoxy is used to affix the brackets to the sides of the stacks.
The interface points of the brackets are called {\it ears} (visible on the left hand side of Figure \ref{fig:mm-0046}).

The alignment strategy is as follows  \cite{Barriere:2019tf,Barriere:2021vp}: the brackets are aligned with respect to each other using a tool providing six degrees of freedom to each bracket. 
The primary stack is then glued to the brackets. 
This alignment relies on mechanical metrology with a coordinate measurement machine (CMM). 
This assembly is then brought to a synchrotron facility where a monochromatic, low-divergence, X-ray pencil beam is available. 
Currently this is done at the X-ray Parallel Beam Facility 2 (XPBF 2) synchrotron beamline, which is installed in the laboratory of the Physikalisch-Technische Bundesanstalt (PTB) at the synchrotron radiation facility BESSY II in Berlin  \cite{handick:2020SPIE}. 
An additional beamline is planned to become operational in 2022 at the ALBA synchrotron facility in Barcelona \cite{Heinis:2021ud}.
The XPBF 2, tailored for the development of Athena's MMs, provides a 1-keV beam monochromatized and collimated by a toroidal mirror with multi-layer coating, allowing beam dimensions between 0.05$\times$0.05~mm$^2$ and 7$\times$7~mm$^2$.
The detector is placed at 12~m$\,\pm\,$0.5~m from the sample, and its position is monitored by a laser tracker.

At the XPBF 2, the direction of the incident beam is used to define the optical axis for the future MM.
The primary stack, already glued within the brackets, is aligned to bring its optical axis parallel to the beam. 
Then the secondary stack is brought in and the rest of the alignment is done in double reflection. 
The secondary stack is aligned with respect to the first stack to minimize the intersection of the ribs and membrane (i.e. maximize throughput of the XOU), bring the point spread function (PSF) at the nominal position, and obtain the best possible optical quality (co-align the stacks' optical axes).

Pencil beam raster scans are used to determine the three-axis orientation of the primary stack with respect to the beam, as well as of the secondary stack with respect to the primary stack \cite{Barriere:2019tf}. 
The position of the incident beam onto the primary stack as well as the position of the beamspot recorded on the camera are reconstructed in a common coordinates system. 
An analytical three-dimensional model of the optic is used in a ray-trace function that transforms incident position into position on the camera. 
This function is used to fit the orientation of a stack in single reflection or double reflection. 
This method allows precise and deterministic alignment.

\subsection{Ruggedisation}
\label{sec:ruggedization}

As SPO development progresses, it is necessary to validate the technology under launch and orbit environmental conditions \cite{10.1117/12.2594230}.
Payloads undergo high levels of vibrations and shocks during a rocket launch and fairing separation.
It is therefore crucial to check that the MMs are robust enough to withstand such conditions and ensure that their mechanical stability and optical performance will remain stable before, during, and after launch.
Environmental testing also allows identifying potential design issues and correcting them.
A ruggedisation process is a key element of the SPO development. 

\begin{figure}[tbp]
    \centering
    \includegraphics[width=0.9\textwidth]{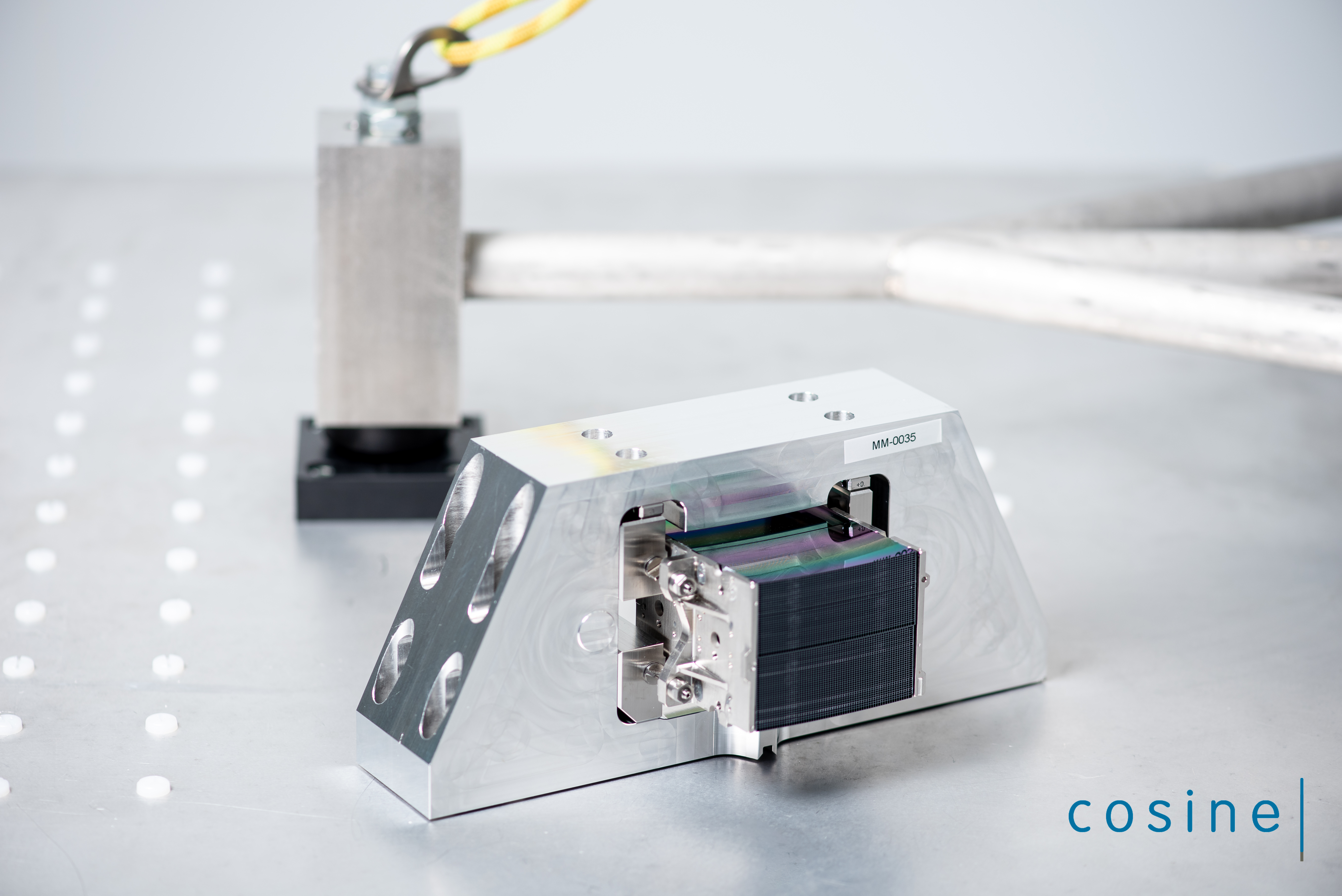}
    \caption{
    MM integrated into a device under test (DUT), laying on a shock table.
    The hammer can be seen resting on an anvil in the background.
    Credit: cosine.
    }
    \label{fig:shock}
\end{figure}

Thanks to its modular nature, SPO can be tested at MM level.
MMs are submitted to vibration and shock tests, in accordance with Athena's launch vehicle and mission requirements.
The MMs are characterized with X-rays before and after each test to verify that the performance remains unchanged (Section \ref{sec:xray-characterization}).
MMs are first integrated into a device under test (DUT, see Figure~\ref{fig:shock}) that is representative of a mirror structure and enables interfacing the MM with test equipment (shock table, shaker, vacuum chamber). 

Vibration testing consists of two steps performed successively: sine qualification and random qualification tests.
Two triaxial acceleration sensors (pilots) are mounted to the test mount to control the shaker input to the MM.
The dynamic response of the MM is monitored on both brackets, close to the MM center of mass.
Two single axis accelerometers are also used to monitor the dynamic transfer over the mechanical shims.
These mechanical shims are used to preload the dowel pin (DP) flexures used to mount the MM and to statically mimic the expected dynamic structural distortions of the mirror structure during low frequency (sine) and vibro-acoustic (random) mission phases.
The Athena sine qualification test levels are shown in Table~\ref{tab:sine-qual}.

For the random test, Athena qualification and notched levels that have been used are presented in Table~\ref{tab:random-qual}.
All random runs at qualification level are preceded by a low level ramp up, meaning a sequence of runs at -12~dB, -9~dB, -6~dB, and -3~dB for 30 seconds each.
These runs are used to predict the 0~dB response levels at the MM center of mass and to check that the design limit load (DLL) values are not exceeded.

Frequency signature test runs (also called resonance searches) are performed before and after each high-level qualification test run to detect any potential changes in structural behavior of the MM or test setup.
A test run is considered successful if the change in frequency is less than 5\%.

\begin{table}[tbp]
\caption{
Athena sine qualification test levels for in-plane and out-of-plane directions.
}
\label{tab:sine-qual}
\begin{center}
\begin{tabular}{|c|c|c|c|}
\hline
\rule[-1ex]{0pt}{3.5ex} {\bf Excitation direction} & {\bf Frequency} & {\bf Amplitude} & {\bf Sweep rate} \\
\hline
\multirow{3}{*}{In-Plane (X, Y)} & 5~Hz & $\pm$11~mm & \multirow{3}{*}{2 Oct/min}\\
& 17~Hz & 13~g &\\
& 100~Hz & 13~g &\\ \hline
\multirow{3}{*}{Out-of-plane (Z)} & 5~Hz & $\pm$11~mm & \multirow{3}{*}{2 Oct/min}\\
& 19~Hz & 16~g & \\
& 100~Hz & 16~g &\\ \hline
\end{tabular}
\end{center}
\end{table}

\begin{table}[tbp]
\caption{
Athena random qualification test levels with notching. Each test lasts 2~min per axis.
}
\label{tab:random-qual}
\begin{center}
\begin{tabular}{|c|l|c|c|c|}
\hline
\rule[-1ex]{0pt}{3.5ex} {\bf Direction} & {\bf Frequency} & {\bf Power Spectral Density (PSD)} & {\bf Overall input} \\
\hline
\multirow{5}{*}{X} & 20-100~Hz & +10~dB/Oct &\\
& 100~Hz & 0.05~g$^2$/Hz & \\
& 100-200~Hz & 0.20~g$^2$/Hz & 9.3~g-rms\\
& 200-400~Hz & 0.05~g$^2$/Hz & \\
& 400-2000~Hz & -15~dB/Oct & \\ \hline
\multirow{5}{*}{Y} & 20-100~Hz & +10~dB/Oct & \\
& 100~Hz & 0.05~g$^2$/Hz & \\
& 100-200~Hz & 0.20~g$^2$/Hz & 9.3~g-rms \\
& 200-400~Hz & 0.05~g$^2$/Hz & \\
& 400-2000~Hz & -15~dB/Oct & \\ \hline
\multirow{5}{*}{Z} & 20-100~Hz & +10~dB/Oct & \\
& 100~Hz & 0.05~g$^2$/Hz & \\
& 100-200~Hz & 0.70~g$^2$/Hz & 6.0~g-rms \\
& 200-400~Hz & 0.05~g$^2$/Hz & \\
& 400-2000~Hz & -15~dB/Oct & \\ \hline
\end{tabular}
\end{center}
\end{table}

\begin{figure}[tbp]
\begin{center}
\includegraphics[trim={0 0 0 1.5cm}, clip, width=0.7\textwidth]{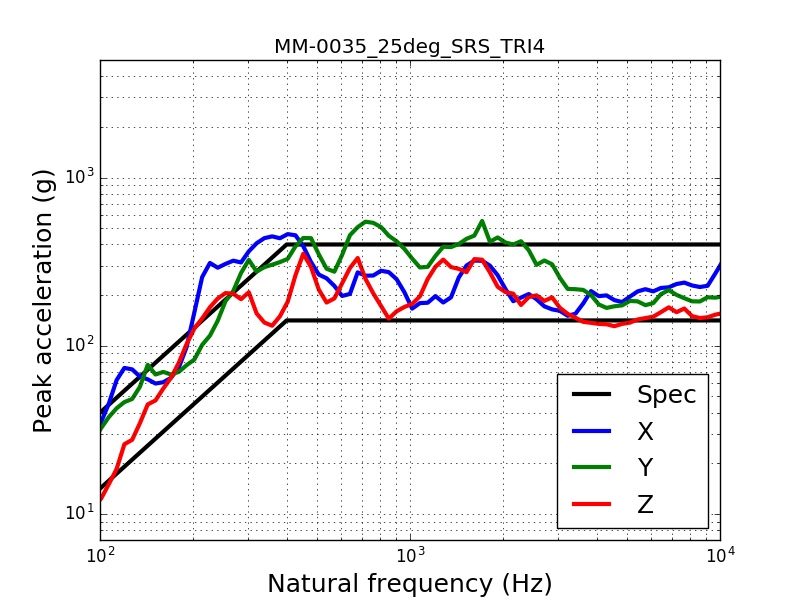}
\end{center}
\caption
{
Shock Response Spectrum (SRS) of the shock applied to a MM.
The qualification level with a margin of +6~dB and -3~dB define a specification (spec) band delimited by the black lines in the plot.
Despite the peak acceleration being outside of the specification band for some frequencies, this SRS is considered excellent, and the test successful.
Adapted from \cite{Girou_2021}.
}
\label{fig:srs}
\end{figure}

Shock testing is performed with a shock table and shock levels are defined by a shock response spectrum (SRS).
Before the test, a calibration of the ringing plate is performed using a dummy middle radius MM. 

Mechanical shims are used to preload the DP flexures similarly to the vibration test.
There, the preloading and SRS level simulate mirror structure distortions during one of the most dramatic event of the launch sequence as seen by the payload corresponding to the the release of the launch locks that will hold the mirror assembly during launch and are release once in space (shown in Figure \ref{fig:Petal_ISM}).
Figure \ref{fig:srs} displays a SRS as measured on the test mount (also called input SRS) of a MM.
Note that the SRS is isotropic, enabling the testing of all three axis in a single shock.
Two high-speed cameras, each looking at one side along the optical axis, are used to assist in identifying failures, as well as their modes.
Resonance searches are also performed to check on the structural integrity of the MM.

SPO stacks have been tested as standalone items, then again once assembled into MM for the different radii used for development (representative of the inner, middle and outer radii of Athena). At the time of writing, the actual Athena flight configuration is also starting to be tested with the first row-08 MM having passed both vibrations and shock tests at qualification levels (Figure \ref{fig:mm-0046}).
Such testing campaigns at qualification levels will continue and extend to all rows of Athena, further ruggedizing SPO technology in general.
For the flight model of Athena, each MM will undergo a workmanship test consisting of random vibration tests at acceptance level; additional environmental tests will be conducted at the mirror assembly level once the MMs are integrated into the mirror structure.

\subsection{X-ray characterization}
\label{sec:xray-characterization}
Optical metrology of the SPO stacks taken during manufacturing produces a wealth of information on the eventual X-ray performance of the optics. 
However, regular X-ray characterization remains an important part of the development process, as it is done at the same energy and incidence angle used during the actual use of the optics during flight.

\subsubsection{SPO stack characterization}
\label{subsec:stack_characterization}

Given the number of SPO stacks that are made, and the fact that individual stacks are not imaging systems, it has long been recognized that a dedicated facility with a high flux of X-rays is required to meaningfully support the development of the optics \cite{Collon:2006wz,krumrey_x-ray_2010,ackermann_performance_2009,krumrey_2016}. 
Characterization by means of a low-divergence ($\approx1\,\mathrm{arcsec}$) pencil beam of X-rays, as opposed to the more customary full-flood illumination, makes it possible to probe the surface of each plate in a stack with high spatial resolution, and arrive at the characterization of the surface slope errors and of the millimeter-scale defects contributing to the widening of the beam \cite{Vacanti:2019wr}.
At the time of writing, two dedicated beamlines are available in the PTB laboratory at the BESSY\,II synchrotron \cite{krumrey_x-ray_2010,krumrey_2016}, and an additional beamline is being assembled at ALBA \cite{Heinis:2021ud} (Section \ref{sec:mm}).

\begin{figure}[tbp]
    \centering
    \includegraphics[width=0.6\textwidth]{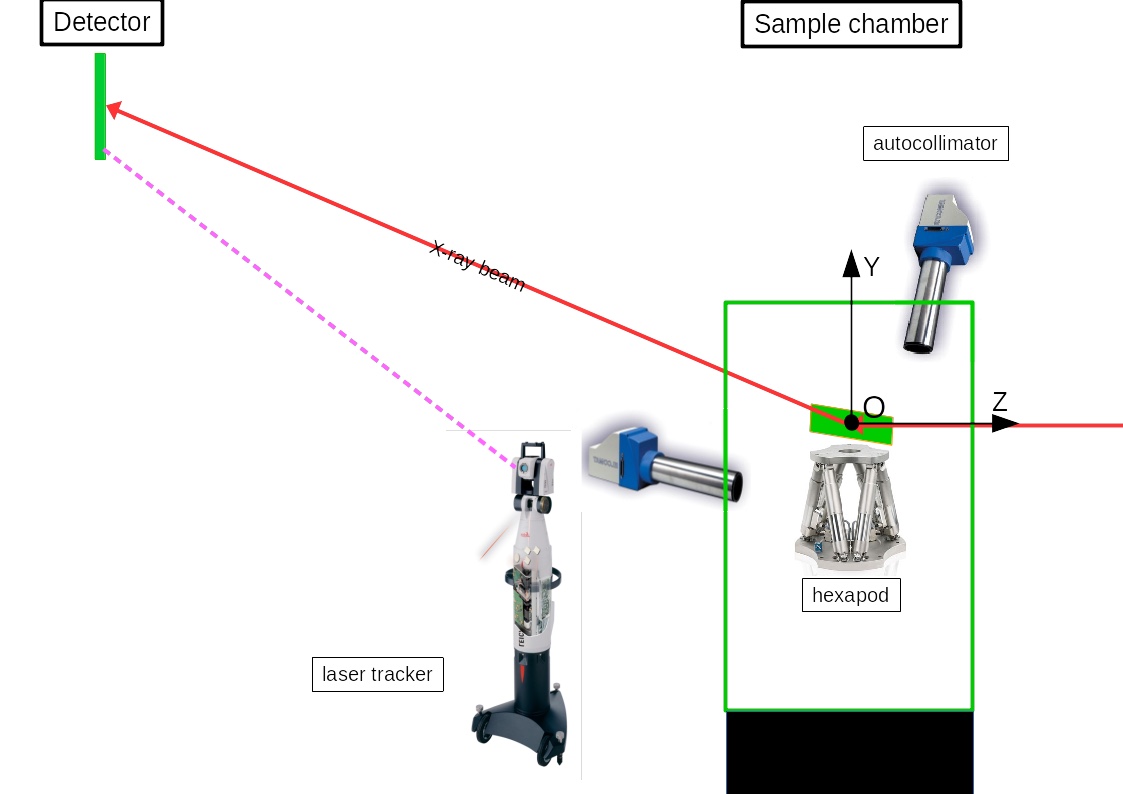}
    \caption{Logical view of a beamline dedicated to the assembly and characterization of SPO. The synchrotron X-ray beam travels from positive to negative Z (red). The sample is loaded inside a vacuum chamber (green) on a platform with six degrees of freedom (hexapod) and moved in front of the X-ray beam. During a measurement the incidence angle of the X-ray is kept constant through the use of two autocollimators. The reflected beam is intercepted by a detector, whose position is monitored with a laser tracker.
    Adapted from \cite{Vacanti:2019wr}.} 
    \label{fig:beamline-logical-02}
\end{figure}

For their characterization, the samples are moved in a rectangular pattern of columns and rows through a $100 \times 100\,\mathrm{\mu m}^2$ X-ray beam. 
In Figure~\ref{fig:beamline-logical-02} we show how the samples are placed in the beam and measured. Scan columns are spaced usually by $\approx 2\,\mathrm{mm}$, and scan rows are spaced by $100\,\mathrm{\mu m}$. 
This so-called \textit{standard scan pattern} probes only a fraction (about $5\,\%$) of the total reflecting area in an SPO stack. 
Tests have shown that the results obtained in this manner are representative of the complete surface. 
That is, making the scan columns more densely spaced, or using overlapping beams, does not add any new information on the properties of the stack derived from the data.

The data collected during a standard scan are processed to identify and isolate the reflected X-ray beam, then characterized by its centroid and RMS length and width.
After this morphological characterization, the positions of the images and the sample in the laboratory frame are determined and converted to a common coordinate frame for further analysis. 
Each reflected beam probes between $5$ and $10\,\mathrm{mm}$ along each of the pores, depending on the type of stack and the angle of incidence used in the measurement. 
Maps of the variation of the RMS length of the images already reveal good and bad areas on each plate.

A global quality indicator of the properties of the reflecting surface is derived by superimposing all the reflected images on their barycenters and calculating the half energy width (HEW) of the resulting image. 
This indicator is called the \textit{on centroid performance} of the sample and is usually expressed in units of the HEW of the direct beam, and it represents the average surface quality on length scales of the order of a few to several millimeters.

\begin{figure}[tbp]
    \centering
    \includegraphics[width=0.17\textwidth]{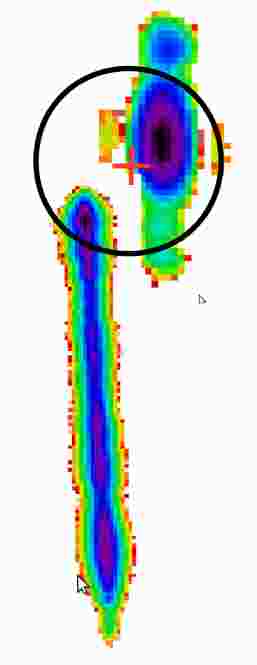}
    \caption{Illustration of the algorithm that leads to the construction of the on model centroid image for a stack. Shown
are two of the reflected beam images. The cross at the center indicates the place where the barycenters of the reflected
beams are expected to fall if the sample behaved ideally. A synthetic image, called the on model centroid image, is
obtained by adding each reflected beam to it after applying the translation calculated during the analysis. Credit: cosine.}
    \label{fig:omc}
\end{figure}

More complex metrics can be obtained by processing the data further. 
A model of the stack is introduced, and the expected position of the barycenters of the reflected X-ray beam is calculated. 
By comparing the expected and the measured positions, a displacement vector is obtained for each reflected image.
All the images are then superimposed on the common barycenter, where they would all end up if the SPO stack behaved ideally, and shifted by the displacement vector calculated. 
The HEW of the resulting image (Figure~\ref{fig:omc}) is a metric of the optical quality of the stack, and it represents the contribution of the stack to the optical quality of a pair of stacks arranged to form an imaging system. This indicator is called the \textit{on-model centroid performance} of the sample.

\begin{figure}[tbp]
  \centering
  \includegraphics[width=0.6\textwidth]{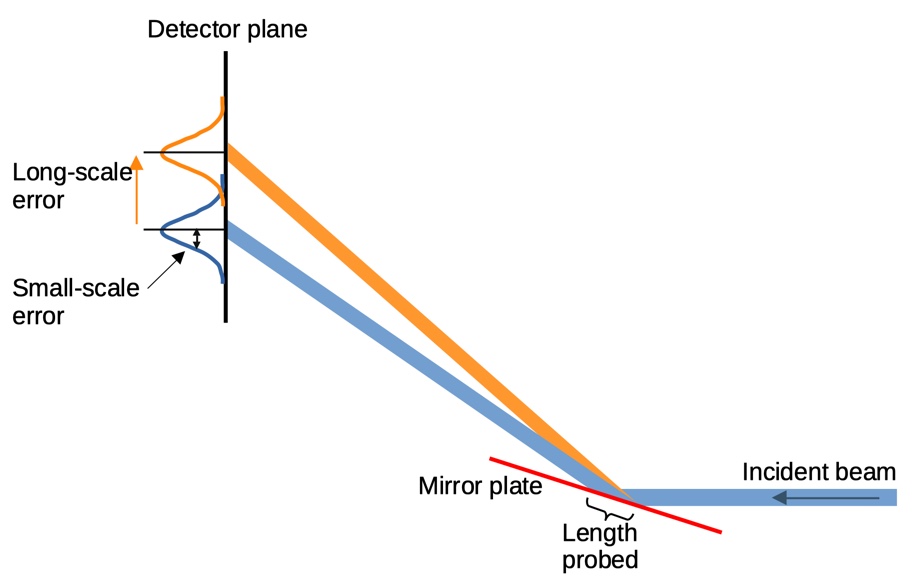}
  \caption{Depiction of the measurement process and two performance indicators derived from it.
  The X-ray beam, which has a finite cross section, impinges the mirror plate whose surface has some defects at grazing incidence angle.
  These defects shift the reflected beam compared to its expected position (long-scale figure errors) and broadens it (short length-scale errors). Credit: cosine.
  }
  \label{fig:stack-performance-indicators}
\end{figure}

Data from the raster measurement can be used to derive a variety of performance indicators related to the optical quality of the stack. 
We mention here three of them, illustrated in Figure~\ref{fig:stack-performance-indicators}, they are:
\begin{itemize}
\item The broadening of the reflected beam, a marker of short length-scale errors;
\item The (best-fit) meridional curvature, a marker of global figure errors;
\item The difference between the measured and expected position of the reflected beam taking into account the best-fit meridional curvature, a marker of long length-scale figure errors.
\end{itemize}

\begin{figure}[tbp]
    \centering
    \includegraphics[width=\textwidth]{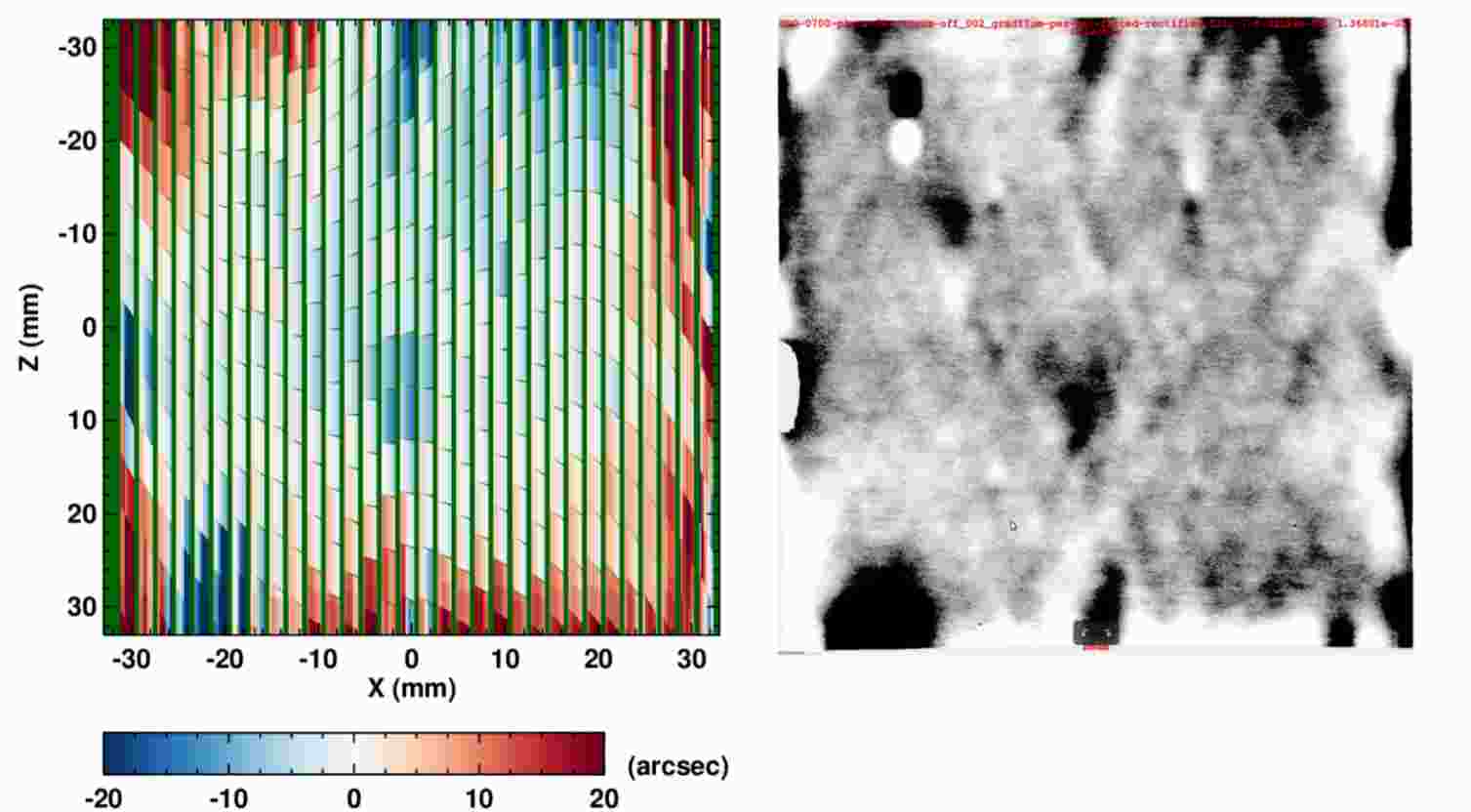}
    \caption{An example of a synoptic slope error map derived from X-ray data (left), compared to the cleanroom surface metrology for one of the
plates in the same stack. 
The two independent datasets are in excellent agreement.
Reproduced from \cite{Vacanti:2017wc}.}
    \label{fig:synoptic-slope-error-map}
\end{figure}

The same analysis that leads to the calculation of the displacement vectors is also able to estimate the local surface normal seen by the X-ray beam, and the difference between ideal and actual surface normal.
This information is coded in a \textit{synoptic slope error map}, an example of which is shown in Figure~\ref{fig:synoptic-slope-error-map}.
The map codes information for the entire stack: each block is made of a number of smaller rectangles, one for each plate, color coding the local slope error experienced by the X-ray beam (left to right for increasing plate number). 
These maps can be compared to similar per-plate maps obtained during manufacturing of the stack.
It can be seen in the Figure that the same features are visible in both maps.
This proves that cleanroom metrology provides information that is immediately relevant for the X-ray behavior of the optics. 
The map also shows that the error experienced by the X-ray beam at each location is nearly the same irrespective of plate number. 
The shape of the first plate in the stack is almost exactly reproduced across the entire stack.

\subsubsection{XOU and MM characterization}
\begin{figure}[tbp]
    \centering
    \includegraphics[width=\textwidth]{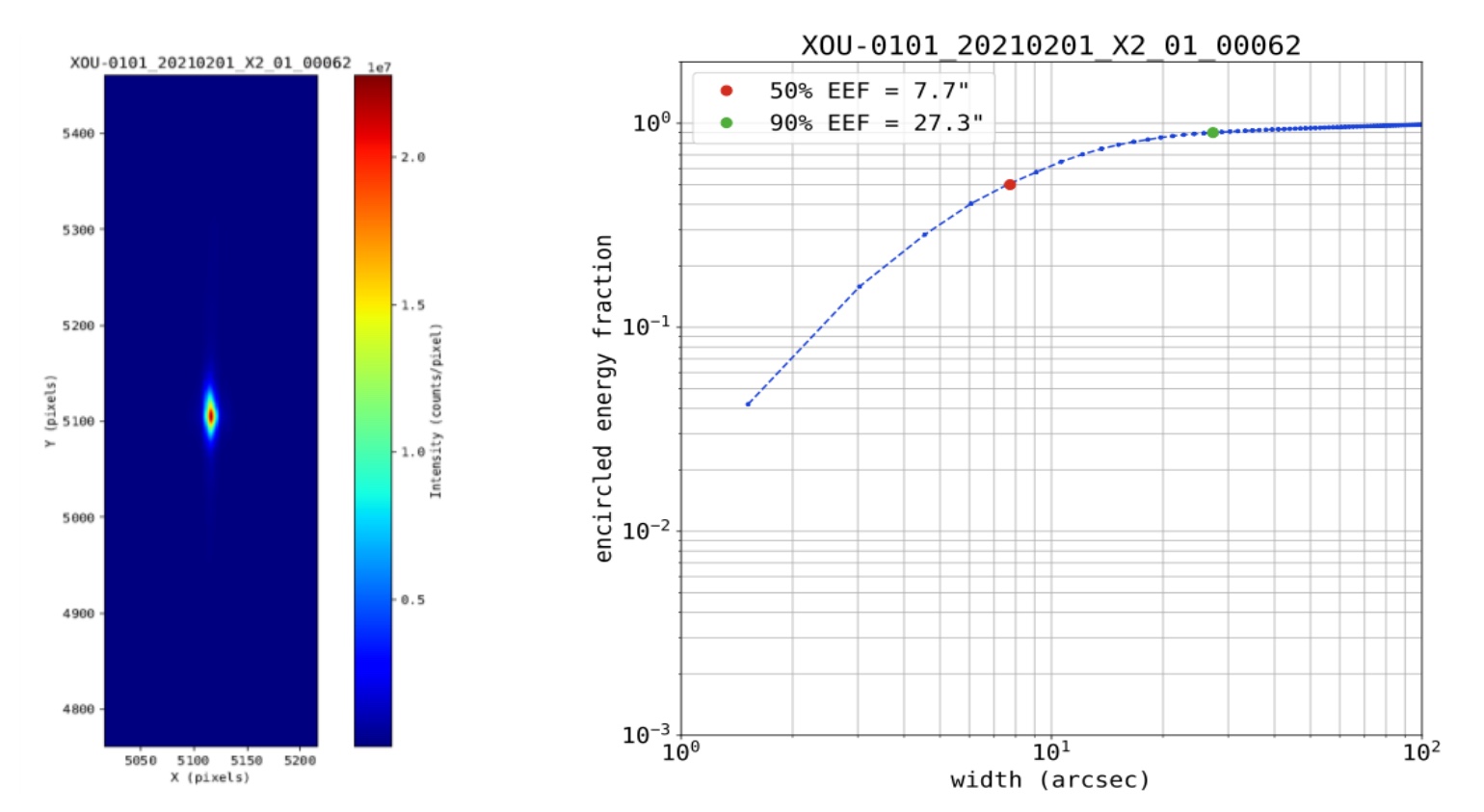}
    \caption{A sample PSF obtained by the algorithmic superposition of several thousands individual pencil-beam images. This particular image refers to the primary-secondary pair called XOU-0101, and it shows the result obtained for the central 70\,\% of the width of the sample (this has been a common selection used during the development of SPO). Reproduced from \cite{Collon:2021tk}.}
    \label{fig:psf}
\end{figure}

Once assembled into an imaging system (XOU), a primary-secondary stack doublet can be characterized with the same standard scan used for the characterization of single stacks. 
The analysis of the data proceeds along the same lines described in the previous section, but now the individual images can be combined to arrive at an estimate of the HEW of the imaging system (Figure \ref{fig:psf}). 
Owing to the use of a pencil beam, it is now possible to analyze the performance of the optics along different axes, for instance by looking at how the HEW changes with plate number. 
Other spatial selections are possible, leading to the ability to compare the performance of different regions of the optics, that can be related to the indicators derived from the cleanroom metrology.

\section{Athena}
\label{sec:athena}

\subsection{Introduction} 
\label{subsec:Athena_Intro}
The Athena mission – Advanced Telescope for High-ENergy Astrophysics - is an X-ray observatory designed to address the Cosmic Vision science theme 'The Hot and Energetic Universe'. The mission was selected as the second L-class mission in ESA's Cosmic Vision 2015–25 plan. The mission is in its final preparation stage (Phase B1) at the time of writing and is expect to proceed into implementation soon.
Athena will be an X-ray observatory with a wide energy range (0.2 -15 keV) and will include two innovative instruments to accomplish its science goals \cite{Nandra:2013ue}.

The first instrument, the X-ray Integral Field Unit (X-IFU) \cite{X-IFU_pajot}, provides spatially-resolved high-energy resolution spectroscopy, 2.5 eV, over a field of view of 5 arcmin, specifically aiming at spectroscopy of faint extended sources. This is an extremely advanced instrument requiring a complex cryogenic chain to allow the operation of a focal plane array based on transition edge sensors cooled down to 50~mK.

The second instrument, the Wide Field Imager  \cite{WFI_Norbert}, is a silicon-based detector using DEPFET active pixel sensor technology with around 150~eV energy resolution over a very large field of view of 40 arcmin, aiming at deep surveys.

\begin{figure}[h]
    \centering
    \includegraphics[width=\textwidth]{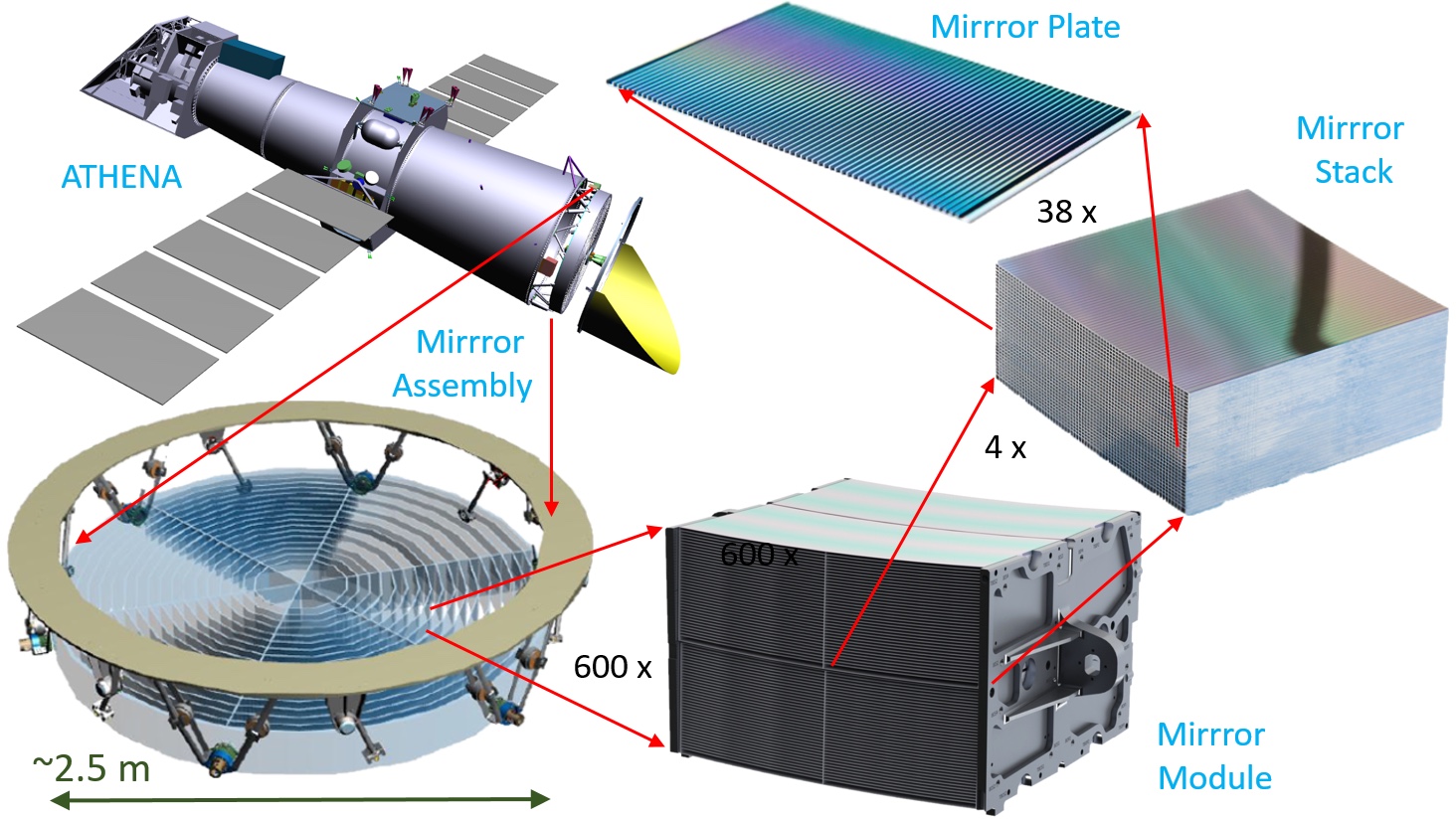}
    \caption 
    {
    The Athena Mirror Assembly (MA) is mounted to the spacecraft by a hexapod system, which allows the optics to be tilted in order to select one of the two focal plane instruments.
    The MA consists of a metallic structure supporting 600 Mirror Modules (MMs, each of which in turn is an assembly of 4 stacks of 38 mirror plates).
    This modular approach makes it possible and efficient to produce the optics using automated machines.
    Credit: ESA.
    }
   \label{fig:athena_overview} 
\end{figure}

To provide signal to these instruments, Athena requires the largest X-ray telescope ever built, targeting an effective around 1.4 m$^2$ at 1 keV and 5 arcsec HEW angular resolution (driven by the faint confusion limited survey mode for the WFI instrument). 
As illustrated in Figure~\ref{fig:athena_overview}, the Athena optics is coupled to the science instruments via  the spacecraft metering structure.
Supported by an hexapod system, the optics can be tilted to select the desired instrument, the X-IFU or the WFI.
The Athena Mirror Assembly (MA) consists of 600~SPO MMs mounted in metallic structure (details in Section~\ref{subsec:MA_design}).

Creating the optics technology for Athena is a formidable task, being addressed in a holistic and structured approach.
From the very beginning the performance requirements were considered together with the programmatic constraints, in particular time and cost to build the flight optics, and the compliance with the environmental conditions of launch and space operations. 
Modularisation and serial production are essential for the production of the Athena optics, given its size, and MMs are rugged by design.
Even the first optics demonstrator had to be compatible with mass production requirements of hundreds of mirror modules and hundred thousand mirror plates. 

The development of the Athena optics is funded and led by ESA, in very close collaboration with industrial and institutional partners.
With increasing maturity a well-coordinated consortium was formed, led by cosine Measurement Systems, which ensures a gradual, evolutionary transition from the study and development to the implementation phase of the Athena mission.

The size of the Athena optics also requires the creation of novel dedicated facilities to:
\begin{itemize}
    \item Assemble the MMs, which requires  several specialised synchrotron radiation beamlines (Section~\ref{sec:mm}),
    \item Populate the MA with the 600 MMs in the allocated time, with arcsecond and micron accuracy (Section \ref{subsec: MM_alignment}),
    \item Verify and calibrate the X-ray performance at the MA level (Section \ref{subsec:MA_x-ray_characterisation}).
\end{itemize}

We discuss hereafter the design and performances of the Athena optics as known at the time of writing, knowing that evolution is still likely to take place. 
The coordinate system in use is as follows: the Z axis is along the optical axis, with its direction along the propagation of the X-rays. The Y axis is radial, pointing outwards. X completes the orthogonal coordinates system.

\subsection{Optical design}
\label{subsec:Optical_design}

With SPO, a given optical design is realized by incrementally copying and modifying the figure provided by a pair of mandrels, one for the primary and one for the secondary. 
The mandrel defines the figure of a putative "zero plate" in the stack: each new plate copies the figure of the previous plate, and at the same time it introduces small changes in the figure, so as to keep the aberrations to a minimum.
The optical design of the Athena is based on the Wolter-Schwarzschild \cite{Chase:73} design of X-ray optics. 
The Wolter-Schwarzschild condition is only met exactly by one primary-secondary plate pair, called the reference plate pair. Once the relative orientation of this pair is set, the orientation of the remaining plates is given by their physical properties, and this leads to small departures from the ideal configuration.

As described above (Section \ref{sec:stacking}), a stack is built from larger to smaller radii. 
The shape of the plate at the larger radius is given by a mandrel that is designed to realize the ideal optical design at that radius. 
From there on, as the stack grows, and under the assumption that within a stack the plates have all the same geometry, each new plate takes a smaller sagittal radius until the desired number of plates has been stacked.

\begin{figure}
    \centering
    \includegraphics[width=0.65
\textwidth]{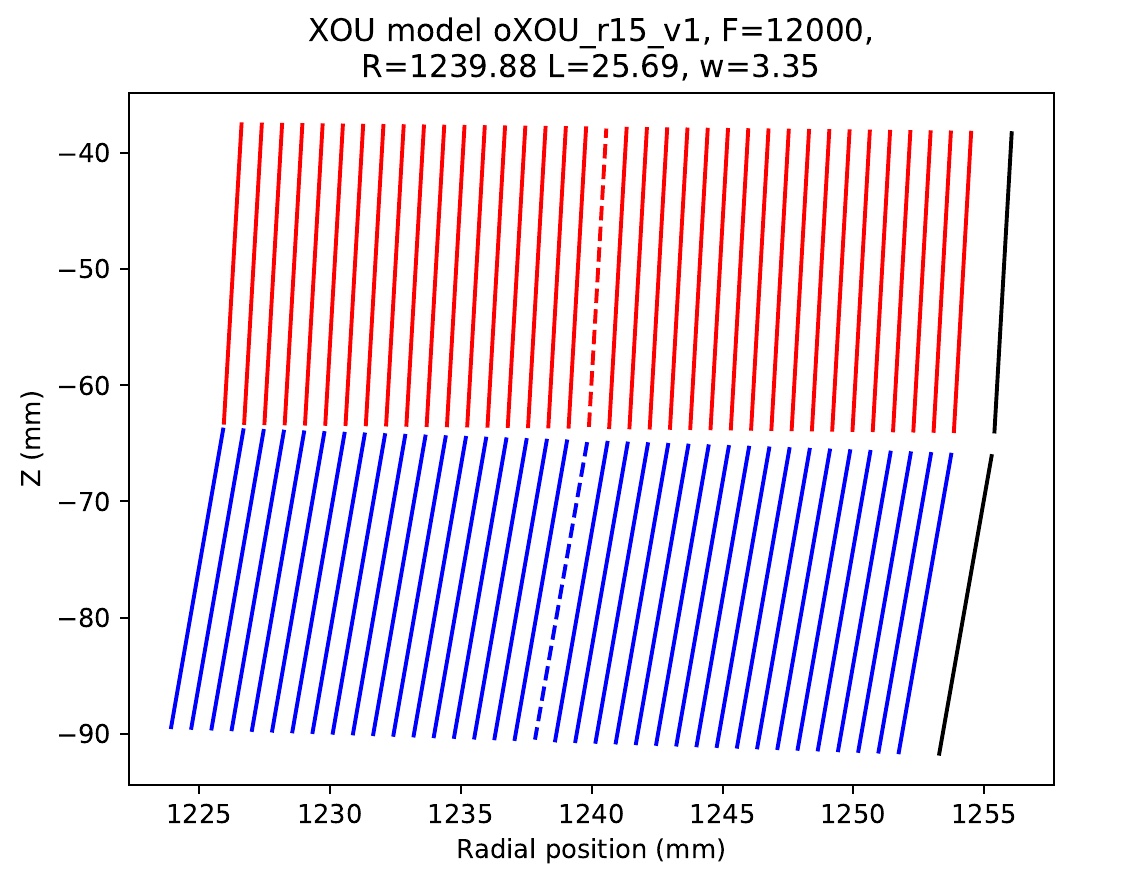}
    \caption{A cut through the central meridional plane of a primary-secondary stack pair from the outer XOU of row-15 Athena MM (see Table \ref{tab:MM_layout}). 
    The reference plate pair is indicated by the dashed segments, its radius at the (virtual) primary-secondary intersection is 1239.88 mm. The plate length $L$ in mm, and the wedge angle $w$ in arc-seconds are indicated in the title. The focal plane of the system at $Z=-12000\,\mathrm{mm}$ in this view. 
    As this is a W-S system, its Z position is closer to the focal plane by about $65\,\mathrm{mm}$.
    Credit: cosine.}
    \label{fig:xou-model}
\end{figure}

We have therefore the situation described in Figure~\ref{fig:xou-model}, where each reflecting surface has a different sagittal radius. The figure shows a cut through the central meridional plane of a primary and secondary stack. For the reference plate pair (dashed) the Wolter-Schwarzschild (W-S) condition holds:
\begin{equation}
  \label{eq:1}
  \alpha_r = \frac{1}{4}\arcsin\left(\frac{R_r}{F}\right)
\end{equation}
and
\begin{equation}
  \label{eq:2}
  \beta_r = 3 \alpha_r
\end{equation}
where $\alpha_r$ and $\beta_r$ are the angle that the primary and secondary plates form with the optical axis.

The W-S condition shows that plates must be placed at different angles with respect to the optical axis, depending on the (sagittal) radius they are at. For the primary stack the angle variation is then
\begin{equation}
  \label{eq:3}
  \Delta\alpha = \frac{\Delta R}{4F}\cos(4\alpha_r)
\end{equation}
and similarly for the secondary stack. This angle change is implemented in the plates, and is called the wedge of the plates (Section~\ref{subsec:plates}). 
If we take $F=12000$~mm, $R=750$~mm and $\Delta R = 0.775$~mm we arrive at values of the wedge of the order of 3.3 arcsec for the primary plates, and 3 times as large for the secondary plates. Other wedge combinations are possible \cite{ackerman_patent} and they are discussed in more detail elsewhere \cite{vacanti_2022}.

Improved optical performance can be obtained if instead of the conical approximation a meridional curvature is introduced in the profile of the reflecting surfaces.
This must be done by manufacturing suitable mandrels. Just as for conventional replicated shell optics, one can decide to introduce parabolic and hyperbolic profiles \cite{vanspeybroeck_design_1972}, or two spherical profiles \cite{saha_equal-curvature_2003}, or one conical profile and one spherical one \cite{willingale_improving_2010}. Which one to use depends on the parameters of the system and the tolerances that can be achieved in the manufacturing of the mandrels. For Athena the choice has been made to place all the meridional curvature in the secondary stacks, and across the aperture this leads to a meridional radius of curvature of  1 to 4 $\,\mathrm{km}$.

Even with the introduction of a meridional curvature, there are practical manufacturing considerations that contribute to the degradation of the angular resolution. These are discussed in detail elsewhere \cite{vacanti_2022}. Here we mention that when building a large segmented X-ray mirror with hundreds of stacks and more than $90\,000$ individual plates, it becomes necessary to reduce complexity by manufacturing standard plates; in particular, stacks of the same central sagittal radius will be manufactured to have plates of the same length and with the same wedge, although an ideal design would call for gradual changes in both parameters. These small approximations imply that the best on-axis optical performance that can be achieved by a perfect optical design is of the order of 0.1--0.2~arcsec.

\subsection{Design and expected performance for the Athena optics}
\label{subsec:MA_design}

\begin{figure}[p]
    \centering
    \includegraphics[width=0.6\textwidth]{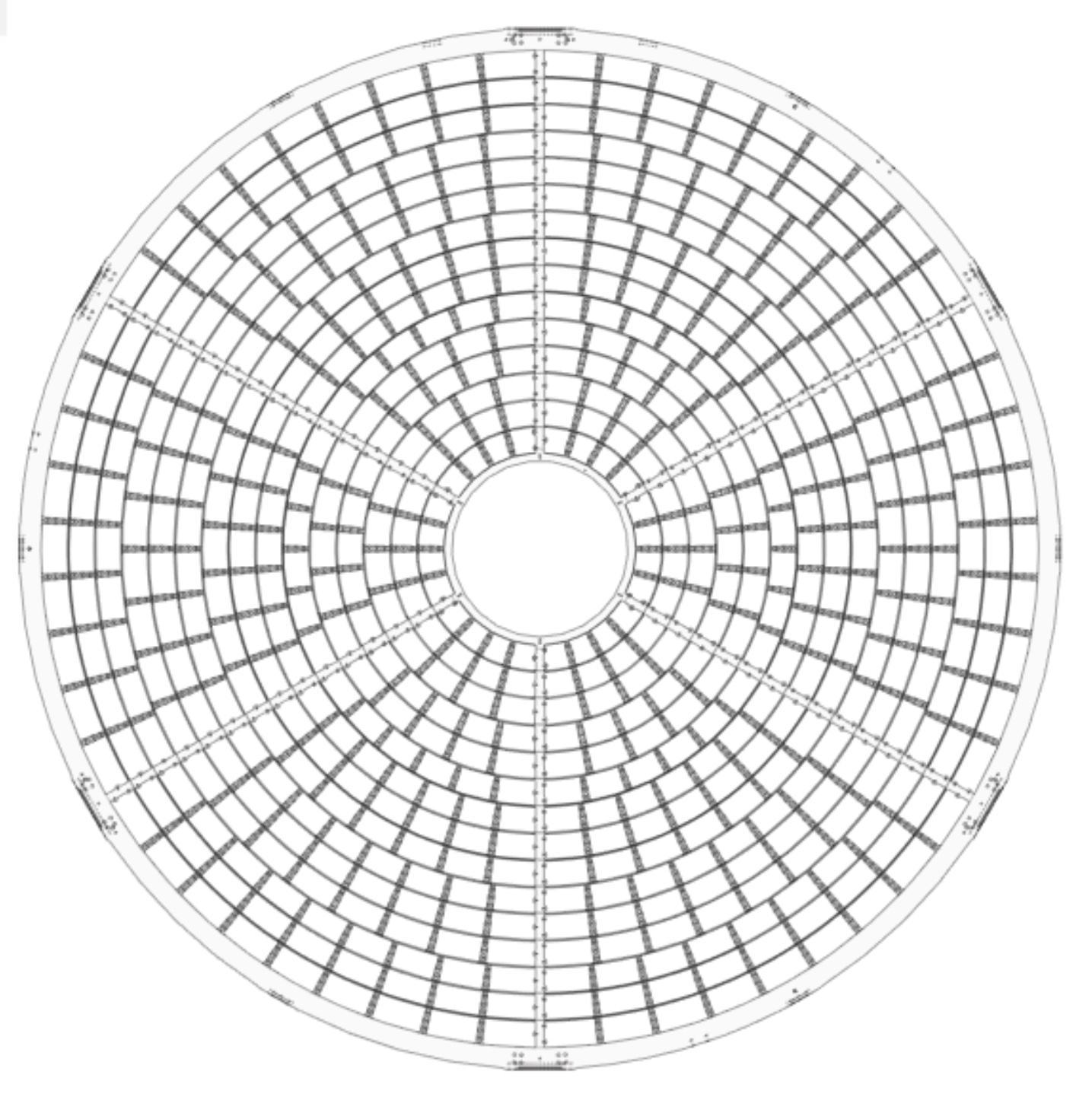}
    \caption 
    { 
    Athena Mirror Assembly (MA) 15-row layout in which each of the six sectors is populated with 100 MMs.
    Credit: ESA.
    }
    \label{fig:mirror_assembly} 
\end{figure}

The 600 MMs of the Athena MA are organized in 15 radial rows.
The assembly is modular to the row level, where the MM design for a particular row follows the design rules described in Section~\ref{subsec:Optical_design}.
The MMs are placed onto a supporting monolithic Mirror Structure (MS) that has many pockets designed to accommodate the different geometries of the MMs. There is a six-fold symmetry (six petals): structural reinforcement is introduced with six radial spokes running between an inner and an outer rim as shown in Figure~\ref{fig:mirror_assembly}.
The MMs are mounted using sets of DPs which are isostatic mounts that prevent the distortions on the MS to percolate onto the MMs.
These DPs are glued to both the MS and the MM brackets.

\begin{figure}[p]
    \centering
    \begin{tabular}{c}
        \includegraphics[width=9.5cm]{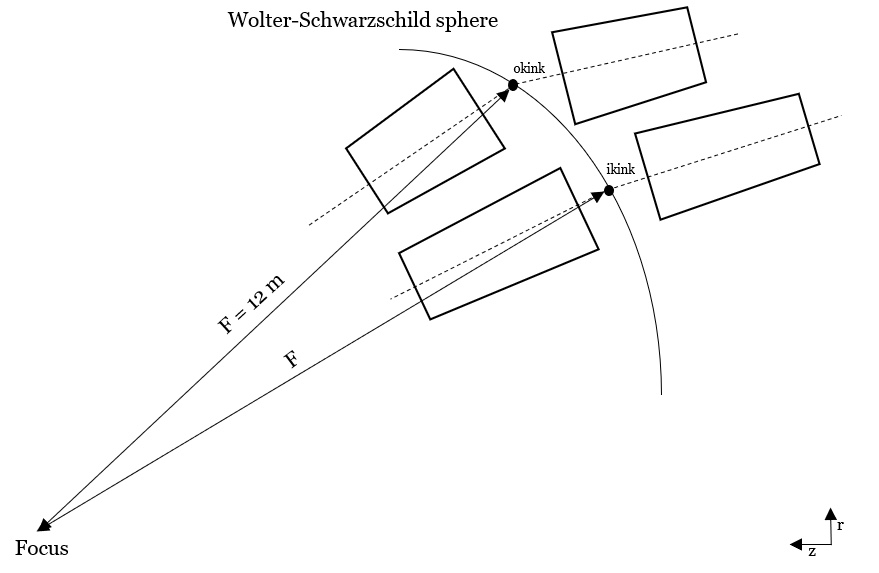}
        \includegraphics[width=1.5cm]{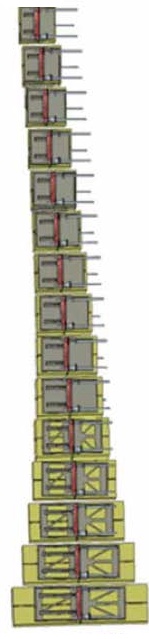}
    \end{tabular}
    \caption 
    { 
    MM layout in Wolter-Schwarzschild configuration.
    (Left) Within a MM, the reference points of the XOUs are placed along a sphere of 12 m centered to the focus. The points {\it ikink} and {\it okink} mark the intersection point between the primary surface and the secondary surface of the reference plates of the inner XOU and outer XOU, respectively.
    (Right) Radial cut-out of the MMs along the aperture (note the shift to place them along the Wolter-Schwarzschild sphere).
    Credit: ESA.
    }
    \label{fig:wolter} 
\end{figure}

The kinks of the reference plates of each XOU (one MM currently includes two XOUs but can be just a single XOU, see Section~\ref{subsec:Optical_design}) are placed on a spherical surface with a radius of 12~m (focal length) centred in the design telescope focus (see Figure~\ref{fig:wolter}), forming an almost perfect Wolter-Schwarzschild configuration  \cite{Chase:73} which significantly reduces coma aberrations and therefore provides better angular resolution averaged over the field of view than Wolter I configuration.

The width and radial position of each MM design is chosen to maximize the effective area, guaranteeing the necessary clearance to accommodate the mirror structure, the necessary thermal equipment, the range needed for the MM alignment, and the dynamic and thermal distortions expected during launch and operations \cite{10.1117/12.2594443}.
An iterative geometric model ensured all the constraints were met which resulted in a layout as described in Table~\ref{tab:MM_layout}.

The MS is thermally controlled with the help of hundreds of heaters and several tens of thermistors.
This thermal control system is needed in order to minimize the thermoelastic distortions (and the corresponding optical performance degradation), and to allow the MMs to be kept close to the temperature set-point at which they were assembled and verified on ground ($20\pm5~^\circ$C).
At the center of the MS, as close as possible to the mirror node, there are a number of star trackers and other metrology elements needed to achieve the required knowledge of the telescope line of sight (absolute knowledge of 10~arcsec, and relative knowledge of 1~arcsec over 100~ksec).

\begin{figure}[tbp]
\centering
    \begin{tabular}{cc}
        \includegraphics[width=6cm]{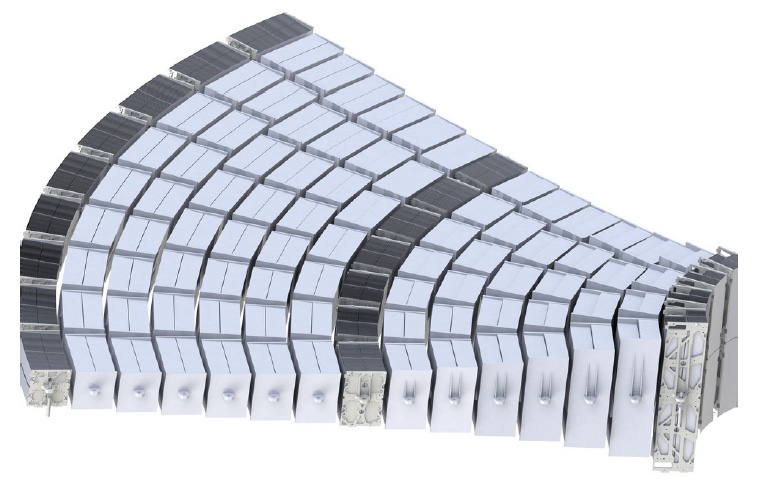}&
        \includegraphics[width=6cm]{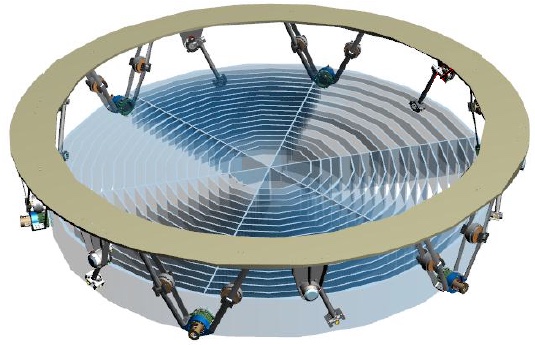}\\
        \multicolumn{2}{c}{}
        \includegraphics[width=12.5cm]{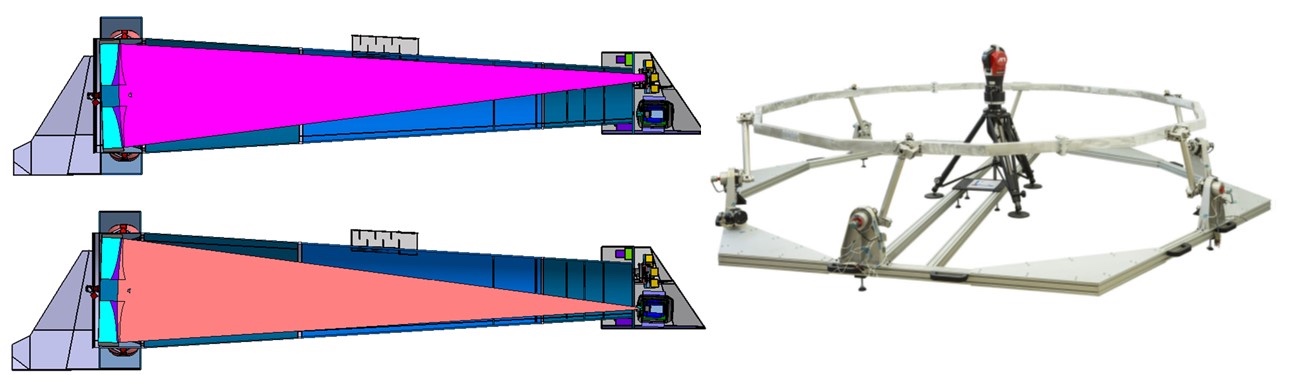}\\
    \end{tabular}
    \caption{
    (Left) Distribution of the SPO MMs along one of the petals.
    Credit: cosine.
    (Right) The Athena MA including the 2.6~m diameter Aluminum Mirror Structure (MS) with the 600~pockets, mounted on the launch locks (HDRMs) and the Instrument Switching Mechanism (ISM).
    Credit: ESA/SENER Sp. z o.o.
    (Bottom) The ISM allows switching the telescope focus between the two science instruments of Athena (the WFI above and the X-IFU below).
    Credit: ESA/SENER Sp. z o.o.
    }
    \label{fig:Petal_ISM}
\end{figure}

The MA consisting of the MS, SPO MMs, thermal control system, and other elements is held during launch with the help of six launch lock bipods able to hold up to 60-100~kN each (Figure~\ref{fig:mirror_assembly}).
Once in orbit, these launch locks, which are Hold Down and Release Mechanisms (HDRMs), are released.
The MA can then be steered with the help of a set of very accurate actuators in a hexapod configuration, referred to as the Instrument Switching Mechanism (ISM) (Figure~\ref{fig:Petal_ISM}).

\begin{table}[tbp]
\centering
\caption{
Main parameters of the MM distribution along the MA aperture. iXOU and oXOU refer to the inner XOU and the outer XOU, respectively, of the mirror module of a given row. The radii are given at the kink of the reference plate (noted ref. pl.), i.e. at the (virtual) intersection of the primary and and secondary plates. The angles are given for the reference plate of the primary stack.
}
\label{tab:MM_layout}
\resizebox{\textwidth}{!}{
\begin{tabular}{|c|c|c|c|c|c|c|c|c|}
\hline
\textbf{Row} &
\textbf{\begin{tabular}[c]{@{}c@{}}Number \\ of \\ MMs
\end{tabular}} &
\textbf{\begin{tabular}[c]{@{}c@{}}iXOU \\ radius ref. pl. \\ (mm)\end{tabular}} & \textbf{\begin{tabular}[c]{@{}c@{}}iXOU \\ pl. length \\ (mm)\end{tabular}} & \textbf{\begin{tabular}[c]{@{}c@{}}iXOU \\ angle ref. pl. \\ (deg)\end{tabular}} & \textbf{\begin{tabular}[c]{@{}c@{}}oXOU \\ radius ref. pl. \\ (mm)\end{tabular}} & \textbf{\begin{tabular}[c]{@{}c@{}}oXOU \\ pl. length \\ (mm)\end{tabular}} & \textbf{\begin{tabular}[c]{@{}c@{}}oXOU \\ angle ref. pl. \\ (deg)\end{tabular}} & \textbf{\begin{tabular}[c]{@{}c@{}}iXOU, oXOU \\  pl. width \\ (mm)\end{tabular}} \\ \hline
1 & 24 & 259.5 & 122.84 & 0.31 & 289.61 & 110.08 & 0.346 & 40.22 \\ \hline
2 & 24 & 327.7 & 97.29 & 0.392 & 357.81 & 89.11 & 0.428 & 58.08 \\ \hline
3 & 24 & 395.9 & 80.84 & 0.473 & 426.01 & 74.85 & 0.509 & 75.94 \\ \hline
4 & 30 & 464.1 & 68.71 & 0.555 & 494.21 & 64.52 & 0.59 & 71.14 \\ \hline
5 & 30 & 532.31 & 59.91 & 0.636 & 562.42 & 56.7 & 0.672 & 85.42 \\ \hline
6 & 36 & 600.51 & 53.1 & 0.718 & 630.62 & 50.56 & 0.753 & 79.84 \\ \hline
7 & 42 & 668.72 & 47.68 & 0.799 & 698.82 & 45.63 & 0.835 & 75.85 \\ \hline
8 & 42 & 736.92 & 43.26 & 0.881 & 767.03 & 41.57 & 0.917 & 86.06 \\ \hline
9 & 42 & 805.12 & 39.6 & 0.962 & 835.23 & 38.17 & 0.998 & 96.26 \\ \hline
10 & 48 & 872.57 & 36.43 & 1.043 & 902.67 & 35.31 & 1.079 & 90.62 \\ \hline
11 & 48 & 940.02 & 33.91 & 1.124 & 970.12 & 32.85 & 1.16 & 99.45 \\ \hline
12 & 48 & 1007.47 & 31.63 & 1.204 & 1037.56 & 30.71 & 1.24 & 108.28 \\ \hline
13 & 54 & 1074.91 & 29.64 & 1.285 & 1105.01 & 28.83 & 1.321 & 101.93 \\ \hline
14 & 54 & 1142.35 & 27.89 & 1.366 & 1172.44 & 27.17 & 1.402 & 109.77 \\ \hline
15 & 54 & 1209.79 & 26.33 & 1.447 & 1239.88 & 25.69 & 1.483 & 117.62 \\ \hline
\end{tabular}
}
\end{table}

The ISM enables the focusing of the telescope onto one of the two instruments, and even to different locations within their respective detector arrays in a large range of motion (including movements along the line of sight from +5 to -35~mm).
This flexibility is also key to compensate for misalignments introduced during launch or due to thermal effects.
The requirement is to be able to position the focal point of the telescope with an accuracy of $\pm250$ $\mu$m onto the detector plane (with a much tighter stability of only $\pm11$ $\mu$m over 100~ks) and $\pm50$ $\mu$m along the line of sight.

\subsubsection{Effective area}
\label{subsec:MA_effective_area}

\begin{figure}[p]
\centering
    \includegraphics[width=0.8\textwidth]{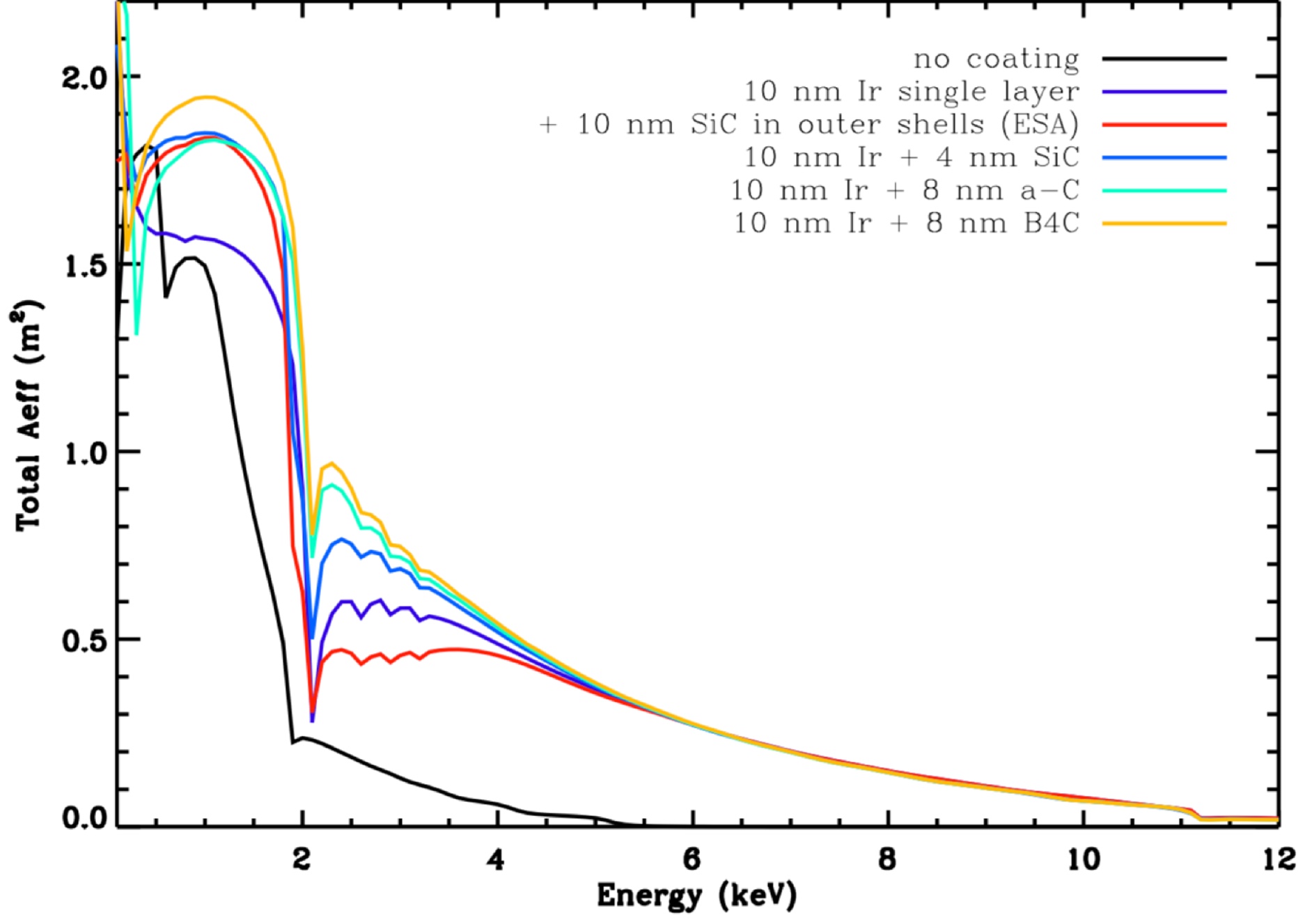}
    \caption{
    Total effective area of the MA as a function of energy for different coating configurations (The 10 to 20\% expected losses discussed in the text are not accounted for in these curves).
    Note that a significant increase in effective area can be expected at 1~keV if e.g. a B$_4$C overcoat is used.
    Reproduced from \cite{10.1117/12.2313275}.
    }
    \label{fig:A_eff_coatings}
\end{figure}

The reference telescope design has been simulated using a ray tracing software \cite{10.1117/12.2530682} and is estimated to deliver the effective area performance shown in Table \ref{tab:aeff_table_per_row}.
This performance estimation assumes that the mirror plates are coated with a recipe consisting of iridium and an overcoat of silicon carbide (SiC).
This coating recipe is still assumed as the baseline at the time of writing due to the demonstrated compatibility of both layers with the SPO process, in particular cleaning and activation  \cite{10.1117/12.2562430}. 
Nevertheless, there is a considerable effort ongoing to verify other overcoats like B$_4$C or carbon, with very promising results.
Figure~\ref{fig:A_eff_coatings} shows the differences in effective area for different coating recipes, showing that a significant improvement can be expected at the low energies for overcoats of either B$_4$C or carbon (see also Section~\ref{subsec:development-of-coatings}).
Multilayer recipes will also be explored to provide effective area enhancement close to the Fe line, which will be particularly relevant to the X-IFU instrument.

\begin{table}[tbp]
\centering
\caption{
Effective area as a function of MM row and as a function of energy (10-20\% losses still need to be added on top – see below) \cite{10.1117/12.2594443}.}
\label{tab:aeff_table_per_row}
\begin{tabularx}{0.8\textwidth} {|>{\centering\arraybackslash}X|>{\centering\arraybackslash}X|>{\centering\arraybackslash}X|>{\centering\arraybackslash}X|>{\centering\arraybackslash}X|>{\centering\arraybackslash}X|>{\centering\arraybackslash}X|}
\hline
\textbf{Row} & \textbf{\begin{tabular}[c]{@{}c@{}}0.2 keV \\ (cm$^2$)\end{tabular}} & \textbf{\begin{tabular}[c]{@{}c@{}}0.35 keV \\ (cm$^2$)\end{tabular}} & \textbf{\begin{tabular}[c]{@{}c@{}}1 keV \\ (cm$^2$)\end{tabular}} & \textbf{\begin{tabular}[c]{@{}c@{}}2.5 keV \\ (cm$^2$)\end{tabular}} & \textbf{\begin{tabular}[c]{@{}c@{}}7 keV \\ (cm$^2$)\end{tabular}} & \textbf{\begin{tabular}[c]{@{}c@{}}10 keV \\ (cm$^2$)\end{tabular}} \\ \hline
1 & 320 & 313 & 311 & 242 & 271 & 263 \\ \hline
2 & 463 & 450 & 447 & 325 & 364 & 305 \\ \hline
3 & 599 & 579 & 573 & 390 & 422 & 90 \\ \hline
4 & 700 & 672 & 665 & 420 & 405 & 6 \\ \hline
5 & 827 & 790 & 780 & 456 & 279 & 1 \\ \hline
6 & 916 & 869 & 856 & 461 & 62 & 0 \\ \hline
7 & 1001 & 944 & 928 & 456 & 11 & 0 \\ \hline
8 & 1114 & 1045 & 1024 & 456 & 3 & 0 \\ \hline
9 & 1259 & 1210 & 1348 & 374 & 4 & 0 \\ \hline
10 & 1335 & 1279 & 1434 & 258 & 1 & 0 \\ \hline
11 & 1432 & 1367 & 1536 & 158 & 0 & 0 \\ \hline
12 & 1523 & 1449 & 1636 & 88 & 0 & 0 \\ \hline
13 & 1584 & 1501 & 1697 & 54 & 0 & 0 \\ \hline
14 & 1665 & 1571 & 1775 & 48 & 0 & 0 \\ \hline
15 & 1740 & 1636 & 1844 & 55 & 0 & 0 \\ \hline
\textbf{Total} & \textbf{16478} & \textbf{15675} & \textbf{16854} & \textbf{4241} & \textbf{1822} & \textbf{665} \\ \hline
\end{tabularx}
\end{table}

\begin{figure}[p]
    \centering
    \includegraphics[width=0.7\textwidth]{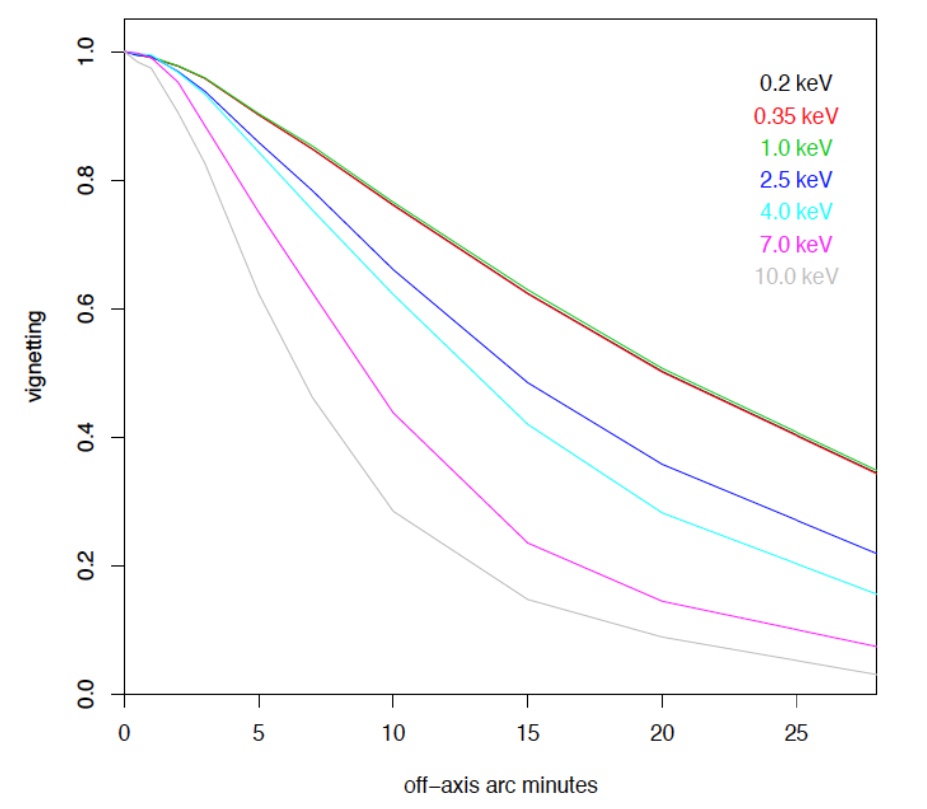}
    \caption{
    MA vignetting curves at the relevant energies. These curves are made by normalizing the effective area to 1 for on-axis illumination.
    Reproduced from \cite{10.1117/12.2594443}.
    }
    \label{fig:MA_vignetting}
\end{figure}

The above estimation includes the losses due to the following effects:
\begin{itemize}
    \item fixed length of plates within a stack,
    \item -1/+1 wedge optical configuration,
    \item alignment and figure errors,
    \item surface roughness of the reflecting surfaces,
    \item variable axial gap between the primary and secondary stack,
    \item parallel ribs,
    \item plate stacking rotation error.
\end{itemize}

All of these losses (included in the numbers in Table~\ref{tab:aeff_table_per_row}) add up to around 10\% at 1~keV but go up to 17.5\% at 7~keV.

However, there are some additional losses that are not included in the effective area numbers reported here which should be applied on top of the effective area numbers presented. These include:
\begin{itemize}
    \item coating imperfections,
    \item particulate contamination (initial and accumulated until launch),
    \item mirror module rotations around X,Y axes,
    \item misalignment between MA optical axis and line of sight,
    \item misalignment between line of sight and target,
    \item margins (other causes not identified yet).
\end{itemize}

The amount to be allocated for these additional losses is still uncertain in many cases.
It is expected to add up to $\sim$ 10-20\% loss depending on the energy, to be subtracted from the numbers in Table~\ref{tab:aeff_table_per_row}.

\subsubsection{Vignetting}
The estimated vignetting factor for the MA was calculated for different energies (Figure~\ref{fig:MA_vignetting}).
The steeper vignetting curve at higher energies reflects the fact that the main contributor for the effective area at these energies are the inner radii MMs.
These inner MMs have longer plates and therefore are more susceptible to shadowing by both the membranes and the ribs with off-axis illumination.
The need to limit the effective area loss at these high energies is the driver for the alignment of the optical axis with respect to the line of sight, which needs to be accurate to the level of a few arc-seconds.

\subsection{Mirror module alignment}
\label{subsec: MM_alignment}

In order to achieve the 5~arcsec HEW angular resolution for the Athena MA, a novel method was developed to enable the alignment of the MMs into the MS with an accuracy better than 1.5~arcsec (1~arcsec as a goal).
There are a number of other critical requirements to address during MM integration: the cleanliness level needs to be maintained to ISO5/6 level, the process has to allow a rate of 2~MMs/day (in order to be compatible with the schedule allocation for the flight programme), it should be possible to set an arbitrary sequence of the integration and there should be the possibility to remove, re-align or replace any MM along the aperture.

\begin{figure}[h]
\centering
    \begin{tabular}{lr}
    \includegraphics[width=5.75cm]{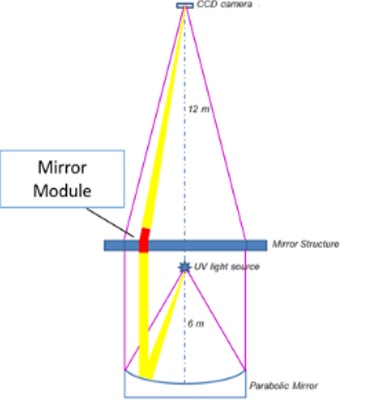} & 
    \includegraphics[width=5.75cm]{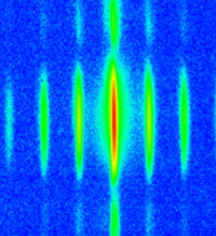}\\
    \multicolumn{2}{c}{
    \includegraphics[width=11.5cm]{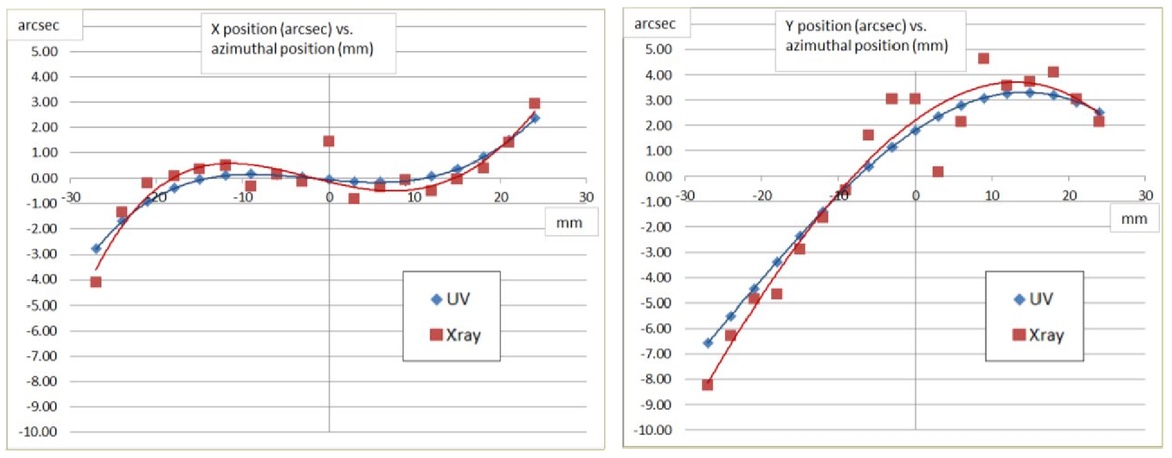}}
    \end{tabular}
    \caption{
    (Top left) Concept of using a UV collimator to generate a highly collimated wavefront to align the MMs onto the MA.
    (Top right) Measured diffracted PSF obtained with UV illumination on a MM.
    (Bottom) Very good agreement between the centroid of the PSF obtained with UV and X-ray illumination (X-ray curve obtained at the PANTER facility, and UV curve obtained with UV facility \cite{10.1117/12.2272997}.
    Credit: Media Lario.
    }
    \label{fig:AIT_process}
\end{figure}

The MM alignment concept was developed at Media Lario (Bosisio Parini, Italy). It consists of a vertical optical bench to capture the focal plane image of each SPO MM while illuminated by a reference plane wave at a wavelength of 218~nm (UV).
This is possible because it has been confirmed by simulations and experimentally that the centroid of the PSF is at the same location in UV and X-rays  \cite{10.1117/12.2272997} (Figure~\ref{fig:AIT_process}).
Working with UV illumination simplifies the alignment operation and allows the use of space-qualified gluing processes in air.

The light emitted by the UV source is reflected by a parabolic mirror to generate a beam collimated to better than 95~km effective source to MM distance, thus approximating illumination from a source at infinity.
Each MM focuses the collimated beam onto a CCD camera placed at the focal position and the acquired PSF is processed in real time to calculate the PSF centroid position and intensity parameters (Figure~\ref{fig:AIT_process}).
This information is then used to guide the robot-assisted alignment sequence, which makes use of a manipulator providing six degrees of freedom.
Four of those result in a change of the position of the PSF (correction of in-plane and focus errors), and the other two result in a change of the intensity (effective area).
A robotic arm holds the MM during glue curing and can correct the alignment during curing using real-time feedback from laser trackers. 

\begin{figure}[htpb]
\centering
    \includegraphics[width=0.92\textwidth]{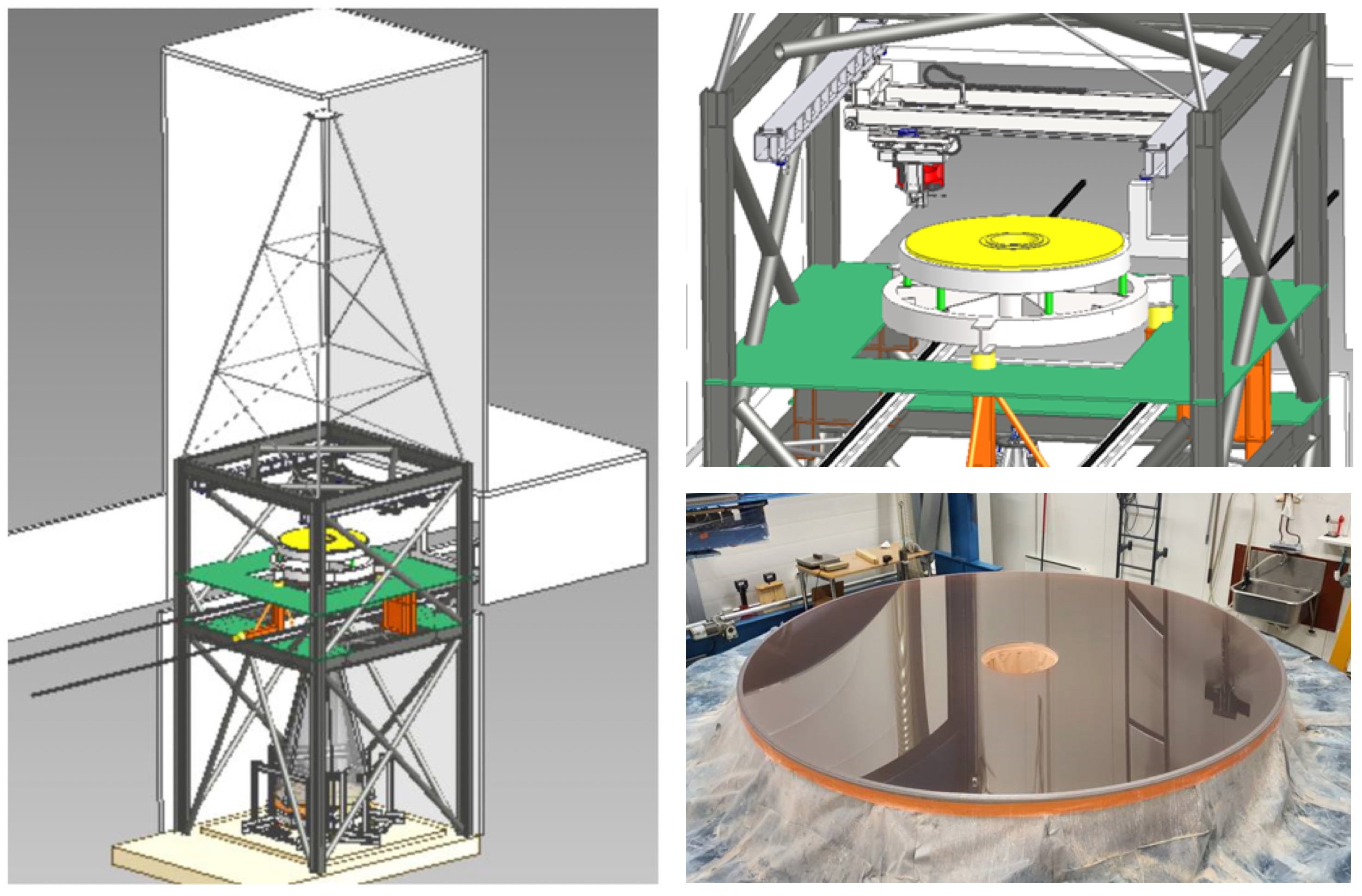}
    \caption{
    (Left) Facility where the MMs are integrated in the mirror structure using a collimated vertical UV beam. The 2.6~m diameter UV collimator is at the bottom (6.5~m below ground floor), the MA (in yellow) is placed at the level of the ground floor, and the UV detector is at the top, 12~m away from the MA.
    (Bottom right) The 2.6~m diameter zerodur UV collimator being polished.
    (Top right) close-up view of the MA with isostatic mounts and force actuators to minimise sagging. 
    Note also the robotic system for the alignment of the MMs (in red).
    Credit: Media Lario, ADS International, BCV Progetti, Opteon.
    }
    \label{fig:SPO_AIT}
\end{figure}

The implementation of this concept is ongoing at the time of writing.
The facility requires a building able to accommodate a large UV collimator with 2.5~m diameter (to cover the entire Athena aperture), the MA, and the detector plane necessary for the verification of the PSF centroid position.
This is accomplished in a building where the UV collimator is placed around 6.5~m below the ground floor, while the robotic integration and glue injection occur at the level of the ground floor, and the detector is placed 12~m away (total building height of 17~m) to accommodate other necessary equipment (Figure~\ref{fig:SPO_AIT}).

The UV detector is held in a position 12~m away from the MA, it can be mounted directly on the walls of the clean room of the building or supported by a CFRP structure.
Since there is the ability to monitor real time the position of the detector and the MA with the help of an accurate laser tracker system, any relative movements introduced by the building can be monitored and removed in post processing (drifts can also be corrected with the help of a linear actuator stages on which the detector is mounted).

\begin{figure}[htbp]
\centering
    \begin{tabular}{lcr}
        \includegraphics[width=7.6cm]{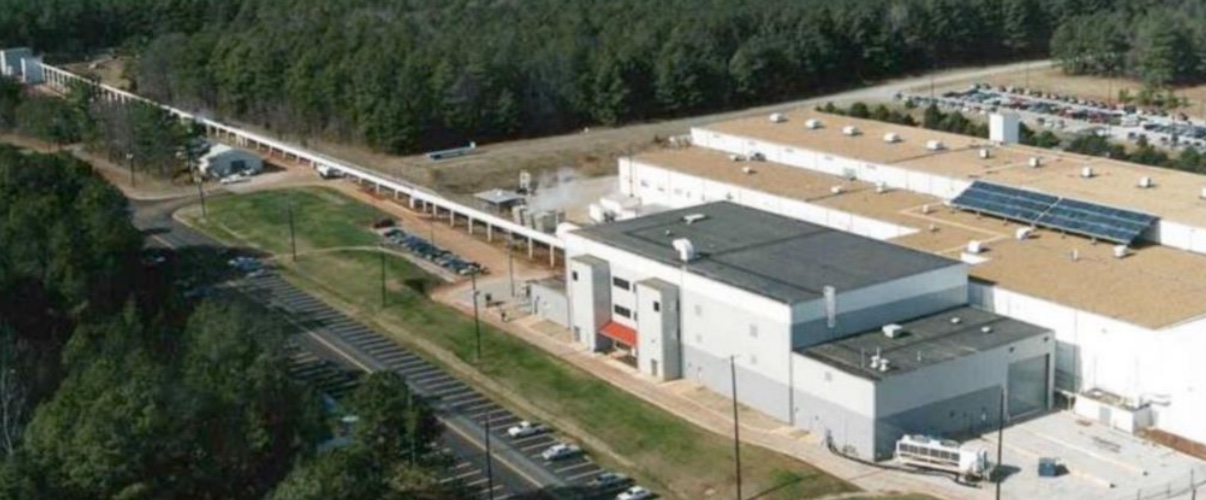} &
        \includegraphics[width=3.4cm]{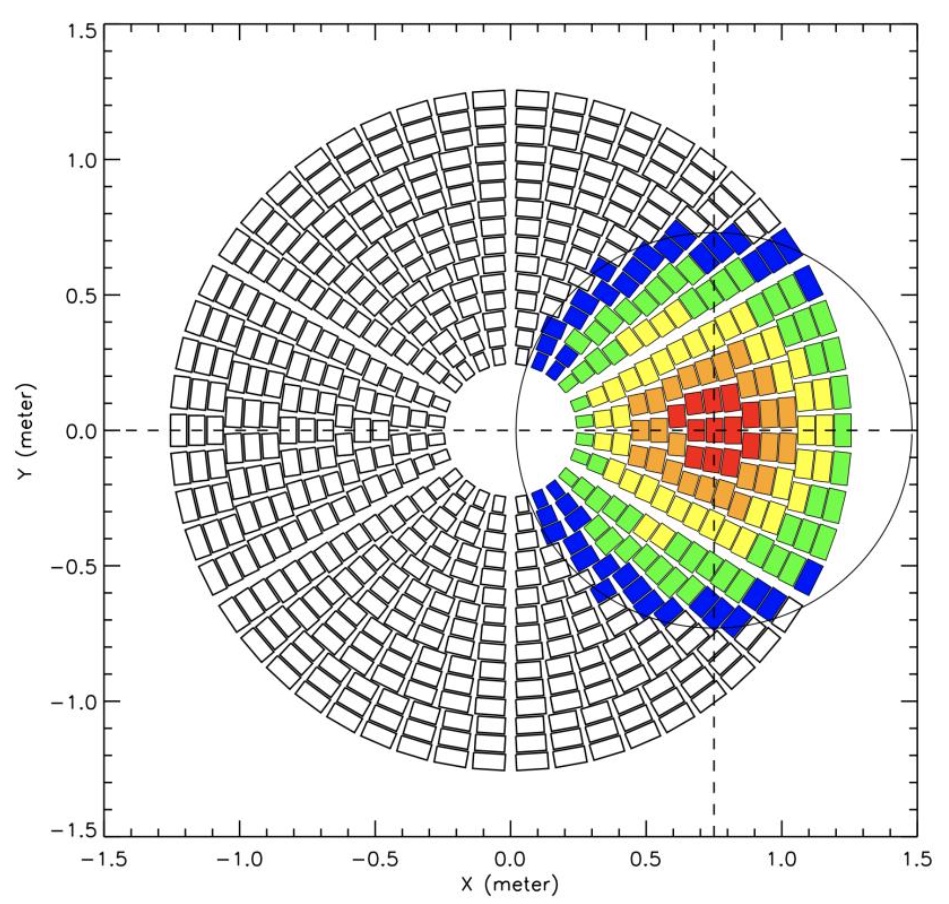} &
        \includegraphics[width=2.8cm]{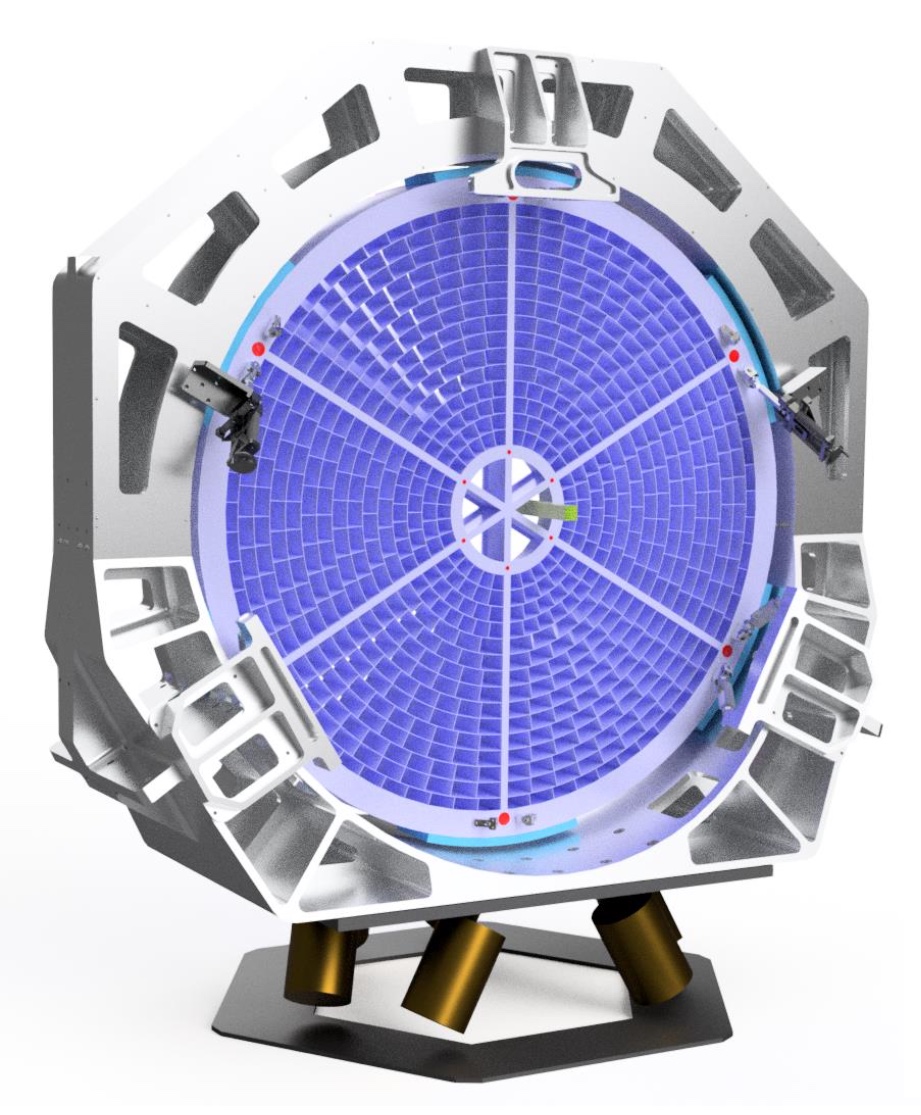}
    \end{tabular}
    \caption{
    (Left) The NASA XRCF facility at the NASA Marshall Space Flight Center with a 518.2~m-long vacuum tube.
    (Center) The expected divergence when illumination a petal of the Athena MA, red = 0-1~arcmin, blue = 4-5~arcmin.
    (Right) The support structure with 1-g sagging mitigation that is being designed and built to accommodate the Athena MA in a vertical configuration on top of an hexapod system once at XRCF.
    Credit: NASA and Max Planck Institute for extraterrestrial Physics.
    }
    \label{fig:XRCF}
\end{figure}

\begin{figure}[htbp]
\centering
    \begin{tabular}{lcr}
        \includegraphics[width=3.5cm]{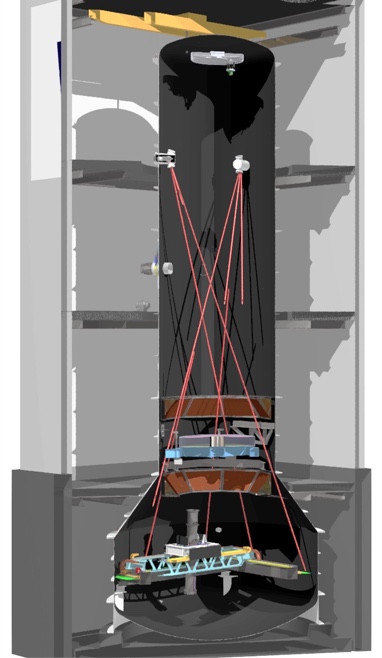} &
        \includegraphics[width=4.5cm]{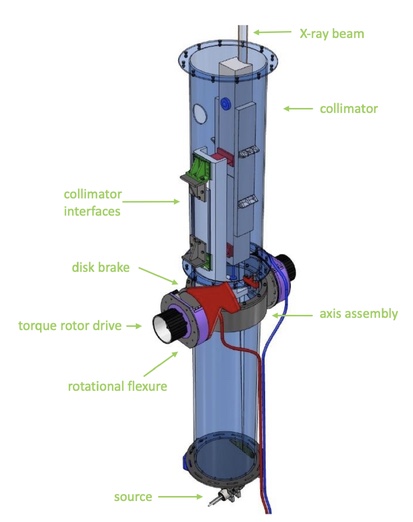} &
        \includegraphics[width=6.5cm]{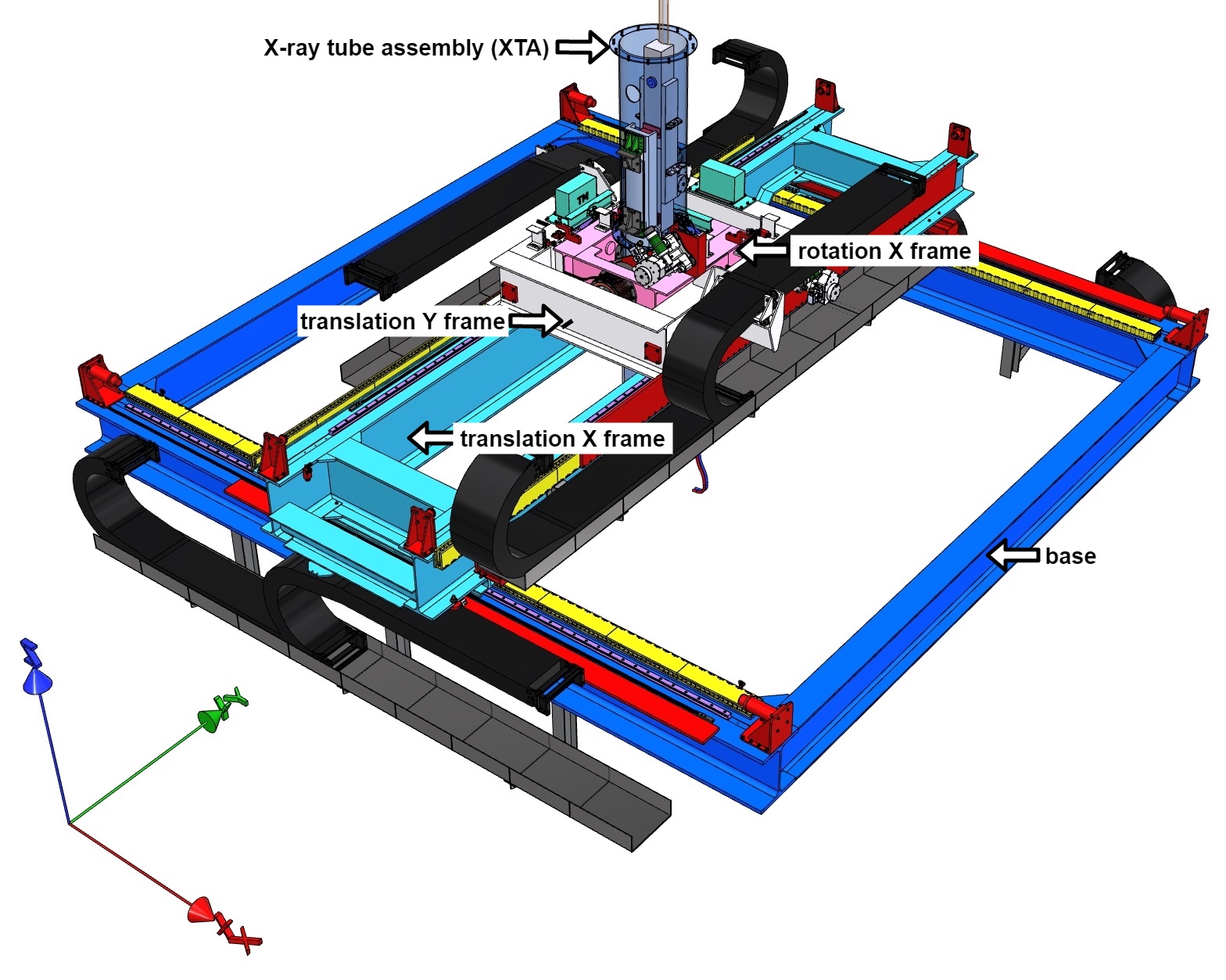}
    \end{tabular}
    \caption{
    (Left) The VERT-X facility consisting of a vacuum vessel with a height of around 18~m.
    This facility shall be placed next to the MM alignment facility (see Section~\ref{subsec: MM_alignment}).
    (Center) X-ray tube able to generate a small X-ray beam collimated to better than 1~arcsec consisting of a microfocus X-ray source held together to a fused silica Wolter I mirror.
    (Right) The raster-scan mechanism used to scan the X-ray tube along the MA aperture, consisting of 2 linear stages and 2 alt-stages. This element is mounted at the ground level of the vacuum vessel - shown on the left.
    Credit: INAF, EIE, Media Lario, BCV Progetti and GPAP.
    }
    \label{fig:VERT-X}
\end{figure}

\subsection{Mirror assembly X-ray characterization}
\label{subsec:MA_x-ray_characterisation}

The need to comply with the tight verification/calibration requirements for the Athena MA is a major challenge, due to its large size.
In order to ensure that the uncertainties due to the divergence of the beam are compatible with the accuracy requirements of the calibrations, the source should be positioned at a minimum distance of 300~m.
Since the largest X-ray calibration facility in Europe is the MPE Panter facility with a 120~m vacuum tube \cite{Panter}, this requirement can only be satisfied with either the NASA X-ray \& Cryogenic Facility \cite{XRCF} or a new concept for an X-ray facility, the VERT-X facility  \cite{2019SPIE11180E..25P}.
Both of these approaches are being followed, and the verification and calibration of the Athena MA is expected to be achieved by using both of these facilities.

The NASA XRCF facility has a 518.2~m vacuum tube built for HEAO-B and then rebuilt for the Chandra mirror X-ray characterisation. More recently it has been recommissioned to perform the cryogenic testing of the JWST primary mirrors panels \cite{2009AAS...21342619L}. 
It is now being prepared to perform the first X-ray characterisation with the Athena MA demonstrator (Figure~\ref{fig:XRCF}).
The size of the X-ray beam allows for the testing of one petal of the MA at a time.
However, despite the distant location of the X-ray source, there is a significant change in the divergence of the beam within the petal (and even within a MM) that will affect both the effective area and angular resolution characterisation.
This will be addressed with post processing enabled by accurate metrology.
The MA is mounted vertically in a dedicated structure that limits the effects of sagging due to gravity. 
This structure interfaces with an hexapod system which enables alignment of the MA in the beam, off-axis characterisation, and defocussing.

The VERT-X facility relies on a small, highly collimated, vertical X-ray beam that is mounted on a raster scan mechanism, which is able to move along two translations on the horizontal plane and two tilts about the two horizontal axes. In this facility, the source is placed at the bottom, the MA, which is horizontal, is at the ground floor level, and an X-ray detector is placed 12~m from the MA at the top (Figure~\ref{fig:VERT-X}).

The collimating X-ray mirror of VERT-X is based on a Wolter I configuration of about 1.1~m in length and an average grazing incidence angle of about 0.4 degree.
The main driving requirements that led to the definition of the design are related to the needs for high reflectivity in the spectral range between 0.2-12~keV, and the limitation to $<1$~arcsec for the divergence error of the collimated beam produced by the mirror.
The mirror substrate is made of fused silica, and it is held together with a micro-focus X-ray source in a common tube.
This assembly generates a collimated X-ray beam of about 6$\times$1~cm$^2$ cross section.

This X-ray source is moved within the MA aperture with a nominal scan velocity of 30~mm/s.
Covering entirely the MA will require a total path length of about 100~m, which is expected to take $\approx$1~hr accounting for $\approx$15~s of settling time at the start and end of each row.
The tube can be tilted not only to cover all the field of view, but also to point up to 3~deg off-axis in order to simulate the stray-light contamination from an off-axis source (Figure~\ref{fig:VERT-X}).
Dedicated metrology and a control system is used to correct in real time the misalignments that occur during scanning.
The main factors contributing to the wobble of the beam during a measurement are the thermal deformation due to temperature gradient along the raster scan and the non-repeatable runout of the trolley linear guide support system.
In both cases the metrology is expected to provide a correction leaving a residual uncertainty of only 0.03~arcsec.

In addition to the collimation advantage (divergence on the order of 1~arcsec, compared to arcmin level at XRCF) VERT-X allows measuring the MA horizontally, which reduces gravity induced sagging errors.
The critical items of this facility, the X-ray source and the raster scan mechanism, are being implemented at the time of this writing.
The plan is to build the facility next to the MM alignment facility at Media Lario (see Section~\ref{subsec: MM_alignment}). 
This will significantly ease the logistics associated with the X-ray verification, particularly for intermediate alignment checks in X-ray \cite{10.1117/12.2593670}.

\section{Summary and conclusions}
SPO uses technology developed over multiple decades by the semiconductor and automotive industries, which have massively invested in the fabrication and processing of silicon wafers. 
It has become a very mature technology, thanks to the continuous development efforts to prepare the industrial production of the largest X-ray optics yet to be launched into space.
SPO is an enabling technology for Athena, and is on-track for meeting the observatory's ambitous requirements for effective area, angular resolution, and mass.
The optics for Athena are under intense development, aiming to start the manufacturing the first flight MMs in 2026. 
In parallel, facilities for the assembly and testing of the MMs, co-alignment of the MMs in the MA, and ground calibration of the MA are being implemented.

It is already recognized that SPO can be applied to applications other than Wolter I optics for space-borne telescopes. 
The technique of replication of a mandrel in stand alone stacks of mirror plates can be used for any medium to large series manufacturing of mirrors with pretty much any kind of figure. 
Whether all the plates are used as mirrors or only the top one is a matter of design. Kirckpatrick-Baez designs or lobster eyes could also be implemented using a similar process. SPO technology also has the potential to create advanced gamma-ray focusing elements via the use of diffraction in the volume of the crystalline plates.


\clearpage
\bibliography{bibliography}

\begin{thebibliography}{10}
\providecommand{\url}[1]{{#1}}
\providecommand{\urlprefix}{URL }
\expandafter\ifx\csname urlstyle\endcsname\relax
  \providecommand{\doi}[1]{DOI~\discretionary{}{}{}#1}\else
  \providecommand{\doi}{DOI~\discretionary{}{}{}\begingroup
  \urlstyle{rm}\Url}\fi

\bibitem{ackerman_patent}
Ackermann, M., Vacanti, G., Vervest, M.: {Method} for assembling an imaging
  x-ray optic.
\newblock NL1041110A;NL1041110B1 (2014)

\bibitem{ackermann_performance_2009}
{Ackermann}, M.D., {Collon}, M.J., {Guenther}, R., {Partapsing}, R., {Vacanti},
  G., {Buis}, E.J., {Krumrey}, M., {M{\"u}ller}, P., {Beijersbergen}, M.W.,
  {Bavdaz}, M., {Wallace}, K.: {Performance prediction and measurement of
  silicon pore optics}.
\newblock In: S.L. {O'Dell}, G.~{Pareschi} (eds.) Optics for EUV, X-Ray, and
  Gamma-Ray Astronomy IV, \emph{Society of Photo-Optical Instrumentation
  Engineers (SPIE) Conference Series}, vol. 7437, p. 74371N (2009).
\newblock \doi{10.1117/12.826840}

\bibitem{Barriere:2021vp}
{Barri{\`e}re}, N.M., {Babi{\'c}}, L., {Bayerle}, A., {Castiglione}, L.,
  {Collon}, M.J., {Eenkhoorn}, N., {Girou}, D., {G{\"u}nther}, R., {Hauser},
  E., {Jenkins}, Y., {Landgraf}, B., {Keek}, L., {Okma}, B., {Mendoza Serano},
  G., {Thete}, A., {Vacanti}, G., {Verhoeckx}, S., {Vervest}, M., {Voruz}, L.,
  {Beijersbergen}, M.W., {Bavdaz}, M., {Wille}, E., {Ferreira}, I., {Fransen},
  S., {Olde Riekerink}, M., {Haneveld}, J., {Schurink}, B., {Start}, R., {van
  Baren}, C., {Handick}, E., {Krumrey}, M., {Valsecchi}, G., {Bradshaw}, M.,
  {Burwitz}, V.: {Assembly of confocal silicon pore optics mirror modules}.
\newblock In: Society of Photo-Optical Instrumentation Engineers (SPIE)
  Conference Series, \emph{Society of Photo-Optical Instrumentation Engineers
  (SPIE) Conference Series}, vol. 11822, p. 1182208 (2021).
\newblock \doi{10.1117/12.2594171}

\bibitem{Barriere:2019tf}
{Barri{\`e}re}, N.M., {Vacanti}, G., {Verhoeckx}, S., {Hauser}, E., {Vervest},
  M., {Keek}, L., {Okma}, B., {Landgraf}, B., {G{\"u}nther}, R., {Voruz}, L.,
  {Girou}, D., {Babi{\'c}}, L., {Collon}, M.J., {Beijersbergen}, M.W.,
  {Bavdaz}, M., {Wille}, E., {Fransen}, S., {Haneveld}, J., {Koelewijn}, A.,
  {Start}, R., {Wijnperle}, M., {Lankwarden}, J.J., {van Baren}, C.,
  {Eigenraam}, A., {M{\"u}ller}, P., {Handick}, E., {Krumrey}, M., {Valsecchi},
  G.: {Assembly of confocal silicon pore optic mirror modules for Athena}.
\newblock In: Optics for EUV, X-Ray, and Gamma-Ray Astronomy IX, \emph{Society
  of Photo-Optical Instrumentation Engineers (SPIE) Conference Series}, vol.
  11119, p. 111190J (2019).
\newblock \doi{10.1117/12.2530706}

\bibitem{Bavdaz2005}
Bavdaz, M., Beijersbergen, M.W.: Optical reflector element, its method of
  fabrication, and an optical instrument implementing such elements  (2005)

\bibitem{Bavdaz_SPIE_2021}
{Bavdaz}, M., {Wille}, E., {Ayre}, M., {Ferreira}, I., {Shortt}, B., {Fransen},
  S., {Millinger}, M., {Collon}, M.J., {Vacanti}, G., {Barriere}, N.M.,
  {Landgraf}, B., {Riekerink}, M.O., {Haneveld}, J., {Start}, R., {van Baren},
  C., {Della Monica Ferreira}, D., {Massahi}, S., {Svendsen}, S.,
  {Christensen}, F., {Krumrey}, M., {Handick}, E., {Burwitz}, V., {Bradshaw},
  M., {Pareschi}, G., {Valsecchi}, G., {Vernani}, D., {Kailla}, G., {Mundon},
  W., {Phillips}, G., {Schneider}, J., {Korhonen}, T., {Sanchez}, A., {Heinis},
  D., {Colldelram}, C., {Torti}, M., {Willingale}, R.: {ATHENA x-ray optics
  development and accommodation}.
\newblock In: Society of Photo-Optical Instrumentation Engineers (SPIE)
  Conference Series, \emph{Society of Photo-Optical Instrumentation Engineers
  (SPIE) Conference Series}, vol. 11822, p. 1182205 (2021).
\newblock \doi{10.1117/12.2594689}

\bibitem{Beijersbergen:2004vc}
{Beijersbergen}, M., {Kraft}, S., {Gunther}, R., {Mieremet}, A.L., {Collon},
  M., {Bavdaz}, M., {Lumb}, D.H., {Peacock}, A.J.: {Silicon pore optics: novel
  lightweight high-resolution X-ray optics developed for XEUS}.
\newblock In: G.~{Hasinger}, M.J.L. {Turner} (eds.) UV and Gamma-Ray Space
  Telescope Systems, \emph{Society of Photo-Optical Instrumentation Engineers
  (SPIE) Conference Series}, vol. 5488, pp. 868--874 (2004).
\newblock \doi{10.1117/12.585122}

\bibitem{Panter}
{Burwitz}, V., {Bavdaz}, M., {Wille}, E., {Collon}, M., {Vacanti}, G.,
  {Barriere}, N., {Valsecchi}, G., {Marioni}, F., {Vernani}, D., {Seure}, T.,
  {Blum}, S., {Willingale}, R., {Smith}, R., {de Roo}, C., {Hertz}, E.,
  {Hartner}, G., {La Caria}, M.M., {Pelliciari}, C., {Langmeier}, A., {Hartl},
  S.F.: {X-ray testing at PANTER of optics for the ATHENA and Arcus Missions}.
\newblock In: International Conference on Space Optics \&mdash; ICSO 2018,
  \emph{Society of Photo-Optical Instrumentation Engineers (SPIE) Conference
  Series}, vol. 11180, p. 1118024 (2019).
\newblock \doi{10.1117/12.2535995}

\bibitem{Chase:73}
{Chase}, R.C., {van Speybroeck}, L.P.: {Wolter-Schwarzschild telescopes for
  X-ray astronomy.}
\newblock Applied Optics \textbf{12}, 1042--1044 (1973).
\newblock \doi{10.1364/AO.12.001042}

\bibitem{10.1117/12.894615}
{Christensen}, F.E., {Jakobsen}, A.C., {Brejnholt}, N.F., {Madsen}, K.K.,
  {Hornstrup}, A., {Westergaard}, N.J., {Momberg}, J., {Koglin}, J.,
  {Fabricant}, A.M., {Stern}, M., {Craig}, W.W., {Pivovaroff}, M.J., {Windt},
  D.: {Coatings for the NuSTAR mission}.
\newblock In: S.L. {O'Dell}, G.~{Pareschi} (eds.) Society of Photo-Optical
  Instrumentation Engineers (SPIE) Conference Series, \emph{Society of
  Photo-Optical Instrumentation Engineers (SPIE) Conference Series}, vol. 8147,
  p. 81470U (2011).
\newblock \doi{10.1117/12.894615}

\bibitem{Collon:2013tj}
{Collon}, M.J., {Ackermann}, M., {G{\"u}nther}, R., {Vacanti}, G.,
  {Beijersbergen}, M.W., {Bavdaz}, M., {Wille}, E., {Wallace}, K., {Haneveld},
  J., {Olde Riekerink}, M., {Koelewijn}, A., {van Baren}, C., {M{\"u}ller}, P.,
  {Krumrey}, M., {Burwitz}, V., {Sironi}, G., {Ghigo}, M.: {Aberration-free
  silicon pore x-ray optics}.
\newblock In: S.L. {O'Dell}, G.~{Pareschi} (eds.) Optics for EUV, X-Ray, and
  Gamma-Ray Astronomy VI, \emph{Society of Photo-Optical Instrumentation
  Engineers (SPIE) Conference Series}, vol. 8861, p. 88610M (2013).
\newblock \doi{10.1117/12.2024982}

\bibitem{Collon:2021tk}
{Collon}, M.J., {Babic}, L., {Barri{\`e}re}, N.M., {Bayerle}, A.,
  {Castiglione}, L., {Eenkhoorn}, N., {Girou}, D., {G{\"u}nther}, R., {Hauser},
  E., {Jenkins}, Y., {Landgraf}, B., {Keek}, L., {Okma}, B., {Mendoza Serrano},
  G., {Thete}, A., {Vacanti}, G., {Verhoeckx}, S., {Vervest}, M., {Voruz}, L.,
  {Beijersbergen}, M.W., {Bavdaz}, M., {Wille}, E., {Ferreira}, I., {Fransen},
  S., {Shortt}, B., {Olde Riekerink}, M., {Haneveld}, J., {Koelewijn}, A.,
  {Wijnperle}, M., {Lankwarden}, J.J., {Schurink}, B., {Start}, R., {van
  Baren}, C., {Hieltjes}, P., {den Herder}, J.W., {Handick}, E., {Krumrey}, M.,
  {Bradshaw}, M., {Burwitz}, V., {Massahi}, S., {Svendsen}, S., {Della Monica
  Ferreira}, D.e., {Christensen}, F.E., {Valsecchi}, G., {Kailla}, G.,
  {Mundon}, W., {Philips}, G., {Ball}, K.: {Silicon pore optics x-ray mirror
  development for the Athena telescope}.
\newblock In: Society of Photo-Optical Instrumentation Engineers (SPIE)
  Conference Series, \emph{Society of Photo-Optical Instrumentation Engineers
  (SPIE) Conference Series}, vol. 11822, p. 1182206 (2021).
\newblock \doi{10.1117/12.2593505}

\bibitem{Collon:2011tx}
{Collon}, M.J., {G{\"u}nther}, R., {Ackermann}, M., {Partapsing}, R.,
  {Vacanti}, G., {Beijersbergen}, M.W., {Bavdaz}, M., {Wallace}, K., {Wille},
  E., {Olde Riekerink}, M., {Haneveld}, J., {Koelewijn}, A., {van Baren}, C.,
  {M{\"u}ller}, P., {Krumrey}, M., {Freyberg}, M., {Jakobsen}, A.C.,
  {Christensen}, F.: {Design, fabrication, and characterization of silicon pore
  optics for ATHENA/IXO}.
\newblock In: S.L. {O'Dell}, G.~{Pareschi} (eds.) Society of Photo-Optical
  Instrumentation Engineers (SPIE) Conference Series, \emph{Society of
  Photo-Optical Instrumentation Engineers (SPIE) Conference Series}, vol. 8147,
  p. 81470D (2011).
\newblock \doi{10.1117/12.893418}

\bibitem{Collon:2006wz}
{Collon}, M.J., {Kraft}, S., {G{\"u}nther}, R., {Maddox}, E., {Beijersbergen},
  M., {Bavdaz}, M., {Lumb}, D., {Wallace}, K., {Krumrey}, M., {Cibik}, L.,
  {Freyberg}, M.: {Performance characterization of silicon pore optics}.
\newblock In: M.J.L. {Turner}, G.~{Hasinger} (eds.) Society of Photo-Optical
  Instrumentation Engineers (SPIE) Conference Series, \emph{Society of
  Photo-Optical Instrumentation Engineers (SPIE) Conference Series}, vol. 6266,
  p. 62661T (2006)

\bibitem{Collon:2017vj}
{Collon}, M.J., {Vacanti}, G., {Barri{\`e}re}, N.M., {Landgraf}, B.,
  {G{\"u}nther}, R., {Vervest}, M., {van der Hoeven}, R., {Dekker}, D.,
  {Chatbi}, A., {Girou}, D., {Sforzini}, J., {Beijersbergen}, M.W., {Bavdaz},
  M., {Wille}, E., {Fransen}, S., {Shortt}, B., {Haneveld}, J., {Koelewijn},
  A., {Booysen}, K., {Wijnperle}, M., {van Baren}, C., {Eigenraam}, A.,
  {M{\"u}ller}, P., {Krumrey}, M., {Burwitz}, V., {Pareschi}, G., {Massahi},
  S., {Christensen}, F.E., {Della Monica Ferreira}, D., {Valsecchi}, G.,
  {Oliver}, P., {Checquer}, I., {Ball}, K., {Zuknik}, K.H.: {Development of
  ATHENA mirror modules}.
\newblock In: Society of Photo-Optical Instrumentation Engineers (SPIE)
  Conference Series, \emph{Society of Photo-Optical Instrumentation Engineers
  (SPIE) Conference Series}, vol. 10399, p. 103990C (2017).
\newblock \doi{10.1117/12.2273704}

\bibitem{Collon:2019ww}
{Collon}, M.J., {Vacanti}, G., {Barri{\`e}re}, N.M., {Landgraf}, B.,
  {G{\"u}nther}, R., {Vervest}, M., {Voruz}, L., {Verhoeckx}, S., {Babi{\'c}},
  L., {Keek}, L., {Girou}, D., {Okma}, B., {Hauser}, E., {Beijersbergen}, M.W.,
  {Bavdaz}, M., {Wille}, E., {Fransen}, S., {Shortt}, B., {Ferreira}, I.,
  {Haneveld}, J., {Koelewijn}, A., {Start}, R., {Wijnperle}, M., {Lankwarden},
  J.J., {van Baren}, C., {Hieltjes}, P., {den Herder}, J.W., {M{\"u}ller}, P.,
  {Handick}, E., {Krumrey}, M., {Bradshaw}, M., {Burwitz}, V., {Pareschi}, G.,
  {Massahi}, S., {Svendsen}, S., {Della Monica Ferreira}, D., {Christensen},
  F.E., {Valsecchi}, G., {Oliver}, P., {Chequer}, I., {Ball}, K.: {Status of
  the silicon pore optics technology}.
\newblock In: Optics for EUV, X-Ray, and Gamma-Ray Astronomy IX, \emph{Society
  of Photo-Optical Instrumentation Engineers (SPIE) Conference Series}, vol.
  11119, p. 111190L (2019).
\newblock \doi{10.1117/12.2530696}

\bibitem{Collon:2015wv}
{Collon}, M.J., {Vacanti}, G., {G{\"u}nther}, R., {Yanson}, A., {Barri{\`e}re},
  N., {Landgraf}, B., {Vervest}, M., {Chatbi}, A., {Beijersbergen}, M.W.,
  {Bavdaz}, M., {Wille}, E., {Haneveld}, J., {Koelewijn}, A., {Leenstra}, A.,
  {Wijnperle}, M., {van Baren}, C., {M{\"u}ller}, P., {Krumrey}, M., {Burwitz},
  V., {Pareschi}, G., {Conconi}, P., {Christensen}, F.E.: {Silicon pore optics
  development for ATHENA}.
\newblock In: Society of Photo-Optical Instrumentation Engineers (SPIE)
  Conference Series, \emph{Society of Photo-Optical Instrumentation Engineers
  (SPIE) Conference Series}, vol. 9603, p. 96030K (2015).
\newblock \doi{10.1117/12.2188988}

\bibitem{Conti:1994uc}
{Conti}, G., {Mattaini}, E., {Santambrogio}, E., {Sacco}, B., {Cusumano}, G.,
  {Citterio}, O., {Braeuninger}, H.W., {Burkert}, W.: {X-ray characteristics of
  the Italian X-Ray Astronomy Satellite (SAX) flight mirror units}.
\newblock In: R.B. {Hoover}, A.B. {Walker} (eds.) Advances in Multilayer and
  Grazing Incidence X-Ray/EUV/FUV Optics, \emph{Society of Photo-Optical
  Instrumentation Engineers (SPIE) Conference Series}, vol. 2279, pp. 101--109
  (1994).
\newblock \doi{10.1117/12.193179}

\bibitem{de-Korte:1981ts}
{de Korte}, P.A.J., {Bleeker}, J.A.M., {den Boggende}, A.J.F.,
  {Branduardi-Raymont}, G., {Brinkman}, A.C., {Culhane}, J.L., {Gronenschild},
  E.H.B.M., {Mason}, I., {McKechnie}, S.P.: {The X-Ray Imaging Telescopes on
  EXOSAT}.
\newblock \ssr \textbf{30}(1-4), 495--511 (1981).
\newblock \doi{10.1007/BF01246070}

\bibitem{10.1117/12.2313275}
{Della Monica Ferreira}, D., {Svendsen}, S., {Massahi}, S., {Jafari}, A., {Vu},
  L.M., {Korman}, J., {Gellert}, N.C., {Christensen}, F.E., {Kadkhodazadeh},
  S., {Kasama}, T., {Shortt}, B., {Bavdaz}, M., {Collon}, M.J., {Landgraf}, B.,
  {Krumrey}, M., {Cibik}, L., {Schreiber}, S., {Schubert}, A.: {Performance and
  stability of mirror coatings for the ATHENA mission}.
\newblock In: J.W.A. {den Herder}, S.~{Nikzad}, K.~{Nakazawa} (eds.) Space
  Telescopes and Instrumentation 2018: Ultraviolet to Gamma Ray, \emph{Society
  of Photo-Optical Instrumentation Engineers (SPIE) Conference Series}, vol.
  10699, p. 106993K (2018).
\newblock \doi{10.1117/12.2313275}

\bibitem{10.1117/12.2594443}
{Ferreira}, I., {Bavdaz}, M., {Ayre}, M., {Fransen}, S., {Pacros}, A.,
  {Linder}, M., {Stefanescu}, A., {Branco}, M., {Guainazzi}, M., {Ness}, J.U.,
  {Oosterbroek}, T., {Willingale}, R.: {ATHENA reference telescope design and
  recent mission level consolidation}.
\newblock In: Society of Photo-Optical Instrumentation Engineers (SPIE)
  Conference Series, \emph{Society of Photo-Optical Instrumentation Engineers
  (SPIE) Conference Series}, vol. 11822, p. 1182204 (2021).
\newblock \doi{10.1117/12.2594443}

\bibitem{Girou_2021}
Girou, D., Ford, E., Wade, C., van Aarle, C., Uliyanov, A., Hanlon, L.,
  Tomsick, J.A., Zoglauer, A., Collon, M.J., Beijersbergen, M.W.,
  Barri{\`{e}}re, N.M.: Design and modeling of a laue lens for radiation
  therapy with hard x-ray photons.
\newblock Physics in Medicine {\&} Biology \textbf{66}(24), 245007 (2021).
\newblock \doi{10.1088/1361-6560/ac3840}

\bibitem{Girou2020}
{Girou}, D., {Massahi}, S., {Ferreira}, D.D.M., {Christensen}, F.E.,
  {Landgraf}, B., {Shortt}, B., {Collon}, M., {Beijersbergen}, M.: {Plasma
  etching for the compatibility of thin film metallic coatings and direct
  bonding of silicon pore optics}.
\newblock Journal of Applied Physics \textbf{128}(9), 095302 (2020).
\newblock \doi{10.1063/5.0010212}

\bibitem{10.1117/12.2594230}
{Girou}, D.A., {van Baren}, C., {te Kloeze}, I., {Castiglione}, L., {Okma}, B.,
  {Vacanti}, G., {Hauser}, E., {Babic}, L., {Bayerle}, A., {Barri{\`e}re},
  N.M., {Eenkhoorn}, N., {G{\"u}nther}, R., {Jenkins}, Y., {Landgraf}, B.,
  {Keek}, L., {Thete}, A., {Verhoeckx}, S., {Vervest}, M., {Voruz}, L.,
  {Collon}, M., {Beijersbergen}, M., {Fransen}, S., {Campoli}, G., {Ferreira},
  I., {Wille}, E., {Bavdaz}, M., {Marioni}, F., {Valsecchi}, G.: {Environmental
  testing of the Athena telescope mirror modules}.
\newblock In: Society of Photo-Optical Instrumentation Engineers (SPIE)
  Conference Series, \emph{Society of Photo-Optical Instrumentation Engineers
  (SPIE) Conference Series}, vol. 11822, p. 1182209 (2021).
\newblock \doi{10.1117/12.2594230}

\bibitem{Gondoin:1994wo}
{Gondoin}, P., {van Katwijk}, K., {Aschenbach}, B.R., {Schulz}, N., {Boerret},
  R., {Glatzel}, H., {Citterio}, O.: {X-ray spectroscopy mission (XMM)
  telescope development}.
\newblock In: M.G. {Cerutti-Maori}, P.~{Roussel} (eds.) Space Optics 1994:
  Earth Observation and Astronomy, \emph{Society of Photo-Optical
  Instrumentation Engineers (SPIE) Conference Series}, vol. 2209, pp. 438--450
  (1994).
\newblock \doi{10.1117/12.185278}

\bibitem{Gorenstein2010}
{Gorenstein}, P.: {Focusing X-Ray Optics for Astronomy}.
\newblock X-Ray Optics and Instrumentation \textbf{2010}, 109740 (2010).
\newblock \doi{10.1155/2010/109740}

\bibitem{Gunther:2006uy}
{G{\"u}nther}, R., {Collon}, M., {Kraft}, S., {Beijersbergen}, M., {Bavdaz},
  M., {Lumb}, D., {Peacock}, A., {Wallace}, K.: {Production of silicon pore
  optics}.
\newblock In: M.J.L. {Turner}, G.~{Hasinger} (eds.) Society of Photo-Optical
  Instrumentation Engineers (SPIE) Conference Series, \emph{Society of
  Photo-Optical Instrumentation Engineers (SPIE) Conference Series}, vol. 6266,
  p. 626619 (2006).
\newblock \doi{10.1117/12.673248}

\bibitem{gosele}
Gösele, U., Tong, Q.Y.: Semiconductor wafer bonding.
\newblock Annual Review of Materials Science \textbf{28}(1), 215--241 (1998).
\newblock \doi{10.1146/annurev.matsci.28.1.215}

\bibitem{handick:2020SPIE}
{Handick}, E., {Cibik}, L., {Krumrey}, M., {M{\"u}ller}, P., {Barri{\`e}re},
  N., {Collon}, M., {Hauser}, E., {Vacanti}, G., {Verhoeckx}, S., {Bavdaz}, M.,
  {Wille}, E.: {Upgrade of the x-ray parallel beam facility XPBF 2.0 for
  characterization of silicon pore optics}.
\newblock In: Society of Photo-Optical Instrumentation Engineers (SPIE)
  Conference Series, \emph{Society of Photo-Optical Instrumentation Engineers
  (SPIE) Conference Series}, vol. 11444, p. 114444G (2020).
\newblock \doi{10.1117/12.2561236}

\bibitem{Harrison:2013wl}
{Harrison}, F.A., {Craig}, W.W., {Christensen}, F.E., {Hailey}, C.J., {Zhang},
  W.W., {Boggs}, S.E., {Stern}, D., {Cook}, W.R., {Forster}, K., {Giommi}, P.,
  {Grefenstette}, B.W., {Kim}, Y., {Kitaguchi}, T., {Koglin}, J.E., {Madsen},
  K.K., {Mao}, P.H., {Miyasaka}, H., {Mori}, K., {Perri}, M., {Pivovaroff},
  M.J., {Puccetti}, S., {Rana}, V.R., {Westergaard}, N.J., {Willis}, J.,
  {Zoglauer}, A., {An}, H., {Bachetti}, M., {Barri{\`e}re}, N.M., {Bellm},
  E.C., {Bhalerao}, V., {Brejnholt}, N.F., {Fuerst}, F., {Liebe}, C.C.,
  {Markwardt}, C.B., {Nynka}, M., {Vogel}, J.K., {Walton}, D.J., {Wik}, D.R.,
  {Alexander}, D.M., {Cominsky}, L.R., {Hornschemeier}, A.E., {Hornstrup}, A.,
  {Kaspi}, V.M., {Madejski}, G.M., {Matt}, G., {Molendi}, S., {Smith}, D.M.,
  {Tomsick}, J.A., {Ajello}, M., {Ballantyne}, D.R., {Balokovi{\'c}}, M.,
  {Barret}, D., {Bauer}, F.E., {Blandford}, R.D., {Brandt}, W.N., {Brenneman},
  L.W., {Chiang}, J., {Chakrabarty}, D., {Chenevez}, J., {Comastri}, A.,
  {Dufour}, F., {Elvis}, M., {Fabian}, A.C., {Farrah}, D., {Fryer}, C.L.,
  {Gotthelf}, E.V., {Grindlay}, J.E., {Helfand}, D.J., {Krivonos}, R., {Meier},
  D.L., {Miller}, J.M., {Natalucci}, L., {Ogle}, P., {Ofek}, E.O., {Ptak}, A.,
  {Reynolds}, S.P., {Rigby}, J.R., {Tagliaferri}, G., {Thorsett}, S.E.,
  {Treister}, E., {Urry}, C.M.: {The Nuclear Spectroscopic Telescope Array
  (NuSTAR) High-energy X-Ray Mission}.
\newblock \apj \textbf{770}(2), 103 (2013).
\newblock \doi{10.1088/0004-637X/770/2/103}

\bibitem{Heinis:2021ud}
{Heinis}, D., {Carballedo}, A., {Colldelram}, C., {Cun{\'\i}}, G., {Valls
  Vidal}, N., {Matilla}, {\'O}., {Marcos}, J., {S{\'a}nchez}, A., {Casas}, J.,
  {Nicol{\`a}s}, J., {Barri{\`e}re}, N., {Collon}, M.J., {Vacanti}, G.,
  {Handick}, E., {M{\"u}ller}, P., {Krumrey}, M., {Ferreira}, I., {Bavdaz}, M.:
  {X-ray facility for the characterization of the Athena mirror modules at the
  ALBA synchrotron}.
\newblock In: Society of Photo-Optical Instrumentation Engineers (SPIE)
  Conference Series, \emph{Society of Photo-Optical Instrumentation Engineers
  (SPIE) Conference Series}, vol. 11852, p. 1185222 (2021).
\newblock \doi{10.1117/12.2599350}

\bibitem{10.1117/1.JATIS.6.3.034005}
{Jafari}, A., {Ferreira}, D.D.M., {Kadkhodazadeh}, S., {Kasama}, T., {Massahi},
  S., {Svendsen}, S., {Vu}, L.M., {Henriksen}, P.L., {Balogh}, Z.M., {Krumrey},
  M., {Cibik}, L., {Christensen}, F.E., {Shortt}, B.: {Long-term performance
  and durability of Ir/B$_{4}$C multilayer x-ray mirrors: focusing on
  composition, structure, and reflectivity properties}.
\newblock Journal of Astronomical Telescopes, Instruments, and Systems
  \textbf{6}, 034005 (2020).
\newblock \doi{10.1117/1.JATIS.6.3.034005}

\bibitem{jansen.2001vn}
{Jansen}, F., {Lumb}, D., {Altieri}, B., {Clavel}, J., {Ehle}, M., {Erd}, C.,
  {Gabriel}, C., {Guainazzi}, M., {Gondoin}, P., {Much}, R., {Munoz}, R.,
  {Santos}, M., {Schartel}, N., {Texier}, D., {Vacanti}, G.: {XMM-Newton
  observatory. I. The spacecraft and operations}.
\newblock \aap \textbf{365}, L1--L6 (2001).
\newblock \doi{10.1051/0004-6361:20000036}

\bibitem{XRCF}
Kegley, J.: XRCF Handbook.
\newblock NASA Marshall Space Flight Center Huntsville, AL United States
  (2016).
\newblock
  \urlprefix\url{https://ntrs.nasa.gov/api/citations/20160003530/downloads/20160003530.pdf}

\bibitem{10003367909}
KERN, W.: Cleaning solutions based on hydrogen peroxide for use in silicon
  semiconductor technology.
\newblock RCA Review \textbf{31}, 187--206 (1970).
\newblock \urlprefix\url{https://ci.nii.ac.jp/naid/10003367909/en/}

\bibitem{1948JOSA...38..766K}
{Kirkpatrick}, P., {Baez}, A.V.: {Formation of optical images by x-rays}.
\newblock Journal of the Optical Society of America (1917-1983) \textbf{38}(9),
  766 (1948)

\bibitem{krumrey_x-ray_2010}
Krumrey, M., Cibik, L., Müller, P., Bavdaz, M., Wille, E., Ackermann, M.,
  Collon, M.J.: X-ray pencil beam facility for optics characterization.
\newblock p. 77324O. San Diego, California, USA (2010).
\newblock \doi{10.1117/12.857335}

\bibitem{krumrey_2016}
Krumrey, M., Müller, P., Cibik, L., Collon, M., Barrière, N., Vacanti, G.,
  Bavdaz, M., Wille, E.: New x-ray parallel beam facility {XPBF} 2.0 for the
  characterization of silicon pore optics.
\newblock p. 99055N. Edinburgh, United Kingdom (2016).
\newblock \doi{10.1117/12.2231687}

\bibitem{Landgraf:2021tf}
{Landgraf}, B., {Babi{\'c}}, L., {Barri{\`e}re}, N.M., {Bayerle}, A.,
  {Castiglione}, L., {Collon}, M.J., {Eenkhoorn}, N., {Girou}, D.,
  {G{\"u}nther}, R., {Hauser}, E., {Jenkins}, Y., {Keek}, L., {Okma}, B.,
  {Serano}, G.M., {Thete}, A., {Vacanti}, G., {Verhoeckx}, S., {Vervest}, M.,
  {Voruz}, L., {Beijersbergen}, M.W., {Bavdaz}, M., {Wille}, E., {Ferreira},
  I., {Fransen}, S., {Shortt}, B., {Riekerink}, M.O., {Haneveld}, J.,
  {Koelewijn}, A., {Wijnperl{\'e}}, M., {Lankwarden}, J.J., {Schurink}, B.,
  {Start}, R., {van Baren}, C., {Hieltjes}, P., {den Herder}, J.W., {Handick},
  E., {Krumrey}, M., {Bradshaw}, M., {Burwitz}, V., {Massahi}, S., {Svendsen},
  S., {Ferreira}, D.D.M., {Christensen}, F.E., {Valsecchi}, G., {Kailla}, G.,
  {Phillips}, G., {Mundon}, W., {Chequer}, I., {Ball}, K.: {SPO mirror plate
  production and coating}.
\newblock In: Society of Photo-Optical Instrumentation Engineers (SPIE)
  Conference Series, \emph{Society of Photo-Optical Instrumentation Engineers
  (SPIE) Conference Series}, vol. 11822, p. 1182207 (2021).
\newblock \doi{10.1117/12.2594234}

\bibitem{Landgraf:2019uz}
{Landgraf}, B., {Collon}, M.J., {Vacanti}, G., {Barri{\`e}re}, N.M.,
  {G{\"u}nther}, R., {Vervest}, M., {Voruz}, L., {Verhoeckx}, S., {Babi{\'c}},
  L., {Keek}, L., {Girou}, D., {Okma}, B., {Beijersbergen}, M.W., {Bavdaz}, M.,
  {Wille}, E., {Fransen}, S., {Shortt}, B., {Ferreira}, I., {Haneveld}, J.,
  {Koelewijn}, A., {Start}, R., {Wijnperl{\'e}}, M., {Lankwarden}, J.J., {van
  Baren}, C., {Hieltjes}, P., {den Herder}, J.W., {Burwitz}, V., {Pareschi},
  G., {Massahi}, S., {Della Monica Ferreira}, D., {Christensen}, F.E.,
  {Valsecchi}, G., {Oliver}, P., {Chequer}, I., {Ball}, K.: {Development and
  manufacturing of SPO X-ray mirrors}.
\newblock In: Optics for EUV, X-Ray, and Gamma-Ray Astronomy IX, \emph{Society
  of Photo-Optical Instrumentation Engineers (SPIE) Conference Series}, vol.
  11119, p. 111190E (2019).
\newblock \doi{10.1117/12.2530941}

\bibitem{2009AAS...21342619L}
{Lightsey}, P., {Gallagher}, B., {Chaney}, D., {Brown}, B.: {James Webb Space
  Telescope Primary Mirror Manufacturing}.
\newblock In: American Astronomical Society Meeting Abstracts \#213,
  \emph{American Astronomical Society Meeting Abstracts}, vol. 213, p. 426.19
  (2009)

\bibitem{10.1117/12.2528351}
{Massahi}, S., {Christensen}, F.E., {Ferreira}, D.D.M., {Jafari}, A.,
  {Svendsen}, S., {Henriksen}, P.L., {Shortt}, B., {Ferreira}, I., {Bavdaz},
  M., {Collon}, M., {Landgraf}, B., {Girou}, D., {Langer}, A.,
  {Sch{\"o}nberger}, W., {Wellner}, T., {Krumrey}, M., {Cibik}, L.:
  {Installation and commissioning of the silicon pore optics coating facility
  for the ATHENA mission}.
\newblock In: Optics for EUV, X-Ray, and Gamma-Ray Astronomy IX, \emph{Society
  of Photo-Optical Instrumentation Engineers (SPIE) Conference Series}, vol.
  11119, p. 111190F (2019).
\newblock \doi{10.1117/12.2528351}

\bibitem{Maszara1988}
{Maszara}, W.P., {Goetz}, G., {Caviglia}, A., {McKitterick}, J.B.: {Bonding of
  silicon wafers for silicon-on-insulator}.
\newblock Journal of Applied Physics \textbf{64}(10), 4943--4950 (1988).
\newblock \doi{10.1063/1.342443}

\bibitem{WFI_Norbert}
{Meidinger}, N., {Albrecht}, S., {Beitler}, C., {Bonholzer}, M., {Emberger},
  V., {Frank}, J., {Lederhuber}, A., {M{\"u}ller-Seidlitz}, J., {Nandra}, K.,
  {Oser}, J., {Ott}, S., {Plattner}, M., {Strecker}, R.: {Development status of
  the wide field imager instrument for Athena}.
\newblock In: Society of Photo-Optical Instrumentation Engineers (SPIE)
  Conference Series, \emph{Society of Photo-Optical Instrumentation Engineers
  (SPIE) Conference Series}, vol. 11444, p. 114440T (2020).
\newblock \doi{10.1117/12.2560507}

\bibitem{Merloni:2012ug}
{Merloni}, A., {Predehl}, P., {Becker}, W., {B{\"o}hringer}, H., {Boller}, T.,
  {Brunner}, H., {Brusa}, M., {Dennerl}, K., {Freyberg}, M., {Friedrich}, P.,
  {Georgakakis}, A., {Haberl}, F., {Hasinger}, G., {Meidinger}, N., {Mohr}, J.,
  {Nandra}, K., {Rau}, A., {Reiprich}, T.H., {Robrade}, J., {Salvato}, M.,
  {Santangelo}, A., {Sasaki}, M., {Schwope}, A., {Wilms}, J., {German eROSITA
  Consortium}, t.: {eROSITA Science Book: Mapping the Structure of the
  Energetic Universe}.
\newblock arXiv e-prints arXiv:1209.3114 (2012)

\bibitem{10.1117/12.2593670}
{Moretti}, A., {Pareschi}, G., {Basso}, S., {Spiga}, D., {Ghigo}, M.,
  {Tagliaferri}, G., {Sironi}, G., {Civitani}, M., {Cotroneo}, V., {La
  Palombara}, N., {Uslenghi}, M., {Tordi}, M., {Delorenzi}, S., {Valsecchi},
  G., {Zocchi}, F., {Marioni}, F., {Vernani}, D., {Amisano}, F., {Parissenti},
  G., {Parodi}, G., {Ottolini}, M., {Corradi}, P., {Bavdaz}, M., {Ferreira},
  I.: {The VERT-X calibration facility: development of the most critical
  parts}.
\newblock In: Society of Photo-Optical Instrumentation Engineers (SPIE)
  Conference Series, \emph{Society of Photo-Optical Instrumentation Engineers
  (SPIE) Conference Series}, vol. 11822, p. 118220K (2021).
\newblock \doi{10.1117/12.2593670}

\bibitem{Nandra:2013ue}
{Nandra}, K., {Barret}, D., {Barcons}, X., {Fabian}, A., {den Herder}, J.W.,
  {Piro}, L., {Watson}, M., {Adami}, C., {Aird}, J., {Afonso}, J.M.,
  {Alexander}, D., {Argiroffi}, C., {Amati}, L., {Arnaud}, M., {Atteia}, J.L.,
  {Audard}, M., {Badenes}, C., {Ballet}, J., {Ballo}, L., {Bamba}, A.,
  {Bhardwaj}, A., {Stefano Battistelli}, E., {Becker}, W., {De Becker}, M.,
  {Behar}, E., {Bianchi}, S., {Biffi}, V., {B{\^\i}rzan}, L., {Bocchino}, F.,
  {Bogdanov}, S., {Boirin}, L., {Boller}, T., {Borgani}, S., {Borm}, K.,
  {Bouch{\'e}}, N., {Bourdin}, H., {Bower}, R., {Braito}, V., {Branchini}, E.,
  {Branduardi-Raymont}, G., {Bregman}, J., {Brenneman}, L., {Brightman}, M.,
  {Br{\"u}ggen}, M., {Buchner}, J., {Bulbul}, E., {Brusa}, M., {Bursa}, M.,
  {Caccianiga}, A., {Cackett}, E., {Campana}, S., {Cappelluti}, N., {Cappi},
  M., {Carrera}, F., {Ceballos}, M., {Christensen}, F., {Chu}, Y.H.,
  {Churazov}, E., {Clerc}, N., {Corbel}, S., {Corral}, A., {Comastri}, A.,
  {Costantini}, E., {Croston}, J., {Dadina}, M., {D'Ai}, A., {Decourchelle},
  A., {Della Ceca}, R., {Dennerl}, K., {Dolag}, K., {Done}, C., {Dovciak}, M.,
  {Drake}, J., {Eckert}, D., {Edge}, A., {Ettori}, S., {Ezoe}, Y., {Feigelson},
  E., {Fender}, R., {Feruglio}, C., {Finoguenov}, A., {Fiore}, F., {Galeazzi},
  M., {Gallagher}, S., {Gandhi}, P., {Gaspari}, M., {Gastaldello}, F.,
  {Georgakakis}, A., {Georgantopoulos}, I., {Gilfanov}, M., {Gitti}, M.,
  {Gladstone}, R., {Goosmann}, R., {Gosset}, E., {Grosso}, N., {Guedel}, M.,
  {Guerrero}, M., {Haberl}, F., {Hardcastle}, M., {Heinz}, S., {Alonso
  Herrero}, A., {Herv{\'e}}, A., {Holmstrom}, M., {Iwasawa}, K., {Jonker}, P.,
  {Kaastra}, J., {Kara}, E., {Karas}, V., {Kastner}, J., {King}, A., {Kosenko},
  D., {Koutroumpa}, D., {Kraft}, R., {Kreykenbohm}, I., {Lallement}, R.,
  {Lanzuisi}, G., {Lee}, J., {Lemoine-Goumard}, M., {Lobban}, A., {Lodato}, G.,
  {Lovisari}, L., {Lotti}, S., {McCharthy}, I., {McNamara}, B., {Maggio}, A.,
  {Maiolino}, R., {De Marco}, B., {de Martino}, D., {Mateos}, S., {Matt}, G.,
  {Maughan}, B., {Mazzotta}, P., {Mendez}, M., {Merloni}, A., {Micela}, G.,
  {Miceli}, M., {Mignani}, R., {Miller}, J., {Miniutti}, G., {Molendi}, S.,
  {Montez}, R., {Moretti}, A., {Motch}, C., {Naz{\'e}}, Y., {Nevalainen}, J.,
  {Nicastro}, F., {Nulsen}, P., {Ohashi}, T., {O'Brien}, P., {Osborne}, J.,
  {Oskinova}, L., {Pacaud}, F., {Paerels}, F., {Page}, M., {Papadakis}, I.,
  {Pareschi}, G., {Petre}, R., {Petrucci}, P.O., {Piconcelli}, E.,
  {Pillitteri}, I., {Pinto}, C., {de Plaa}, J., {Pointecouteau}, E., {Ponman},
  T., {Ponti}, G., {Porquet}, D., {Pounds}, K., {Pratt}, G., {Predehl}, P.,
  {Proga}, D., {Psaltis}, D., {Rafferty}, D., {Ramos-Ceja}, M., {Ranalli}, P.,
  {Rasia}, E., {Rau}, A., {Rauw}, G., {Rea}, N., {Read}, A., {Reeves}, J.,
  {Reiprich}, T., {Renaud}, M., {Reynolds}, C., {Risaliti}, G., {Rodriguez},
  J., {Rodriguez Hidalgo}, P., {Roncarelli}, M., {Rosario}, D., {Rossetti}, M.,
  {Rozanska}, A., {Rovilos}, E., {Salvaterra}, R., {Salvato}, M., {Di Salvo},
  T., {Sanders}, J., {Sanz-Forcada}, J., {Schawinski}, K., {Schaye}, J.,
  {Schwope}, A., {Sciortino}, S., {Severgnini}, P., {Shankar}, F., {Sijacki},
  D., {Sim}, S., {Schmid}, C., {Smith}, R., {Steiner}, A., {Stelzer}, B.,
  {Stewart}, G., {Strohmayer}, T., {Str{\"u}der}, L., {Sun}, M., {Takei}, Y.,
  {Tatischeff}, V., {Tiengo}, A., {Tombesi}, F., {Trinchieri}, G., {Tsuru},
  T.G., {Ud-Doula}, A., {Ursino}, E., {Valencic}, L., {Vanzella}, E.,
  {Vaughan}, S., {Vignali}, C., {Vink}, J., {Vito}, F., {Volonteri}, M.,
  {Wang}, D., {Webb}, N., {Willingale}, R., {Wilms}, J., {Wise}, M., {Worrall},
  D., {Young}, A., {Zampieri}, L., {In't Zand}, J., {Zane}, S., {Zezas}, A.,
  {Zhang}, Y., {Zhuravleva}, I.: {The Hot and Energetic Universe: A White Paper
  presenting the science theme motivating the Athena+ mission}.
\newblock arXiv e-prints arXiv:1306.2307 (2013)

\bibitem{Okajima:2016ts}
{Okajima}, T., {Soong}, Y., {Serlemitsos}, P., {Mori}, H., {Olsen}, L.,
  {Robinson}, D., {Koenecke}, R., {Chang}, B., {Hahne}, D., {Iizuka}, R.,
  {Ishida}, M., {Maeda}, Y., {Sato}, T., {Kikuchi}, N., {Kurashima}, S.,
  {Nakaniwa}, N., {Hayashi}, T., {Ishibashi}, K., {Miyazawa}, T., {Tachibana},
  K., {Tamura}, K., {Furuzawa}, A., {Tawara}, Y., {Sugita}, S.: {First peek of
  ASTRO-H Soft X-ray Telescope (SXT) in-orbit performance}.
\newblock In: J.W.A. {den Herder}, T.~{Takahashi}, M.~{Bautz} (eds.) Space
  Telescopes and Instrumentation 2016: Ultraviolet to Gamma Ray, \emph{Society
  of Photo-Optical Instrumentation Engineers (SPIE) Conference Series}, vol.
  9905, p. 99050Z (2016).
\newblock \doi{10.1117/12.2231705}

\bibitem{X-IFU_pajot}
{Pajot}, F., {Barret}, D., {Lam-Trong}, T., {den Herder}, J.W., {Piro}, L.,
  {Cappi}, M., {Huovelin}, J., {Kelley}, R., {Mas-Hesse}, J.M., {Mitsuda}, K.,
  {Paltani}, S., {Rauw}, G., {Rozanska}, A., {Wilms}, J., {Barbera}, M.,
  {Douchin}, F., {Geoffray}, H., {den Hartog}, R., {Kilbourne}, C., {Le Du},
  M., {Macculi}, C., {Mesnager}, J.M., {Peille}, P.: {The Athena X-ray Integral
  Field Unit (X-IFU)}.
\newblock Journal of Low Temperature Physics \textbf{193}(5-6), 901--907
  (2018).
\newblock \doi{10.1007/s10909-018-1904-5}

\bibitem{2019SPIE11180E..25P}
{Pareschi}, G., {Moretti}, A., {Salmaso}, B., {Sironi}, G., {Tagliaferri}, G.,
  {Uslenghi}, M., {Fiorini}, M., {Attin{\`a}}, P., {Bressan}, R., {Marchiori},
  G., {Tordi}, M., {Marioni}, F., {Valsecchi}, G., {Zocchi}, F.: {A vertical
  facility based on raster scan configuration for the x-ray scientific
  calibrations of the ATHENA optics}.
\newblock In: International Conference on Space Optics \&mdash; ICSO 2018,
  \emph{Society of Photo-Optical Instrumentation Engineers (SPIE) Conference
  Series}, vol. 11180, p. 1118025 (2019).
\newblock \doi{10.1117/12.2535996}

\bibitem{saha_equal-curvature_2003}
Saha, T.T., Zhang, W.: Equal-curvature grazing-incidence x-ray telescopes.
\newblock Applied Optics \textbf{42}(22), 4599 (2003).
\newblock \doi{10.1364/AO.42.004599}

\bibitem{Serlemitsos:1995ti}
{Serlemitsos}, P.J., {Jalota}, L., {Soong}, Y., {Kunieda}, H., {Tawara}, Y.,
  {Tsusaka}, Y., {Suzuki}, H., {Sakima}, Y., {Yamazaki}, T., {Yoshioka}, H.,
  {Furuzawa}, A., {Yamashita}, K., {Awaki}, H., {Itoh}, M., {Ogasaka}, Y.,
  {Honda}, H., {Uchibori}, Y.: {The X-Ray Telescope on board ASCA}.
\newblock \pasj \textbf{47}, 105--114 (1995)

\bibitem{Shimbo1986}
{Shimbo}, M., {Furukawa}, K., {Fukuda}, K., {Tanzawa}, K.: {Silicon-to-silicon
  direct bonding method}.
\newblock Journal of Applied Physics \textbf{60}(8), 2987--2989 (1986).
\newblock \doi{10.1063/1.337750}

\bibitem{Randall_SPIE}
{Smith}, R.K.: {The Arcus soft x-ray grating spectrometer explorer}.
\newblock In: Society of Photo-Optical Instrumentation Engineers (SPIE)
  Conference Series, \emph{Society of Photo-Optical Instrumentation Engineers
  (SPIE) Conference Series}, vol. 11444, p. 114442C (2020).
\newblock \doi{10.1117/12.2576047}

\bibitem{10.1117/12.2562430}
{Svendsen}, S., {Ferreira}, D.D.M., {Massahi}, S., {Jafari}, A., {Gellert},
  N.C., {Christensen}, F.E., {Henriksen}, P.L., {Vu}, L.M., {Jegers}, A.S.,
  {Shortt}, B., {Landgraf}, B., {Girou}, D.A., {Collon}, M.J., {Cibik}, L.,
  {Handick}, E., {Krumrey}, M.: {Status of the Ir and Ir/SiC coating
  development for the Athena optics}.
\newblock In: Society of Photo-Optical Instrumentation Engineers (SPIE)
  Conference Series, \emph{Society of Photo-Optical Instrumentation Engineers
  (SPIE) Conference Series}, vol. 11444, p. 114444K (2020).
\newblock \doi{10.1117/12.2562430}

\bibitem{Truemper:1982uu}
{Truemper}, J.: {The ROSAT mission}.
\newblock Advances in Space Research \textbf{2}(4), 241--249 (1982).
\newblock \doi{10.1016/0273-1177(82)90070-9}

\bibitem{Vacanti:2017wc}
{Vacanti}, G., {Barri{\`e}re}, N., {Bavdaz}, M., {Chatbi}, A., {Collon}, M.,
  {Dekker}, D., {Girou}, D., {G{\"u}nther}, R., {van der Hoeven}, R.,
  {Krumrey}, M., {Landgraf}, B., {M{\"u}ller}, P., {Schreiber}, S., {Vervest},
  M., {Wille}, E.: {Measuring silicon pore optics}.
\newblock In: Society of Photo-Optical Instrumentation Engineers (SPIE)
  Conference Series, \emph{Society of Photo-Optical Instrumentation Engineers
  (SPIE) Conference Series}, vol. 10399, p. 103990N (2017).
\newblock \doi{10.1117/12.2274357}

\bibitem{Vacanti:2021wh}
{Vacanti}, G., {Barri{\`e}re}, N., {Collon}, M.J.: {On the optical design of a
  large X-ray mirror based on silicon pore optics}.
\newblock In: Society of Photo-Optical Instrumentation Engineers (SPIE)
  Conference Series, \emph{Society of Photo-Optical Instrumentation Engineers
  (SPIE) Conference Series}, vol. 11822, p. 118220E (2021).
\newblock \doi{10.1117/12.2594228}

\bibitem{Vacanti:2019wr}
{Vacanti}, G., {Barri{\`e}re}, N.M., {Collon}, M.J., {Hauser}, E., {Babi{\'c}},
  L., {Bayerle}, A., {Girou}, D., {G{\"u}nther}, R., {Keek}, L., {Landgraf},
  B., {Okma}, B., {Verhoeckx}, S., {Vervest}, M., {Voruz}, L., {Bavdaz}, M.,
  {Wille}, E., {Krumrey}, M., {M{\"u}ller}, P., {Handick}, E.: {X-ray testing
  of silicon pore optics}.
\newblock In: Optics for EUV, X-Ray, and Gamma-Ray Astronomy IX, \emph{Society
  of Photo-Optical Instrumentation Engineers (SPIE) Conference Series}, vol.
  11119, p. 111190I (2019).
\newblock \doi{10.1117/12.2530977}

\bibitem{vacanti_2022}
Vacanti, G., et~al.: {The} optical design of x-ray mirrors based on {Silicon}
  {Pore} {Optics}  (2022)

\bibitem{10.1117/12.2272997}
{Valsecchi}, G., {Marioni}, F., {Bianucci}, G., {Zocchi}, F.E., {Gallieni}, D.,
  {Parodi}, G., {Ottolini}, M., {Collon}, M., {Civitani}, M., {Pareschi}, G.,
  {Spiga}, D., {Bavdaz}, M., {Wille}, E.: {Optical integration of SPO mirror
  modules in the ATHENA telescope}.
\newblock In: Society of Photo-Optical Instrumentation Engineers (SPIE)
  Conference Series, \emph{Society of Photo-Optical Instrumentation Engineers
  (SPIE) Conference Series}, vol. 10399, p. 103990E (2017).
\newblock \doi{10.1117/12.2272997}

\bibitem{van-Speybroeck:1979wp}
{van Speybroeck}, L.P.: {Einstein Observatory /HEAO-B/ mirror design and
  performance}.
\newblock In: M.~{Weisskopf} (ed.) Space optics: Imaging X-ray optics workshop,
  \emph{Society of Photo-Optical Instrumentation Engineers (SPIE) Conference
  Series}, vol. 184, pp. 2--11 (1979).
\newblock \doi{10.1117/12.957428}

\bibitem{vanspeybroeck_design_1972}
VanSpeybroeck, L.P., Chase, R.C.: Design {Parameters} of
  {Paraboloid}-{Hyperboloid} {Telescopes} for {X}-ray {Astronomy}.
\newblock Applied Optics \textbf{11}(2), 440 (1972).
\newblock \doi{10.1364/AO.11.000440}

\bibitem{Weisskopf:2002wb}
{Weisskopf}, M.C., {Brinkman}, B., {Canizares}, C., {Garmire}, G., {Murray},
  S., {Van Speybroeck}, L.P.: {An Overview of the Performance and Scientific
  Results from the Chandra X-Ray Observatory}.
\newblock \pasp \textbf{114}(791), 1--24 (2002).
\newblock \doi{10.1086/338108}

\bibitem{10.1117/12.2530682}
{Willingale}, R.: {Stray x-ray flux in the Athena Mirror}.
\newblock In: Optics for EUV, X-Ray, and Gamma-Ray Astronomy IX, \emph{Society
  of Photo-Optical Instrumentation Engineers (SPIE) Conference Series}, vol.
  11119, p. 111190Q (2019).
\newblock \doi{10.1117/12.2530682}

\bibitem{willingale_improving_2010}
Willingale, R., Spaan, F.H.P.: Improving the angular resolution of the conical
  {Wolter}-{I} silicon pore optics ({SPO}) mirror design for the
  {International} {X}-ray {Observatory} ({IXO}).
\newblock p. 773241. San Diego, California, USA (2010).
\newblock \doi{10.1117/12.857027}

\bibitem{Wolter:1952vf}
{Wolter}, H.: {Spiegelsysteme streifenden Einfalls als abbildende Optiken
  f{\"u}r R{\"o}ntgenstrahlen}.
\newblock Annalen der Physik \textbf{445}(1), 94--114 (1952).
\newblock \doi{10.1002/andp.19524450108}

\end{thebibliography}
\bibliographystyle{spmpsci.bst}

\end{document}